\documentstyle[12pt]{article}
\def\Journal#1#2#3#4{{#1} {\bf #2}, #3 (#4)}

\def\NCA{\em Nuovo Cimento}
\def\CR{\em C.R. Acad. Sci. (Paris)}
\def\CP{\em Cahiers de Physique}
\def\PRL{\em Phys. Rev. Lett.}
\def\JMP{\em J. Math. Phys.}
\def\GRG{\em Gen. Rel. Grav.}
\def\CQG{\em Class. Quantum Grav.}
\def\PRD{{\em Phys. Rev.} D}
\def\APP{{\em Acta. Phys. Pol.} B}

\newcommand{\bm}[1]{\mbox{\boldmath $#1$}}
\def\be{\begin{equation}}
\def\ee{\end{equation}}
\def\bea{\begin{eqnarray}}
\def\eea{\end{eqnarray}}
\def\b*{\begin{eqnarray*}}
\def\e*{\end{eqnarray*}}
\def\N{\hfill \rule{2.5mm}{2.5mm}}
\def\R{{\rm I\!R}}
\def\U{\Upsilon}
\def\g{\gamma}
\def\z{\zeta}
\def\S{\Sigma}
\def\p{\phi}
\def\P{{\bf Proof:} \hspace{3mm}}
\font\eightmsb=msbm10 scaled 1200
\def\bbb#1{\hbox{\eightmsb#1}}

\newtheorem{defi}{Definition}[section]
\newtheorem{theo}{Theorem}[section]
\newtheorem{coro}{Corollary}[section]
\newtheorem{prop}{Proposition}[section]
\newtheorem{lem}{Lemma}[section]

\newtheorem{pr}{Property}[section]

\begin{document}

\title{Super-energy tensors}

\author{Jos\'{e} M.M. Senovilla\\
Departamento de F\'{\i}sica Te\'orica, Universidad del Pa\'{\i}s Vasco,\\
Apartado 644, 48080 Bilbao, Spain.\\ and \\
Departament de F\'{\i}sica Fonamental, Universitat de Barcelona,\\
Diagonal 647, 08028 Barcelona, Spain.\\
e-mail: seno@ffn.ub.es
}


\maketitle

\abstract{A simple and purely algebraic construction of super-energy
(s-e) tensors for {\it arbitrary} fields is presented in any dimensions.
These tensors have good {\it mathematical} and {\it physical} properties, and
they can be used in any theory having as basic arena an $n$-dimensional manifold
with a metric of Lorentzian signature.

In general, the completely timelike component of these s-e tensors has the
mathematical features of an energy density: they are positive
definite and satisfy the {\it dominant property}, which provides s-e 
estimates useful for global results and helpful in other matters, such as the
causal propagation of the fields. Similarly, there appear super-momentum
vectors with the mathematical properties of s-e flux vectors.

The classical Bel and Bel-Robinson tensors for the gravitational fields are
included in our general definition. The energy-momentum and super-energy tensors
of physical fields are also obtained, and the procedure will be illustrated by
writing down these tensors explicitly for the cases of scalar, electromagnetic,
and Proca fields. Moreover, higher order (super)$^k$-energy
tensors are defined and shown to be meaningful and in agreement for the
different physical fields. In flat spacetimes, they provide infinitely many
conserved quantities. 

In non-flat spacetimes, the fundamental question of the {\it interchange} of
s-e quantities between different fields is addressed, and answered
affirmatively. Conserved s-e currents are found for any minimally
coupled scalar field whenever there is a Killing vector. Furthermore, the
exchange of gravitational and electromagnetic super-energy is also shown by
studying the propagation of discontinuities. This seems to open the door for
new types of conservation physical laws.
}


\newpage
\section{Introduction}
One of the consequences of the geometrization of gravity is
the lack of a well-posed definition of {\it local energy-momentum tensor}
of the gravitational field. This is sensible from the physical
point of view because the geometrization follows naturally from
Einstein's Principle of Equivalence \cite{MTW}, which is an essential
ingredient in most theories that incorporate the gravitational field.
But due to the Principle of Equivalence, one can always choose a reference
system along any timelike curve such that the local
gravitational field vanishes on the curve, and thus the gravitational
energy density must also vanish there. This is usually referred to as the
{\it non-localizability} of the gravitational energy \cite{MTW}. Therefore,
any theory of the gravitational field in accordance with the Pinciple of
Equivalence cannot include the concept of gravitational energy-momentum
{\it tensor}. 

Nevertheless, there are {\it local} tensors describing the strength of the
gravitational field. For instance, in the prominent
theory of General Relativity, Bel \cite{B1,Beltesis,B2} and independently
Robinson (see for instance the recent Ref.\cite{Rob}) constructed a four-index
tensor for the gravitational field in vacuum by exploiting the mathematical
analogy of this field with the electromagnetic one. The properties of the
now famous Bel-Robinson tensor are similar to that of traditional
energy-momentum tensors: it possesses a positive-definite timelike component
and a causal `momentum' vector; its divergence vanishes (in vacuum); the tensor
is zero if and only if the curvature of the spacetime vanishes, and some others.
The appearance of the Bel-Robinson tensor
seemed more an embarrassment than an achievement, though, and has remained as a
mistery over the years. It ressembles energy-momentum tensors, yet it is not
such a thing (it cannot be, as explained in the first paragraph). It has
four indices, instead of the usual pair. It
looks related {\it somehow} to the energy-momentum properties of the
the gravitational field, but its physical dimensions
are not those of an energy density \cite{B1,Beltesis,B2,BS2,Ro}, and so on
\footnote{Due to all these reasons, Bel \cite{B1,B4,Beltesis,B2} introduced the
name {\it super-energy} to refer to the `enigmatic' quantities described by his
tensor, but wishing to keep some grammatical link with `energy'.}.
Actually, as first proved by Horowitz and Schmidt \cite{HS}, the Bel-Robinson
s-e appears at the first relevant order in the definition of gravitational
quasi-local masses in vacuum. This has been later confirmed many times for the
various definitions of quasi-local masses, see
\cite{Ber0,BerL,BY,BLY,Dou,DM,HS,K,LV,Sza,Sza4}.

In spite of all the interpretational problems of the Bel-Robinson tensor, it is
a fact that it has been succesfully used in many different
applications\footnote{Deser has applied the epithets
{\it ubiquitous} \cite{Des} and {\it immortal} \cite{Des2} to the Bel-Robinson
tensor.}, see
for a selection \cite{calidad,BS2,BFI,Le,L,Lord,MB,MCQ,MTW,PR,Ro,Wald,Z}. 
Of at least similar importance is that the Bel-Robinson tensor arises as a
relevant tool in many mathematical formalisms involving the gravitational field.
Outstanding cases are: the hyperbolic formulations \cite{F,Reu} of the Einstein
field equations \cite{ACY,Bo}, the causal propagation of gravity
\cite{ACY,BerS,BS}, the existence of global solutions of the Cauchy problem
\cite{CY,HE,Ren}, and the study of the global stability of spacetimes
\cite{CK,Ren}. Furthermore, the Bel-Robinson tensor (or some of its
generalizations) appears as a pertinent object deserving attention in many other
theories of the gravitational field, such as supergravity, string theory or the
now fashionable M-theory \cite{Gib}, as has been demonstrated explicitly 
several times, see e.g \cite{Des,Des2,DK,DKS,DY} and references
therein. All in all, it seems inevitable to ask the question of whether or not
the Bel-Robinson tensor and its applications can be translated to more general
fields and theories.

Generalizations of the Bel-Robinson tensor to non-vacuum cases
in General Relativity and to other physical fields have been seeked several
times. To start with, Bel himself provided the generalization to a
non-vacuum spacetime in General Relativity \cite{B4,Beltesis}. This Bel tensor
keeps most of the previous good properties, but it has non-zero divergence in
general \cite{B4,Beltesis,BS2,S,S2}. This is natural if one does believe in the
physical reality of the s-e concept, because the existence of matter currents
in the spacetime implies that other types of super-energies are present. 
Despite this, some other generalizations trying to keep the divergence-free
property were built \cite{Sa,Col,BL}, but they lacked all other good
properties of the Bel and Bel-Robinson tensors, and thus they have been
hardly useful. In fact, as I shall discuss in section \ref{sec:cons}, not 
even their divergence-free property has physical relevance, because
it does not allow to construct any conserved current or quantity. A more
fruitful line was launched by Chevreton \cite{C}, who constructed a s-e
four-index tensor for the electromagnetic field, showing in particular that in
Special Relativity it had all the good properties of the Bel-Robinson tensor,
including the divergence-free one. The importance of Chevreton's idea is that
he tried to carry the s-e concepts to other fields (see also \cite{Kom});
very recently this has also been done for the scalar field in Special
Relativity by Teyssandier \cite{Tey} with interesting results.
Yet another type of Bel-Robinson generalization was put forward by \"Oktem
\cite{Ok}, and a different approach has been pursued by Garecki
\cite{Ga,Ga2,Ga3,Ga4,Ga5,Ga6}, see \cite{Ga7} for a recent compilation,
although their results were only partly useful. Recently, Robinson himself
\cite{Rob} has given some clues of how to generalize the Bel tensor.
Nevertheless, all these results are only of partial relevance and of limited
application, and have not answered the question of how to define s-e tensors in
{\it complete} generality with the required mathematical and physical properties.

One of the main purposes of this paper is to prove that the concept of
super-energy tensor is not exclusive of the gravitational field and that it can
be associated in a unique and precise way to any physical field. To that end,
I put forward what seems to be the correct and {\it universal}
generalization of the traditional s-e tensors. The only pre-requisite is
an $n$-dimensional differentiable manifold with a metric of Lorentzian
signature, and thus it is valid for almost all known theories (including or not
the gravitational field), such as supergravity, string theory, M-theory,
Kaluza-Klein and gauge theories, higher order lagrangians, Brans-Dicke and
scalar-tensor theories, dilaton gravity, General and Special Relativity, and
many others. First of all, I give a fully
general algorithm to decompose {\it any} given tensor field into its `electric'
and `magnetic' parts with respect to any observer (section \ref{sec:EH}). Then,
I present a new and general definition for the basic s-e
tensor of any tensor field (section \ref{sec:set}, Definition
\ref{def:set}). This definition is very simple and purely algebraic and keeps
all typical algebraic good properties of the Bel-Robinson tensor. In particular,
the completely timelike component can be seen as the timelike component of a
s-e flux vector which is causal, and it is positive definite, as
explicitly shown in section \ref{sec:set}. In fact, this component is shown to
be the sum of the squares of all the electric and magnetic parts of the tensor.

A specially remarkable property of the Bel-Robinson tensor that has played an
important role in some of the above-mentioned applications
\cite{Ber,BerS,Bo,BS,CK,PR,S,S2} is that a positive
(resp.\ non-negative) number is obtained when contracting the Bel-Robinson
tensor with any four future-pointing timelike (resp.\ causal) vectors
\cite{PR,BS,Ber,S,S2}. This property will be called the {\it dominant} property
\cite{BS,S,S2} in analogy with the dominant energy condition for the
energy-momentum tensor \cite{HE}. 
In my opinion, this is one of the important requisites to be kept by any good
definition of s-e tensor, even more important than the formulas related to the
divergence of the tensor. One of the main results in this paper is the proof 
that Definition \ref{def:set} achieves this goal, for the
{\it dominant s-e property} is shown to hold in full generality in
section \ref{sec:dsep}. This is very important because the cited results and
applications can be then translated to general fields by using the s-e tensors
of Definition \ref{def:set}. As a relevant example, a recent theorem \cite{BerS}
in which the dominant property of the s-e tensors is essential and that proves
the causal propagation of the gravitational and many other fields in
general spacetimes is included here. 

All the previous properties are also shown to hold for a more general family of
s-e tensors obtained by permutting indices in the basic s-e tensor. This is
explained in section \ref{sec:general}. However, as
I will argue in this paper, only the completely symmetric part of these tensors
(which is unique) plays a role in the physical applications. This can be
seen as indirect proof of the uniqueness of the s-e tensors herein presented.

The physical applications are left for the next sections. In section
\ref{sec:physics}, the Definition \ref{def:set} is used to build up and study
in some depth the s-e tensors of the gravitational, electromagnetic, Proca, and
scalar (with or without mass) fields in general theories. I shall show that
there exists, for any field, a hierarchy of infinitely many
`(super)$^k$-energy' tensors involving the field and its derivatives up to a
given order related to $k$. The first level in the list depends on the
physical characteristics of the particular field so that, for typical fields
the energy-momentum tensor arises naturally as a first step in the hierarchy,
but for the gravitational field the first natural step is at the super-energy
level. This hierarchy can be put in correspondance for the different fields,
and moreover it provides an infinite set of conserved quantities in flat
spacetimes (such as in Special Relativity).

Finally, the fundamental question of whether or not the s-e quantities
can be transferred between different physical fields (such as energy does)
is considered. The results obtained are very promising and illuminating.
In particular, I prove that one can always construct divergence-free
currents in any spacetime with a minimally coupled scalar field and at least
one Killing vector (a symmetry). These are {\it mixed} currents, in the
sense that they involve the sum of the s-e tensors for the gravitational and
the scalar field, so that they prove the exchange of s-e quantities between
these two fields \cite{S2}. Furthermore, these currents reduce to the
corresponding generalized Bel-Robinson conserved currents in the absence of the
scalar field, and to the divergence-free currents of the scalar field
previously built in flat spacetimes if the gravitational field is switched off.
Another enlightening result is obtained by studying the propagation of mixed
electromagnetic-gravitational wave fronts, where appropriate
(super)$^k$-energy {\it mixed} quantities are shown to be conserved along the
wave-front if there are conformal symmetries. These conserved quantities
necessarily involve the gravitational {\it and} the electromagnetic
contributions, and thus they are also a proof of the s-e exchange and its
validity at the various hierarchical levels. All this opens the door to a new
type of conservation physical laws which deserve further investigation in the
different available physical-geometrical theories.

\section{Tensors as $r$-fold forms. Electric-magnetic parts}
\label{sec:EH}
Let $V_n$ be any differentiable $n$-dimensional manifold endowed with a metric
$g$ of Lorentzian signature (--,+,\dots,+). Indices in $V_n$ run from 0 to
$n-1$ and are denoted by Greek small letters. Thus, the components of $g$
in any given basis are written as $g_{\alpha\beta}$. ``Spatial'' indices will
also be needed, so that Latin small letters $i,j,\dots$ will run from 1 to
$n-1$. Similarly, capital letters $I,J,\dots$ stand for indices running from
2 to $n-1$. Boldface letters and arrowed symbols indicate exterior $p$-forms
and vectors, respectively.
The covariant derivative associated to $g$ is denoted by $\nabla$ and
its Riemann and Ricci tensors by $R_{\alpha\beta\lambda\mu}$ and
$R_{\mu\nu}\equiv R^{\rho}{}_{\mu\rho\nu}$, respectively (sign conventions as
in \cite{Ei}).
The canonical volume element $n$-form is denoted by $\bm{\eta}$, with components
$\eta_{\alpha_1\dots\alpha_n}$. Round and square brackets embracing any
number of indices will denote the usual symmetrization and antisymmetrization,
respectively, so that for instance $\eta_{\alpha_1\dots\alpha_n}=
\eta_{[\alpha_1\dots\alpha_n]}$, $R_{\mu\nu}=R_{(\mu\nu)}$.
Equalities by definition are denoted by
$\equiv$, and the end of a proof is signalled by \rule{2.5mm}{2.5mm}.
 
A basic operation in what follows is the standard Hodge dual, denoted by *,
and defined as follows. Let $A^{\{\Omega\}}{}_{\mu_1\dots\mu_p}=
A^{\{\Omega\}}{}_{[\mu_1\dots\mu_p]}$ denote any
tensor with an arbitrary number of indices schematically denoted by
$\{\Omega\}$, plus a set of $p\leq n$ {\it completely antisymmetrical} indices
$\mu_1\dots\mu_p$. Then, the dual of $A$ with respect to $\mu_1\dots\mu_p$
is defined and denoted by
\be
A^{\{\Omega\}}{}_{\stackrel{*}{\mu_{p+1}\dots\mu_n}}\equiv
\frac{1}{p!}\eta_{\mu_1\dots\mu_n}A^{\{\Omega\}}{}^{\mu_1\dots\mu_p}
\label{*}
\ee
where the * is placed over the indices onto which the dual operation acts.
Observe that $A^{\{\Omega\}}{}_{\stackrel{*}{\mu_{p+1}\dots\mu_n}}=
A^{\{\Omega\}}{}_{\stackrel{*}{[\mu_{p+1}\dots\mu_n}]}$. It is easy to
prove that
\b*
A^{\{\Omega\}}{}_{\stackrel{**}{\mu_1\dots\mu_p}}=
(-1)^{p(n-p)+1}\, A^{\{\Omega\}}{}_{\mu_1\dots\mu_p}
\e*
which can be written simply as $**\,=(-1)^{p(n-p)+1}$.
Consider now any tensor $A^{\{\Omega\}}{}_{\mu_1\dots\mu_p}{}_{\nu_1\dots\nu_q}$
which is completely antisymmetric in the sets $\mu_1\dots\mu_p$ and
$\nu_1\dots\nu_q$ and also has any number of other indices signalled by
$\{\Omega\}$. Then one can form the duals with respect to $\mu_1\dots\mu_p$, or
with respect to $\nu_1\dots\nu_q$, or to both. The fundamental identity
concerning the dual operation is given by
\bea
A^{\{\Omega\}}{}_{\stackrel{*}{\mu_{p+1}\dots
\mu_n}}{}^{\stackrel{*}{\nu_{q+1}\dots\nu_n}}+
\frac{1}{q!}\left(\begin{array}{c} n\\ p\end{array}\right)
A^{\{\Omega\}}{}^{[\nu_1\dots\nu_p}{}_{\nu_1\dots\nu_q}
\delta^{\nu_{p+1}\dots\nu_n]}_{\mu_{p+1}\dots\mu_n} = 0 \, , \label{**1}\\
A^{\{\Omega\}}{}_{\stackrel{*}{\mu_{p+1}\dots
\mu_n}}{}^{\stackrel{*}{\nu_{q+1}\dots\nu_n}}+
\frac{1}{p!}\left(\begin{array}{c} n\\ q\end{array}\right)
A^{\{\Omega\}}{}^{\mu_1\dots\mu_p}{}_{[\mu_1\dots\mu_q}
\delta^{\nu_{q+1}\dots\nu_n}_{\mu_{q+1}\dots\mu_n]} = 0 \, , \label{**2}
\eea
where $\delta^{\mu_1\dots\mu_s}_{\nu_1\dots\nu_s}$ denotes the Kronecker
symbol of order $s$, see Appendix. The above relation together with many
others derived from it which will be used later are proven in the Appendix.

Let us now show how to decompose any given
tensor into `electric' and `magnetic' parts.
Take any $m$-covariant tensor $t_{\mu_1\dots\mu_m}$ and consider
the set of indices which are antisymmetric with $\mu_1$, plus $\mu_1$ itself.
This set constitutes a block of (say) $n_1\leq n$ completely antisymmetric
indices, which will be denoted by $[n_1]$. Take then the next index in
$t_{\mu_1\dots\mu_m}$ which is not in $[n_1]$ and do the same, forming
thus a second block $[n_2]$ of $n_2\leq n$ completely antisymmetric indices.
Continue like this until all indices of $t_{\mu_1\dots\mu_m}$ belong to
one of these blocks. Then, we have separated the $m$ indices of
$t_{\mu_1\dots\mu_m}$ into (say) $r$ blocks, each containing $n_{\U}$
($\U=1,\dots,r$) completely antisymmetric indices, with $n_1+\dots +n_r =m$.
In this way, we can consider $t_{\mu_1\dots\mu_m}$ as an
{\it r-fold $(n_1,\dots,n_r)$-form} which will be denoted schematically
by $t_{[n_1],\dots,[n_r]}$ where $[n_{\U}]$ indicates the $\U$-th block with
$n_{\U}$ antisymmetrical indices. This can always be done because, even if
$t_{\mu_1\dots\mu_m}$ has no antisymmetries, it can be seen as an
$m$-fold (1,\dots,1)-form.
Several examples of $r$-fold forms are: the volume element
$\eta_{\mu_1\dots\mu_n}$ is a simple $n$-form;
$F_{\mu\nu}=F_{[\mu\nu]}$ is a simple 2-form, while $\nabla_{\rho}F_{\mu\nu}$
is a double (1,2)-form; the Riemann tensor is a double {\it symmetrical}
(2,2)-form (the pairs can be interchanged) and the Ricci tensor is a double
symmetrical (1,1)-form; a tensor such as $t_{\mu\nu\rho}=t_{(\mu\nu\rho)}$ is
a triple symmetrical (1,1,1)-form, etcetera.

At this stage, one can define all possible duals by using the * acting on each
of the $[n_{\U}]$ blocks, obtaining the following tensors (obvious notation):
\b*
t_{\stackrel{*}{[n-n_1]},\dots,[n_r]}, \dots ,\hspace{2mm}
t_{[n_1],\dots,\stackrel{*}{[n-n_r]}},\hspace{2mm}
t_{\stackrel{*}{[n-n_1]},\stackrel{*}{[n-n_2]},\dots,[n_r]}, \, \dots \dots,
\hspace{2mm}
t_{\stackrel{*}{[n-n_1]},\stackrel{*}{[n-n_2]},\dots,\stackrel{*}{[n-n_r]}} \, .
\e*
There are
$1+ \left(\begin{array}{c} r\\ 1\end{array}\right)+
\dots + \left(\begin{array}{c} r\\ r\end{array}\right)=2^r$
tensors in this set (including $t_{[n_1],\dots,[n_r]}$), each of which is
an $r$-fold form (except when $n_{\U}=n$ for some $\U$, but these special
cases will be treated as self-evident). All these tensors
allow to construct a canonical
``electric-magnetic'' decomposition of $t_{\mu_1\dots\mu_m}$ relative to
any observer. To see it, take any timelike unit vector $\vec{u}$,
$u_{\mu}u^{\mu}=-1$, and contract all the above duals with $r$ copies
of $\vec{u}$, each one with the first index
of each block. Whenever $\vec{u}$ is contracted with 
a `starred' block $\stackrel{*}{[n-n_{\U}]}$, we obtain a `magnetic part' in
that block, and an `electric part' otherwise. Thus, the electric-magnetic parts
are $r$-fold forms\footnote{Of course, if one of the blocks has $n_{\U}=1$,
then when concracting with $\vec{u}$ this block dissapears, so that the
electric parts of 1-blocks are scalars, and analogously for magnetic parts
corresponding to $(n-1)$-blocks. Then, some of the E-H parts may be
$\tilde{r}$-forms with $\tilde{r}<r$, and this is to be understood.}
which can be denoted by 
\b*
({}^t_{\vec{u}}\underbrace{EE\dots E}_r)_{[n_1-1],[n_2-1],\dots,[n_r-1]},\,\, 
({}^t_{\vec{u}}H\underbrace{E\dots E}_{r-1})_{[n-n_1-1],[n_2-1],\dots,[n_r-1]},\,\,
\dots \\
\dots \, ,\, 
({}^t_{\vec{u}}\underbrace{E\dots E}_{r-1}H)_{[n_1],\dots,[n_{r-1}-1][n-n_r-1]}\, ,\,\,
({}^t_{\vec{u}}HH\underbrace{E\dots E}_{r-2})_{[n-n_1-1],[n-n_2-1],\dots,[n_r-1]},\,\, 
\dots\, ,\\
({}^t_{\vec{u}}\underbrace{E\dots E}_{r-2}HH)_{[n_1-1],\dots,[n-n_{r-1}-1],[n-n_r-1]}\, ,
\, \dots ,
({}^t_{\vec{u}}\underbrace{HH\dots H}_r)_{[n-n_1-1],[n-n_2-1],\dots,[n-n_r-1]},
\e*
where, for instance,
\b*
({}^t_{\vec{u}}\underbrace{EE\dots E}_r)_{\mu_2\dots\mu_{n_1},\nu_2\dots\nu_{n_2},
\dots ,\rho_2\dots\rho_{n_r}}\equiv
\tilde{t}_{\mu_1\mu_2\dots\mu_{n_1},\nu_1\nu_2\dots\nu_{n_2},
\dots ,\rho_1\rho_2\dots\rho_{n_r}}u^{\mu_1}u^{\nu_1}\dots u^{\rho_1}\, ,\\
({}^t_{\vec{u}}H\underbrace{E\dots E}_{r-1})_{\mu_{n_1+2}\dots\mu_{n},\nu_2\dots\nu_{n_2},
\dots ,\rho_2\dots\rho_{n_r}}\!\equiv\! 
\tilde{t}_{\stackrel{*}{\mu_{n_1+1}\mu_{n_1+2}\dots\mu_{n}},
\nu_1\nu_2\dots\nu_{n_2},
\dots ,\rho_1\rho_2\dots\rho_{n_r}}u^{\mu_{n_1+1}}u^{\nu_1}\dots u^{\rho_1}
\e*
and so on. Here, $\tilde{t}_{[n_1],\dots,[n_r]}$ denotes the tensor obtained
from $t_{[n_1],\dots,[n_r]}$ by permutting the indices such that the
first $n_1$ indices are those precisely in the block $[n_1]$, the next
$n_2$ indices are those in the block $[n_2]$, and so on.
There are $2^r$ E-H parts, they are {\it spatial} relative to $\vec{u}$
in the sense that they are orthogonal to $\vec{u}$ in any index,
and all of them determine $t_{\mu_1\dots\mu_m}$ uniquely and completely.
Besides, $t_{\mu_1\dots\mu_m}$ vanishes iff all its E-H parts do.
The simplest way to see this is to take any orthonormal basis
$\{\vec{e}_{\mu}\}$ with $\vec{e}_0=\vec{u}$ and decompose all the
components of $t_{[n_1],\dots,[n_r]}$ in sets according to whether they have
a `zero' or not in each of the blocks (notice that blocks cannot have any
repeated index). This provides $2^r$ sets which altogether comprise obviously
all the components of $t_{\mu_1\dots\mu_m}$. Observe now that each of these
sets coincide with one of the E-H parts: blocks with one zero correspond to `E'
and those without zero to `H'.

Of course, this `E-H' decomposition agrees with the classical one
of any electromagnetic field $F_{\mu\nu}$ and its generalizations to the
Riemann and Weyl tensors \cite{XYZZ,Beltesis,B2,MB,Mat,MAWH}. To illustrate the
decomposition, take any simple
$p$-form $\S_{\mu_1\dots\mu_p}=\S_{[\mu_1\dots\mu_p]}$. In general, this
has $\left(\begin{array}{c} n\\ p\end{array}\right)$ independent components.
Given any unit timelike $\vec{u}$, the electric part of $\bm{\S}$ with respect
to $\vec{u}$ is the simple $(p-1)$-form
\b*
\left({}^{\S}_{\vec{u}}E\right)_{\mu_2\dots\mu_p}\equiv
\S_{\mu_1\mu_2\dots\mu_p}u^{\mu_1}, \hspace{5mm}
\left({}^{\S}_{\vec{u}}E\right)_{\mu_2\dots\mu_p}
=\left({}^{\S}_{\vec{u}}E\right)_{[\mu_2\dots\mu_p]},
\hspace{3mm} \left({}^{\S}_{\vec{u}}E\right)_{\mu_2\dots\mu_p}u^{\mu_2}=0.
\e*
Obviously, $\left({}^{\S}_{\vec{u}}\bm{E}\right)$ has
$\left(\begin{array}{c} n-1 \\ p-1 \end{array}\right)$ independent components,
which correspond to those with a `zero' among the components
$\S_{\mu_1\dots\mu_p}$ in any orthonormal basis
$\{\vec{e}_{\mu}\}$ with $\vec{e}_0=\vec{u}$, that is, to $\S_{0i_2\dots i_p}$
(remember that $i,j,\dots =1,\dots, n-1$). Similarly, the magnetic part of
$\bm{\S}$ with respect to $\vec{u}$ is the simple $(n-p-1)$-form
\b*
\left({}^{\S}_{\vec{u}}H\right)_{\mu_{2}\dots\mu_{n-p}}\equiv
\S_{\stackrel{*}{\mu_{1}\mu_{2}\dots\mu_{n-p}}}u^{\mu_{1}},
\left({}^{\S}_{\vec{u}}H\right)_{\mu_{2}\dots\mu_{n-p}}=
\left({}^{\S}_{\vec{u}}H\right)_{[\mu_{2}\dots\mu_{n-p}]},
\left({}^{\S}_{\vec{u}}H\right)_{\mu_{2}\dots\mu_{n-p}}u^{\mu_2}=0.
\e*
Now, $\left({}^{\S}_{\vec{u}}\bm{H}\right)$ has
$\left(\begin{array}{c} n-1 \\ n-p-1 \end{array}\right)=
\left(\begin{array}{c} n-1 \\ p \end{array}\right)$ independent components,
which correspond to those without `zeros' in the basis above, that is,
to $\S_{i_1\dots i_{p}}$. The sum of the components of the electric
part plus the components of the magnetic part gives
\b*
\left(\begin{array}{c} n-1 \\ p-1 \end{array}\right)+
\left(\begin{array}{c} n-1 \\ p \end{array}\right)=
\left(\begin{array}{c} n\\ p\end{array}\right)
\e*
which is the total number of components of $\bm{\S}$. The simple
$p$-form $\bm{\S}$ can be expressed, for any unit timelike $\vec{u}$, in terms
of its electric and magnetic parts as
\b*
\S_{\mu_1\dots\mu_p}=-p\, u_{[\mu_1}({}^{\S}_{\vec{u}}E)_{\mu_2\dots\mu_p]}+
\frac{(-1)^{p(n-p)}}{(n-p)!}\eta_{\rho_1\dots\rho_{n-p}\mu_1\dots\mu_p}
u^{\rho_1}\left({}^{\S}_{\vec{u}}H\right)^{\rho_{2}\dots\rho_{n-p}}\\
\S_{\stackrel{*}{\mu_{p+1}\mu_{p+2}\dots\mu_n}}=
-(n-p)u_{[\mu_{p+1}}({}^{\S}_{\vec{u}}H)_{\mu_{p+2}\dots\mu_n]}-\frac{1}{p!}
\eta_{\mu_1\dots\mu_n}u^{\mu_1}\left({}^{\S}_{\vec{u}}E
\right)^{\mu_2\dots\mu_p}
\e*
which can be written in a compact form as
\b*
\bm{\S}=-\bm{u}\wedge \left({}^{\S}_{\vec{u}}\bm{E}\right)+(-1)^{p(n-p)}
\stackrel{\hspace{-1mm}*}{\left[\bm{u}\wedge \left({}^{\S}_{\vec{u}}\bm{H}
\right)\right]},
\hspace{5mm}
\stackrel{*}{\bm{\S}}=-\bm{u}\wedge \left({}^{\S}_{\vec{u}}\bm{H}\right)-
\stackrel{*}{\left[\bm{u}\wedge \left({}^{\S}_{\vec{u}}\bm{E}\right)\right]}
\e*
where $\wedge$ denotes the exterior product. These formulae show once again
that the E-H parts of $\bm{\S}$ fully determine $\bm{\S}$, and that $\bm{\S}$
vanishes if and only if its E-H parts are zero with respect to some (and
then to all) $\vec{u}$. The invariants asociated with $\bm{\S}$ can also
be expressed in terms of its E-H parts. For instance, from formula (\ref{ap:pp})
in the Appendix one gets
\be
\frac{1}{p!}\S_{\mu_1\dots\mu_p}\S^{\mu_1\dots\mu_p}=
\frac{1}{(n-p-1)!}({}^{\S}_{\vec{u}}H)_{\mu_{p+2}\dots\mu_n}
({}^{\S}_{\vec{u}}H)^{\mu_{p+2}\dots\mu_n}-
\frac{1}{(p-1)!}({}^{\S}_{\vec{u}}E)_{\mu_2\dots\mu_p}
({}^{\S}_{\vec{u}}E)^{\mu_2\dots\mu_p} \label{inv}
\ee
The general decomposition for
arbitrary $r$-fold forms does all this on each $[n_{\U}]$-block.

For any spatial tensor $S_{\mu\nu\dots\rho}$ relative to $\vec{u}$ (that is,
orthogonal to $\vec{u}$ in every index), the following abbreviated notation
\b*
[S]^2\equiv S_{\mu\nu\dots\rho}S^{\mu\nu\dots\rho}>0
\e*
will be used. Notice that this is positive and vanishes if and only if the
whole tensor $S_{\mu\nu\dots\rho}$ is zero. In particular,
$[{}^t_{\vec{u}}\underbrace{EE\dots E}_r]^2$,
$[{}^t_{\vec{u}}H\underbrace{E\dots E}_{r-1}]^2$, etcetera
are all non-negative and any of them vanishes iff the corresponding E-H
part so does.

\section{Definition of the basic super-energy tensor}
\label{sec:set}
In this section I show how to construct a good s-e tensor for any given
tensor $t_{\mu_1\dots\mu_m}$. To that end, let us start by defining the
``semi-square'' $(t_{[n_1],\dots,[n_r]}\times t_{[n_1],\dots,[n_r]})$ of
any $r$-fold $(n_1,\dots,n_r)$-form by contracting all indices but one of
each block in the product of $\tilde{t}$ with itself, that is to say
\be
(t\times t)_{\lambda_1\mu_1\dots\lambda_r\mu_r}\equiv
\left(\prod_{\U=1}^{r}\frac{1}{(n_{\U}-1)!}\right)\,
\tilde{t}_{\lambda_1\rho_2\dots\rho_{n_1},\dots ,
\lambda_r\sigma_2\dots\sigma_{n_r}}
\tilde{t}_{\mu_1\hspace{10mm}\dots ,\mu_r}^{\hspace{2mm}\rho_2\dots\rho_{n_1}
,\hspace{5mm}\sigma_2\dots\sigma_{n_r}} \, .\label{semi-s}
\ee
This is a $2r$-covariant tensor. With this at hand, let us put forward the
following general definition \cite{S,S2}:
\begin{defi}
\label{def:set}
The {\em basic s-e tensor of $t$} is defined as half of the sum of the $2^r$
semi-squares constructed with $t_{[n_1],\dots,[n_r]}$ and all its duals.
Explicitly:
\bea
T_{\lambda_1\mu_1\dots\lambda_r\mu_r}\left\{t\right\}\equiv
\frac{1}{2}\left\{\begin{array}{l} \\ \\ \end{array}\hspace{-4mm}
\left(t_{[n_1],\dots,[n_r]}\times t_{[n_1],\dots,
[n_r]}\right)_{\lambda_1\mu_1\dots\lambda_r\mu_r}+\right. \nonumber \\
+\left(t_{\stackrel{*}{[n-n_1]},\dots,[n_r]}\times
t_{\stackrel{*}{[n-n_1]},\dots,[n_r]}\right)_{\lambda_1\mu_1\dots\lambda_r\mu_r}
+\dots +\nonumber \\
+\dots + \left(t_{[n_1],\dots,\stackrel{*}{[n-n_r]}}\times
t_{[n_1],\dots,\stackrel{*}{[n-n_r]}}\right)_{\lambda_1\mu_1\dots\lambda_r\mu_r}
+ \dots\nonumber \\
+\dots +\left(t_{\stackrel{*}{[n-n_1]},\stackrel{*}{[n-n_2]},\dots,[n_r]}\times
t_{\stackrel{*}{[n-n_1]},\stackrel{*}{[n-n_2]},\dots,[n_r]}
\right)_{\lambda_1\mu_1\dots\lambda_r\mu_r}+\dots +\nonumber\\
\left. +\dots \, \dots +
\left(t_{\stackrel{*}{[n-n_1]},\dots,\stackrel{*}{[n-n_r]}}\times
t_{\stackrel{*}{[n-n_1]},\dots,
\stackrel{*}{[n-n_r]}}\right)_{\lambda_1\mu_1\dots\lambda_r\mu_r}\right\}\, .
\label{set}
\eea
\end{defi}
As we will see, this tensor coincides, sometimes, with the energy-momentum
tensor of some physical fields, and thus the generic name `super-energy'
tensor may be inappropriate. However, in a majority of cases the
tensor will be related to higher order `energies'
(see section \ref{sec:physics}),
so that it is preferable to keep the `s-e' adjective. Concerning the
adjective `basic', this is due to two main reasons: first, for massive physical
fields the {\it physical} s-e tensor will be a combination of several `basic'
s-e tensors (see subsection \ref{subsec:massive}); second, in general a linear
combination of basic s-e tensors permutting the indices with positive
constants will also be a good mathematical s-e tensor. This point
will be treated is section \ref{sec:general}.

Notice that any dual of $t_{[n_1],\dots,[n_r]}$ gives rise to the same basic
s-e tensor (\ref{set}). Therefore, one only needs to consider blocks with
at most $n/2$ indices if $n$ is even, or $(n+1)/2$ if $n$ is odd. This will
be implicitly assumed sometimes in what follows. Observe that
if there is any block $[n_{\U}]$ with $n_{\U}=n$, then this block
`disappears' giving rise to `0-blocks' after dualizing. 

The properties of (\ref{set}) are the following.
\begin{pr}
The tensor (\ref{set}) is symmetric on each $(\lambda_{\U}\mu_{\U})$-pair,
that is,
\b*
T_{\lambda_1\mu_1\dots\lambda_r\mu_r}\left\{t\right\}=
T_{(\lambda_1\mu_1)\dots(\lambda_r\mu_r)}\left\{t\right\}
\e*
\label{pr:sym}
\end{pr}
\P
Notice that all the terms in the definition (\ref{set}) can be separated
in two groups, one of them without duals in the block $[n_{\U}]$ and the
other with duals in that block. Now, due to formula (\ref{ap:sym}) in the
Appendix, the sum of these two groups is symmetric in the corresponding
$(\lambda_{\U}\mu_{\U})$-pair. As this can be done for any block the
result follows.\N
\begin{pr}
If the tensor $t_{[n_1],\dots,[n_r]}$ is symmetric in the interchange of the
block $[n_{\U}]$ with the block $[n_{\U'}]$ ($n_{\U}=n_{\U'}$), then
the s-e tensor (\ref{set}) is symmetric in the interchange of the
corresponding $(\lambda_{\U}\mu_{\U})$- and $(\lambda_{\U'}\mu_{\U'})$-pairs.
\label{pr:sympair}
\end{pr}
\P
Assume that $t_{\dots,[n_{\U}],\dots,[n_{\U'}],\dots}=
t_{\dots,[n_{\U'}],\dots,[n_{\U}],\dots}$ with $n_{\U}=n_{\U'}$. Then it
is obvious that
\b*
t_{\dots,\stackrel{*}{[n-n_{\U}]},\stackrel{*}{\dots,[n-n_{\U'}]},\dots}=
t_{\dots,\stackrel{*}{[n-n_{\U'}]},\stackrel{*}{\dots,[n-n_{\U}]},\dots},
\hspace{5mm}
t_{\dots,[n_{\U}],\stackrel{*}{\dots,[n-n_{\U'}]},\dots}=
t_{\dots,\stackrel{*}{[n-n_{\U'}]},\dots,[n_{\U}],\dots}
\e*
and thus, from the explicit expression (\ref{set}) it follows immediately
that
\b*
T_{\dots\lambda_{\U}\mu_{\U}\dots\lambda_{\U'}\mu_{\U'}\dots}\left\{t\right\}=
T_{\dots\lambda_{\U'}\mu_{\U'}\dots\lambda_{\U}\mu_{\U}\dots}\left\{t\right\}
\e*
\N
\begin{pr}
If $n$ is even, then the s-e tensor (\ref{set}) is traceless in any
$(\lambda_{\U}\mu_{\U})$-pair with $n_{\U}=n/2$.
\end{pr}
\P
Assume that $n=2n_{\U}$ for some $\U$. Then, as in the proof of Property
\ref{pr:sym}, we can separate all the terms in (\ref{set}) into two groups,
one with $[n_{\U}]$-blocks and the other with $\stackrel{*}{[n-n_{\U}]}=
\stackrel{*}{[n_{\U}]}$-blocks. Then, application of formula (\ref{ap:tr}) in
the Appendix to the trace in the $(\lambda_{\U}\mu_{\U})$-pair leads
immediately to
\b*
T_{\dots}{}^{\mu_{{}_{\U}}}{}_{\mu_{{}_{\U}}\dots}\left\{t\right\}=0,
\hspace{3mm} \mbox{if} \hspace{2mm} n=2n_{\U}
\e*
\N

One of the good properties of the s-e tensor (\ref{set}) is that its
completely timelike component has the mathematical features of
an energy density, which can be also seen as the timelike component
of a vector with the typical properties of an $n$-momentum vector.
These good properties are going to be proved partly now,
and the rest will be shown in section \ref{sec:dsep}. However, the physical
dimensions of this timelike component will not be, in general, those of an
energy density. Thus, following Bel in his pionering work on s-e tensors
\cite{B1,B4,Beltesis,B2}, I shall generically use the term `super-energy'
for these concepts.
\begin{defi}
The {\em super-energy density} of the tensor $t$ relative to the timelike
vector $\vec{u}$ is
denoted by $W_t(\vec{u})$ and defined by
\be
W_t\left(\vec{u}\right)\equiv
T_{\lambda_1\mu_1\dots\lambda_r\mu_r}\{t\}
u^{\lambda_1}u^{\mu_1}\dots u^{\lambda_r}u^{\mu_r} \, .\label{sed}
\ee
\end{defi}
The basic properties of $W_t(\vec{u})$ are
\begin{pr}
For all timelike vectors $\vec{u}$, $W_t(\vec{u})$ is non-negative.
Furthermore,
\be
\left\{\exists \vec{u}\hspace{3mm} \mbox{{\em such that}} \hspace{2mm}
W_t\left(\vec{u}\right)=0 \right\} \Longleftrightarrow
T_{\lambda_1\mu_1\dots\lambda_r\mu_r}\{t\}=0 \Longleftrightarrow
t_{\mu_1\dots\mu_m}=0 .\label{cero}
\ee
\label{pr:1}
\end{pr}
\P
From (\ref{set}) and (\ref{sed}), together with the definition of the E-H
parts of $t$ we get
\bea
W_t\left(\vec{u}\right)\equiv \frac{1}{2}\left\{
\left(\prod_{\U=1}^{r}\frac{1}{(n_{\U}-1)!}\right)\,
[{}^t_{\vec{u}}\underbrace{EE\dots E}_r]^2 +\dots +\right.\nonumber\\
\left. + \dots +
\left(\prod_{\U=1}^{r}\frac{1}{(n-n_{\U}-1)!}\right)\,
[{}^t_{\vec{u}}\underbrace{HH\dots H}_{r}]^2 \right\} \geq 0 \label{sq}
\eea
from where the first part is obvious and one sees that $W_t(\vec{u})$ can
vanish for some $\vec{u}$ if and only if all the E-H parts of $t$ vanish.
But as we saw in the previous section, all the electric-magnetic parts of
$t$ vanish iff the tensor $t$ is zero, in which case the whole $T\{t\}$
also vanishes.\N
\begin{pr}
The s-e density is half the sum of the squares of all the components
of $t$ in any orthonormal basis $\{\vec{e}_{\mu}\}$ with $\vec{e}_0=\vec{u}$:
\b*
W_t(\vec{e}_0)=T_{0\dots 0}\{t\}=\frac{1}{2}\,\,\sum_{\mu_1,\dots,\mu_m=0}^{n-1}
|t_{\mu_1\dots\mu_m}|^2 \, .
\e*
\label{pr:1'}
\end{pr}
\P
It is enough to remember that the E-H decompostion is simply a decomposition
according to whether a block has a zero or not in any orthonormal basis
(see previous section), and then use formula (\ref{sq}).\N

In fact, there is another interesting formula for the super-energy density
which is very useful in some calculations, and allows to compare with other
similar positive quantities. Given any unit timelike $\vec{u}$, let us define
the `positive-definite metric' relative to $\vec{u}$ as
\be
h_{\mu\nu}\left(\vec{u}\right)\equiv g_{\mu\nu}+2u_{\mu}u_{\nu}. \label{h}
\ee
As this is positive definite, it allows to construct positive quantities
when contracting with tensors. Actually, we have
\begin{pr}
\label{square}
The s-e density of the tensor $t$ relative to $\vec{u}$ is
proportional to the `square' of $t$ with respect to the
positive-definite metric $h\!\left(\vec{u}\right)$ according to
\b*
W_{t}\left(\vec{u}\right)=\frac{1}{2}\!
\left(\prod_{\U=1}^{r}\frac{1}{n_{\U}!}\right)
\tilde{t}_{\mu_1\dots\mu_{n_1},\dots ,\rho_1\dots\rho_{n_r}}
\tilde{t}_{\nu_1\dots\nu_{n_1},\dots ,\sigma_1\dots\sigma_{n_r}}
h^{\mu_1\nu_1}\dots h^{\mu_{n_1}\nu_{n_1}}\dots
h^{\rho_1\sigma_1}\dots h^{\rho_{n_r}\sigma_{n_r}}.
\e*
\end{pr}
\P The case of simple $p$-forms $\bm{\S}$ is proved in Property \ref{square2}
below. The general case is left for Section \ref{sec:dsep}, see Corollary
\ref{square3}.\N

The s-e density can be seen as the timelike component of some
`momentum' vectors constructed from the s-e tensor. The precise definition is
\begin{defi}
The {\em super-energy flux vectors} of the tensor $t$ relative to the timelike
vector $\vec{u}$ are
denoted by ${}^{\U}\vec{P}_t(\vec{u})$ and defined by
\be
{}^{\U}P^{\nu}_t\left(\vec{u}\right)\equiv
-\, T_{\lambda_1\mu_1\dots\lambda_{\U-1}\mu_{\U-1}}{}^{\nu}{}_{\mu_{\U}
\dots\lambda_r\mu_r}\{t\}u^{\lambda_1}
u^{\mu_1}\dots u^{\lambda_{\U-1}}u^{\mu_{\U-1}}
u^{\mu_{\U}}\dots u^{\lambda_r}u^{\mu_r} \, .\label{sev}
\ee
\end {defi}
There are, in general, $r$ different s-e flux vectors, one for each
$(\lambda_{\U}\mu_{\U})$-pair. However, sometimes some of these vectors
may coincide with each other, such as for instance when the Property
\ref{pr:sympair} holds for some pairs, in which case the corresponding
s-e flux vectors are equal.
The s-e flux vectors can be decomposed with respect to a unit $\vec{u}$ into
their timelike component and the corresponding spatial part as
\b*
{}^{\U}P^{\nu}_t\left(\vec{u}\right)=W_t\left(\vec{u}\right)u^{\nu}+
\left(\delta^{\nu}_{\rho}+u^{\nu}u_{\rho}\right)\,
{}^{\U}P^{\rho}_t\left(\vec{u}\right).
\e*
\begin{pr}
${}^{\U}\vec{P}_t\left(\vec{u}\right)$ are causal vectors with the same time
orientation than $\vec{u}$.
\label{pr:2}
\end{pr}
\P
That ${}^{\U}\vec{P}_t\left(\vec{u}\right)$ is causal, that is, either
timelike or
null, is a particular case of a more general property of basic s-e
tensors to be proven in the next section, see Corollary \ref{Pt}. 
On the other hand, we have from (\ref{sev}) and Property \ref{pr:1}
\b*
{}^{\U}P^{\rho}_t\left(\vec{u}\right)u_{\rho}=-W_t\left(\vec{u}\right)<0
\e*
which proves that ${}^{\U}\vec{P}_t\left(\vec{u}\right)$ is future-pointing if
$\vec{u}$ is future-pointing.\N

Let us present some s-e tensors of type (\ref{set}) explicitly. Starting with
a scalar $f$ we simply have
\b*
T\{f\}=\frac{1}{2}f^2\, .
\e*
For any given simple $p$-form $\S_{\mu_1\dots\mu_p}=\S_{[\mu_1\dots\mu_p]}$,
definition (\ref{set}) produces
\be
T_{\lambda}{}^{\mu}\left\{\S_{[p]}\right\}=
\frac{1}{2}\left(\frac{1}{(p-1)!}\S_{\lambda\rho_2\dots\rho_p}
\S^{\mu\rho_2\dots\rho_p}+\frac{1}{(n-p-1)!}
\S_{\stackrel{*}{\lambda\rho_{p+2}\dots\rho_n}}
\S^{\stackrel{*}{\mu\rho_{p+2}\dots\rho_n}}\right) \label{set1}
\ee
and then using formula (\ref{ap:pp}) in the Appendix we equivalently obtain
\be
T_{\lambda\mu}\left\{\S_{[p]}\right\}=\frac{1}{(p-1)!}
\left(\S_{\lambda\rho_2\dots\rho_p}\S_{\mu}{}^{\rho_2\dots\rho_p}-
\frac{1}{2p}g_{\lambda\mu}
\S_{\rho_1\rho_2\dots\rho_p}\S^{\rho_1\rho_2\dots\rho_p}\right)
\label{set1'}
\ee
so that
\be
T_{\lambda\mu}\left\{\S_{[p]}\right\}=T_{\mu\lambda}\left\{\S_{[p]}\right\},
\hspace{3mm}
T^{\mu}{}_{\mu}\left\{\S_{[p]}\right\}=\left(p-\frac{n}{2}\right)
\frac{1}{p!}\S_{\rho_1\rho_2\dots\rho_p}\S^{\rho_1\rho_2\dots\rho_p}.
\ee
These explicit expressions allow to prove Property \ref{square}
for the case of simple $p$-forms.
\begin{pr}
\label{square2}
The s-e density of the simple p-form $\bm{\S}$ relative to $\vec{u}$ is
proportional to the `square' of $\bm{\S}$ with respect to the
positive-definite metric (\ref{h}), that is
\b*
W_{\S}\left(\vec{u}\right)=\frac{1}{2\, p!}
\S_{\mu_1\dots\mu_p}\S_{\nu_1\dots\nu_p}h^{\mu_1\nu_1}\dots h^{\mu_p\nu_p}.
\e*
\end{pr}
\P
From (\ref{set1'}) and the definition of
$\left({}^{\S}_{\vec{u}}\bm{E}\right)$ it follows
\b*
W_{\S}\left(\vec{u}\right)=\frac{1}{(p-1)!}
\left[\left({}^{\S}_{\vec{u}}E\right)_{\rho_2\dots\rho_p}
\left({}^{\S}_{\vec{u}}E\right)^{\rho_2\dots\rho_p}+
\frac{1}{2p}\S_{\rho_1\rho_2\dots\rho_p}\S^{\rho_1\rho_2\dots\rho_p}\right]
\e*
while using (\ref{h}) together with the properties that 
$\left({}^{\S}_{\vec{u}}\bm{E}\right)$ is spatial relative to $\vec{u}$ and
completely antisymmetric, one gets
\b*
\S_{\mu_1\dots\mu_p}\S_{\nu_1\dots\nu_p}
h^{\mu_1\nu_1}\dots h^{\mu_p\nu_p}=\hspace{3cm}\\
=\left[\S_{\mu_1\mu_2\dots\mu_p}\S^{\mu_1}{}_{\nu_2\dots\nu_p}+2
\left({}^{\S}_{\vec{u}}E\right)_{\mu_2\dots\mu_p}
\left({}^{\S}_{\vec{u}}E\right)_{\nu_2\dots\nu_p}\right]
h^{\mu_2\nu_2}\dots h^{\mu_p\nu_p}=\\
=\left[\S_{\mu_1\mu_2\dots\mu_p}\S^{\mu_1\mu_2}{}_{\nu_3\dots\nu_p}+4
\left({}^{\S}_{\vec{u}}E\right)_{\mu_2\dots\mu_p}
\left({}^{\S}_{\vec{u}}E\right)^{\mu_2}{}_{\nu_3\dots\nu_p}\right]
h^{\mu_3\nu_3}\dots h^{\mu_p\nu_p}=\dots =\\
=\dots\, \dots =
\S_{\mu_1\mu_2\dots\mu_p}\S^{\mu_1\mu_2\dots\mu_p}+2p
\left({}^{\S}_{\vec{u}}E\right)_{\mu_2\dots\mu_p}
\left({}^{\S}_{\vec{u}}E\right)^{\mu_2\dots\mu_p}
\e*
which comparing with the previous equation proves the formula. \N

Observe that introducing (\ref{inv}) into the formula just obtained we can
write
\bea
\frac{1}{p!}\S_{\mu_1\dots\mu_p}\S_{\nu_1\dots\nu_p}
h^{\mu_1\nu_1}\dots h^{\mu_p\nu_p}=
\frac{1}{(n-p-1)!}({}^{\S}_{\vec{u}}H)_{\mu_{p+2}\dots\mu_n}
({}^{\S}_{\vec{u}}H)^{\mu_{p+2}\dots\mu_n}+\nonumber \\
+\frac{1}{(p-1)!}({}^{\S}_{\vec{u}}E)_{\mu_2\dots\mu_p}
({}^{\S}_{\vec{u}}E)^{\mu_2\dots\mu_p} \hspace{1cm} \label{square4}
\eea
which will be useful in the proof of the general case, see Corollary
\ref{square3}.

Consider now any double $(p,q)$-form $K_{[p],[q]}$ and take its corresponding
$\tilde{K}_{[p],[q]}$ with ordered indices:
$\tilde{K}_{\mu_1\dots\mu_p,\nu_1\dots\nu_q}=
\tilde{K}_{[\mu_1\dots\mu_p],[\nu_1\dots\nu_q]}$. Its s-e tensor (\ref{set})
reads
\bea
T_{\alpha}{}^{\beta}{}_{\lambda}{}^{\mu}\left\{K_{[p],[q]}\right\}=
\frac{1}{2}\left(\frac{1}{(p-1)!(q-1)!}\,
\tilde{K}_{\alpha\rho_2\dots\rho_p,\lambda\sigma_2\dots\sigma_q}
\tilde{K}^{\beta\rho_2\dots\rho_p,\mu\sigma_2\dots\sigma_q}+\right.\nonumber\\
+\frac{1}{(p-1)!(n-q-1)!}\,\tilde{K}_{\alpha\rho_2\dots\rho_p,
\stackrel{*}{\lambda\sigma_{q+2}\dots\sigma_n}}
\tilde{K}^{\beta\rho_2\dots\rho_p,\stackrel{*}{\mu\sigma_{q+2}\dots\sigma_n}}
+\nonumber \\
+\frac{1}{(n-p-1)!(q-1)!}\,
\tilde{K}_{\stackrel{*}{\alpha\rho_{p+2}\dots\rho_n},
\lambda\sigma_2\dots\sigma_q}
\tilde{K}^{\stackrel{*}{\beta\rho_{p+2}\dots\rho_n},\mu\sigma_2\dots\sigma_q}
+\nonumber\\
+\left.\frac{1}{(n-p-1)!(n-q-1)!}\,
\tilde{K}_{\stackrel{*}{\alpha\rho_{p+2}\dots\rho_n},
\stackrel{*}{\lambda\sigma_{q+2}\dots\sigma_n}}
\tilde{K}^{\stackrel{*}{\beta\rho_{p+2}\dots\rho_n},
\stackrel{*}{\mu\sigma_{q+2}\dots\sigma_n}}\right) \label{set11}
\eea
which after use of formulae (\ref{ap:1}-\ref{ap:3}) in the Appendix is
equivalent to
\bea
T_{\alpha\beta\lambda\mu}\left\{K_{[p],[q]}\right\}=
\frac{1}{(p-1)!(q-1)!}\left(
\tilde{K}_{\alpha\rho_2\dots\rho_p,\lambda\sigma_2\dots\sigma_q}
\tilde{K}_{\beta}{}^{\rho_2\dots\rho_p,}{}_{\mu}{}^{\sigma_2\dots\sigma_q}
+\right.\hspace{1cm}\nonumber\\
+\tilde{K}_{\alpha\rho_2\dots\rho_p,\mu\sigma_2\dots\sigma_q}
\tilde{K}_{\beta}{}^{\rho_2\dots\rho_p,}{}_{\lambda}{}^{\sigma_2\dots\sigma_q}
-\frac{1}{p}g_{\alpha\beta}
\tilde{K}_{\rho_1\rho_2\dots\rho_p,\lambda\sigma_2\dots\sigma_q}
\tilde{K}^{\rho_1\rho_2\dots\rho_p,}{}_{\mu}{}^{\sigma_2\dots\sigma_q}
-\hspace{1cm} \label{set11'}\\
\left. -\frac{1}{q}g_{\lambda\mu}
\tilde{K}_{\alpha\rho_2\dots\rho_p,\sigma_1\sigma_2\dots\sigma_q}
\tilde{K}_{\beta}{}^{\rho_2\dots\rho_p,\sigma_1\sigma_2\dots\sigma_q}+
\frac{1}{2pq}g_{\alpha\beta}g_{\lambda\mu}
\tilde{K}_{\rho_1\rho_2\dots\rho_p,\sigma_1\sigma_2\dots\sigma_q}
\tilde{K}^{\rho_1\rho_2\dots\rho_p,\sigma_1\sigma_2\dots\sigma_q}
\right)\nonumber
\eea
and thus in general
\be
T_{\alpha\beta\lambda\mu}\left\{K_{[p],[q]}\right\}=
T_{(\alpha\beta)(\lambda\mu)}\left\{K_{[p],[q]}\right\}.\label{sym11}
\ee

One can find similar explicit formulas for the basic s-e tensor
$T_{\alpha\beta\lambda\mu\tau\nu}\left\{A_{[p],[q],[s]}\right\}$ of
general triple $(p,q,s)$-forms in a straightforward way, and so on.
For later use, the form of the s-e tensor for a triple $(p,q,s)$-form
$A_{[p],[q],[s]}$ is presented:
\bea
T_{\alpha\beta\lambda\mu\tau\nu}\left\{A_{[p],[q],[s]}\right\}=
\frac{1}{(p-1)!(q-1)!(s-1)!}\left[\frac{}{}
\tilde{A}_{\alpha\rho_2\dots\rho_p,\lambda\sigma_2\dots\sigma_q,
\tau\z_2\dots\z_s}
\tilde{A}_{\beta}{}^{\rho_2\dots\rho_p,}{}_{\mu}{}^{\sigma_2\dots\sigma_q,}
{}_{\nu}{}^{\z_2\dots\z_s}\right.\nonumber\\
+\tilde{A}_{\beta\rho_2\dots\rho_p,\lambda\sigma_2\dots\sigma_q,
\tau\z_2\dots\z_s}
\tilde{A}_{\alpha}{}^{\rho_2\dots\rho_p,}{}_{\mu}{}^{\sigma_2\dots\sigma_q,}
{}_{\nu}{}^{\z_2\dots\z_s}+\hspace{4cm}\nonumber\\
+\tilde{A}_{\alpha\rho_2\dots\rho_p,\mu\sigma_2\dots\sigma_q,
\tau\z_2\dots\z_s}
\tilde{A}_{\beta}{}^{\rho_2\dots\rho_p,}{}_{\lambda}{}^{\sigma_2\dots\sigma_q,}
{}_{\nu}{}^{\z_2\dots\z_s}+\hspace{3cm}\nonumber\\
+\tilde{A}_{\alpha\rho_2\dots\rho_p,\lambda\sigma_2\dots\sigma_q,
\nu\z_2\dots\z_s}
\tilde{A}_{\beta}{}^{\rho_2\dots\rho_p,}{}_{\mu}{}^{\sigma_2\dots\sigma_q,}
{}_{\tau}{}^{\z_2\dots\z_s}-\hspace{2cm}\nonumber\\
-\frac{1}{p}g_{\alpha\beta}\left(
\tilde{A}_{\rho_1\dots\rho_p,\lambda\sigma_2\dots\sigma_q,
\tau\z_2\dots\z_s}
\tilde{A}^{\rho_1\dots\rho_p,}{}_{\mu}{}^{\sigma_2\dots\sigma_q,}
{}_{\nu}{}^{\z_2\dots\z_s}+\right.\hspace{4cm}\nonumber\\
\left.\tilde{A}_{\rho_1\dots\rho_p,\lambda\sigma_2\dots\sigma_q,
\nu\z_2\dots\z_s}
\tilde{A}^{\rho_1\dots\rho_p,}{}_{\mu}{}^{\sigma_2\dots\sigma_q,}
{}_{\tau}{}^{\z_2\dots\z_s}\right)-\hspace{2cm}\nonumber\\
-\frac{1}{q}g_{\lambda\mu}\left(
\tilde{A}_{\alpha\rho_2\dots\rho_p,\sigma_1\dots\sigma_q,
\tau\z_2\dots\z_s}
\tilde{A}_{\beta}{}^{\rho_2\dots\rho_p,}{}^{\sigma_1\dots\sigma_q,}
{}_{\nu}{}^{\z_2\dots\z_s}+\right.\hspace{4cm}\nonumber\\
\left.\tilde{A}_{\alpha\rho_2\dots\rho_p,\sigma_1\dots\sigma_q,
\nu\z_2\dots\z_s}
\tilde{A}_{\beta}{}^{\rho_2\dots\rho_p,}{}^{\sigma_1\dots\sigma_q,}
{}_{\tau}{}^{\z_2\dots\z_s}\right)-\hspace{2cm}\nonumber\\
-\frac{1}{s}g_{\tau\nu}\left(
\tilde{A}_{\alpha\rho_2\dots\rho_p,\lambda\sigma_2\dots\sigma_q,\z_1\dots\z_s}
\tilde{A}_{\beta}{}^{\rho_2\dots\rho_p,}{}_{\mu}{}^{\sigma_2\dots\sigma_q,}
{}^{\z_1\dots\z_s}+\right.\hspace{4cm}\nonumber\\
\left.
\tilde{A}_{\alpha\rho_2\dots\rho_p,\mu\sigma_2\dots\sigma_q,\z_1\dots\z_s}
\tilde{A}_{\beta}{}^{\rho_2\dots\rho_p,}{}_{\lambda}{}^{\sigma_2\dots\sigma_q,}
{}^{\z_1\dots\z_s}\right)+\hspace{2cm}\nonumber\\
+\frac{1}{pq}g_{\alpha\beta}g_{\lambda\mu}
\tilde{A}_{\rho_1\dots\rho_p,\sigma_1\dots\sigma_q,\tau\z_2\dots\z_s}
\tilde{A}^{\rho_1\dots\rho_p,}{}^{\sigma_1\dots\sigma_q,}
{}_{\nu}{}^{\z_2\dots\z_s}+\hspace{5cm}\nonumber\\
+\frac{1}{ps}g_{\alpha\beta}g_{\tau\nu}
\tilde{A}_{\rho_1\dots\rho_p,\lambda\sigma_2\dots\sigma_q,\z_1\dots\z_s}
\tilde{A}^{\rho_1\dots\rho_p,}{}_{\mu}{}^{\sigma_2\dots\sigma_q,}
{}^{\z_1\dots\z_s}+\hspace{4cm}\nonumber\\
+\frac{1}{qs}g_{\lambda\mu}g_{\tau\nu}
\tilde{A}_{\alpha\rho_2\dots\rho_p,\sigma_1\dots\sigma_q,\z_1\dots\z_s}
\tilde{A}_{\beta}{}^{\rho_2\dots\rho_p,}{}^{\sigma_1\dots\sigma_q,}
{}^{\z_1\dots\z_s}-\hspace{3cm}\nonumber\\
-\left.\frac{1}{2\,pqs}g_{\alpha\beta}g_{\lambda\mu}g_{\tau\nu}
\tilde{A}_{\rho_1\dots\rho_p,\sigma_1\dots\sigma_q,\z_1\dots\z_s}
\tilde{A}^{\rho_1\dots\rho_p,\sigma_1\dots\sigma_q,\z_1\dots\z_s}
\right].\hspace{2cm}\label{set111'}
\eea
It is noteworthy that expressions (\ref{set1'}), (\ref{set11'}) and
(\ref{set111'}) do not depend on the dimension $n$ explicitly. This is in
fact a general property, so that a formula for the s-e tensor (\ref{set})
can be given, in general, without any explicit dependence on $n$. It is enough
to expand all the duals. Actually, this general expression can be easily
guessed for general $r$-fold $(n_1,\dots ,n_r)$-forms from the
cases given in (\ref{set1'}), (\ref{set11'}), (\ref{set111'}). However, this
general formula is cumbersome and adds very little to what has been said, so 
that it will be omitted here.

\section{The dominant super-energy property}
\label{sec:dsep}
Properties \ref{pr:1} and \ref{pr:2} are simple particular cases of a quite
general and much more important property of the basic s-e tensors (\ref{set}),
which I call the {\it dominant super-energy property} (DSEP). The general
definition is
\begin{defi}
A rank-s tensor $T_{\mu_1\dots\mu_s}$ is said to satisfy the {\em dominant
super-energy property} if
\be
T_{\mu_1\dots\mu_s}k^{\mu_1}_1\dots k^{\mu_s}_s \geq 0 \label{dsep}
\ee
for any future-pointing causal vectors $\vec{k}_1,\dots,\vec{k}_s$.
\end{defi}
The simplest examples of tensors with the DSEP are the
{\it past}-directed causal vectors and their tensor products. Another simple
example is {\it minus} the metric tensor, and any tensor products of it
\b*
-g, \hspace{5mm} g\otimes g, \hspace{5mm} -g\otimes g\otimes g,
\hspace{5mm} \dots \dots ,
\hspace{5mm} (-1)^k \underbrace{g\otimes \dots \otimes g}_k \, .
\e*
Actually, we have
\begin{pr}
The tensor product of any number of tensors with the DSEP
satisfies this property too. Furthermore, any linear combination with
{\em non-negative} coefficients of tensors satisfying the DSEP 
has also the DSEP.
\label{pr:alg}
\end{pr}
\P
Trivial.\N

Let us remark that, in some particular cases, even more general linear
combinations (for instance with some negative coefficient) of tensors which
satisfy the DSEP may also keep this property. Property
\ref{pr:alg} shows that the set of all tensors with the DSEP
at any point of $V_n$ has the algebraic structure of a vector `semi'-space,
given that one can only multiply by the scalars of the `semi'-field $\R^+$.

The DSEP is a generalization of the so-called dominant energy
condition for energy-momentum tensors \cite{HE}. The justification for its
name is given in the following
\begin{lem}
If a tensor $T_{\mu_1\dots\mu_s}$ satisfies the DSEP, then
\b*
T_{0\dots 0}\geq |T_{\mu_1\dots\mu_s}| \hspace{1cm} \forall
\mu_1,\dots ,\mu_s=0,\dots, n-1 
\e*
in any orthonormal basis $\{\vec{e}_{\nu}\}$.
\end{lem}
Thus, the completely timelike component of $T_{\mu_1\dots\mu_s}$ (which
may be called its `super-energy' relative to $\vec{e}_0$) `dominates' over
all other components of $T_{\mu_1\dots\mu_s}$.

\P
Let $\{\vec{e}_{\nu}\}$ be any orthonormal basis with a future-pointing
$\vec{e}_0$ and define a set of $(n-1)+(n-1)=2n-2$ future-pointing null
vectors $\vec{\ell}_i^{\pm}$ by means of
\b*
\vec{\ell}_i^{\pm}\equiv \vec{e}_0\pm \vec{e}_i\, , \hspace{15mm}
g\left(\vec{\ell}_i^{+},\vec{\ell}_i^{+}\right)=
g\left(\vec{\ell}_i^{-},\vec{\ell}_i^{-}\right)=0 \, .
\e*
As initial step, from (\ref{dsep}) one immediately obtains in the given basis
\b*
T_{\mu_1\dots\mu_s}e_0^{\mu_1}\dots e_0^{\mu_s}\geq 0 \hspace{5mm}
\Longleftrightarrow \hspace{5mm} T_{0\dots 0}\geq 0 \, .
\e*
Next step is to contract with at most one of the null vectors
$\vec{\ell}^{\pm}_i$. For instance, from (\ref{dsep}) one derives
\b*
T_{\mu_1\dots\mu_s}e_0^{\mu_1}\dots e_0^{\mu_{s-1}}\ell^{\pm\mu_s}_i
\geq 0 \hspace{5mm} \Longleftrightarrow \hspace{5mm}
T_{0\dots 0}\pm T_{0\dots 0i}\geq 0
\hspace{5mm} \Longleftrightarrow \hspace{5mm}
T_{0\dots 0}\geq |T_{0\dots 0i}|\, ,\\
T_{\mu_1\dots\mu_s}e_0^{\mu_1}\dots \ell^{\pm\mu_{s-1}}_i e_0^{\mu_{s}}
\geq 0 \hspace{5mm} \Longleftrightarrow \hspace{5mm}
T_{0\dots 0}\pm T_{0\dots i0}\geq 0
\hspace{5mm} \Longleftrightarrow \hspace{5mm}
T_{0\dots 0}\geq |T_{0\dots i0}|,
\e*
and so on for the components with $(s-1)$ zeros. Next, one computes
\b*
T_{\mu_1\dots\mu_s}e_0^{\mu_1}\dots e_0^{\mu_{s-2}}
\left(\ell^{+\mu_{s-1}}_i\ell^{\pm\mu_s}_j+
\ell^{-\mu_{s-1}}_i\ell^{\mp\mu_s}_j\right) 
\e*
which is non-negative due to (\ref{dsep}) and equivalent in the given basis
to
\b*
T_{0\dots 0}\pm T_{0\dots 0j}+T_{0\dots i0}\pm T_{0\dots ij}+
T_{0\dots 0}\mp T_{0\dots 0j}-T_{0\dots i0}\pm T_{0\dots ij}\geq 0
\Longleftrightarrow \\
\Longleftrightarrow \hspace{15mm} T_{0\dots 0}\pm T_{0\dots ij}\geq 0
\hspace{15mm} \Longleftrightarrow \hspace{15mm}
T_{0\dots 0}\geq |T_{0\dots ij}| 
\e*
and similarly for any component with $(s-2)$ zeros. Next one can calculate
\b*
T_{\mu_1\dots\mu_s}e_0^{\mu_1}\dots e_0^{\mu_{s-3}}\!\left[\ell^{+\mu_{s-2}}_k
\!\left(\ell^{+\mu_{s-1}}_i\ell^{\pm\mu_s}_j\!+\!
\ell^{-\mu_{s-1}}_i\ell^{\mp\mu_s}_j\right)\!+\!
 \ell^{-\mu_{s-2}}_k \left(\ell^{+\mu_{s-1}}_i\ell^{\mp\mu_s}_j\!+\!
\ell^{-\mu_{s-1}}_i\ell^{\pm\mu_s}_j\right)\right]
\e*
which leads by (\ref{dsep}) to
\b*
T_{0\dots 0}\geq |T_{0\dots kij}|
\e*
and analogously for the components with $(s-3)$ zeros. One proceeds in this
manner in order, and at the $(l+1)$-th step one uses the schematic combination
given by 
\b*
\vec{\ell}^+_{i_{l+1}}\left(\mbox{combination of $l$-th step}
\right)_{i_1\dots i_l}+\vec{\ell}^-_{i_{l+1}}
\left(\mbox{same combination interchanging $\pm \leftrightarrow \mp$}
\right)_{i_1\dots i_l}
\e*
plus the necessary number of $\vec{e}_0$'s to complete the $s$ vectors.
This finishes the proof.\N

A possible generalization of definition (\ref{set}) that one is immediately
tempted to adopt is to allow for different positive weights in front of
each of the semi-squares on the righthand side. This will obviously keep
the Property \ref{pr:1}. However, Property \ref{pr:2} and the DSEP would be
lost as follows from the following result, which is also the essential step
in the proof by induction of the fundamental Theorem \ref{th:dsep}.
\begin{lem}
Among all the tensors
\be
T_{\lambda}{}^{\mu}\left\{\S_{[p]};c_1,c_2\right\}\equiv
\frac{c_1}{(p-1)!}\S_{\lambda\rho_2\dots\rho_p}
\S^{\mu\rho_2\dots\rho_p}+\frac{c_2}{(n-p-1)!}
\S_{\stackrel{*}{\lambda\rho_{p+2}\dots\rho_n}}
\S^{\stackrel{*}{\mu\rho_{p+2}\dots\rho_n}}
\ee
with arbitrary constants $c_1,c_2$, the only ones which satisfy
{\em generically} the DSEP are proportional to the basic s-e tensor
(\ref{set1}), that is, precisely those with $c_1=c_2>0$.
\label{r=1}
\end{lem}
\P
By using eq.(\ref{ap:pp}) in the Appendix, a calculation similar to that leading to
(\ref{set1'}) proves that
\be
T_{\lambda\mu}\left\{\S_{[p]};c_1,c_2\right\}=
\frac{(c_1+c_2)}{(p-1)!}\,
\S_{\lambda\rho_2\dots\rho_p}\S_{\mu}{}^{\rho_2\dots\rho_p}-
\frac{c_2}{p!}\,g_{\lambda\mu}\,
\S_{\rho_1\rho_2\dots\rho_p}\S^{\rho_1\rho_2\dots\rho_p}\, .\label{paso}
\ee
Let $\vec{u}$ be any unit future-pointing timelike vector and let $\vec{v}$
be any causal vector with the same time orientation: $u^{\mu}v_{\mu}< 0$.
Obviously, one can write
\be
\vec{v}=a\vec{u}+\vec{b}, \hspace{5mm} a\equiv -u^{\mu}v_{\mu}> 0, \hspace{5mm}
u^{\mu}b_{\mu}=0, \hspace{5mm} b^2\equiv b^{\mu}b_{\mu}\geq 0 \label{v}
\ee
and the condition that $\vec{v}$ be causal reads
\be
v^{\mu}v_{\mu}=-a^2+b^2\leq 0 \hspace{5mm} \Longleftrightarrow \hspace{5mm}
b^2\leq a^2 \, . \label{ba}
\ee
Starting with
\b*
T_{\lambda\mu}\left\{\S_{[p]};c_1,c_2\right\}u^{\lambda}u^{\mu}=
\frac{c_1}{(p-1)!}[{}^{\S}_{\vec{u}}E]^2+
\frac{c_2}{(n-p-1)!}[{}^{\S}_{\vec{u}}H]^2 > 0
\e*
one immediately gets for generic $\bm{\S}$
\b*
c_1> 0, \hspace{1cm} c_2 > 0\, .
\e*
Concerning the general result, by using (\ref{paso}) and (\ref{inv}) we obtain
\b*
T_{\lambda\mu}\left\{\S_{[p]};c_1,c_2\right\}u^{\lambda}v^{\mu}=
\frac{(c_1+c_2)}{(p-1)!} \left({}^{\S}_{\vec{u}}E
\right)_{\rho_2\dots\rho_p}
\left(a\,\left({}^{\S}_{\vec{u}}E\right)^{\rho_2\dots\rho_p}+
b_{\rho_1}\S^{\rho_1\dots\rho_p}\right)+\\
+a c_2\left(\frac{1}{(n-p-1)!}[\left({}^{\S}_{\vec{u}}H\right)]^2-
\frac{1}{(p-1)!}[\left({}^{\S}_{\vec{u}}E\right)]^2\right)=\hspace{1cm}\\
=a\left(\frac{c_1}{(p-1)!}[\left({}^{\S}_{\vec{u}}E\right)]^2+
\frac{c_2}{(n-p-1)!}[\left({}^{\S}_{\vec{u}}H\right)]^2+
\frac{(c_1+c_2)}{(p-1)!}\frac{b^{\rho_1}}{a}\,
\left({}^{\S}_{\vec{u}}E\right)^{\rho_2\dots\rho_p}\S_{\rho_1\dots\rho_p}
\right).
\e*
In order to simplify the calculation
one can choose an orthonormal basis $\left\{\vec{e}_{\mu}\right\}$ with
$\vec{u}=\vec{e}_0$ and $\vec{b}=b\vec{e}_1$ with $b\geq 0$. Keeping in mind
the notation that Latin small letters $i,j,\dots$ run from 1 to
$n-1$ and capital letters $I,J,\dots$ run from 2 to $n-1$ we get
\b*
T_{\lambda\mu}\left\{\S_{[p]};c_1,c_2\right\}u^{\lambda}v^{\mu}=
a\left(c_1\sum_{i_2<\dots <i_p}\left(
\left({}^{\S}_{\vec{u}}E\right)_{i_2\dots i_p}\right)^2+\right.\\
\left.+c_2\sum_{j_2<\dots <j_{n-p}}\left(
\left({}^{\S}_{\vec{u}}H\right)_{j_2\dots j_{n-p}}\right)^2
+\frac{(c_1+c_2)}{(p-1)!}\,
\frac{b}{a}\, \S_{1 I_2\dots I_p}\left({}^{\S}_{\vec{u}}E
\right)^{I_2\dots I_p}\right) =\\
=a\left(c_1\sum_{i_3<\dots <i_p}\left(
\left({}^{\S}_{\vec{u}}E\right)_{1i_3\dots i_p}\right)^2+
c_1\sum_{I_2<\dots <I_p}\left(
\left({}^{\S}_{\vec{u}}E\right)_{I_2\dots I_p}\right)^2+\right.\\
+c_2\sum_{j_3<\dots <j_{n-p}}\left(
\left({}^{\S}_{\vec{u}}H\right)_{1j_3\dots j_{n-p}}\right)^2+
c_2\sum_{J_2<\dots <J_{n-p}}\left(
\left({}^{\S}_{\vec{u}}H\right)_{J_2\dots J_{n-p}}\right)^2+\\
\left.+(c_1+c_2)\frac{b}{a}\,\,\,
\sum_{I_2<\dots <I_p,\, J_2<\dots <J_{n-p}}^{I_2,\dots ,I_p\neq  
J_2,\dots ,J_{n-p}} (-1)^{n+|\sigma |}\left({}^{\S}_{\vec{u}}E
\right)_{I_2\dots I_p}
\left({}^{\S}_{\vec{u}}H\right)_{J_2\dots J_{n-p}}\right)
\e*
where $(-1)^{|\sigma|}$ denotes the sign of the permutation
\b*
\left(\begin{array}{lccccr}
I_2 &\dots &I_p & J_2 &\dots & J_{n-p}\\
2 & . & . & . & . & (n-1)
\end{array}\right).
\e*
By re-orginizing in perfect squares we arrive at
\b*
T_{\lambda\mu}\left\{\S_{[p]};c_1,c_2\right\}u^{\lambda}v^{\mu}=
a\left\{c_1\sum_{i_3<\dots <i_p}\left(
\left({}^{\S}_{\vec{u}}E\right)_{1i_3\dots i_p}\right)^2
+c_2\sum_{j_3<\dots <j_{n-p}}\left(
\left({}^{\S}_{\vec{u}}H\right)_{1j_3\dots j_{n-p}}\right)^2\right.\\
+c_1\sum_{I_2<\dots <I_p, J_2<\dots <J_{n-p}}^{I_2,\dots ,I_p\neq  
J_2,\dots ,J_{n-p}}\left(\left({}^{\S}_{\vec{u}}E\right)_{I_2\dots I_p}+
(-1)^{n+|\sigma |}\frac{b}{2a}\left(1+\frac{c_2}{c_1}\right)
\left({}^{\S}_{\vec{u}}H\right)_{J_2\dots J_{n-p}}\right)^2+\\
+\left.\left[c_2-\frac{b^2}{4a^2}c_1\left(1+\frac{c_2}{c_1}\right)^2\right]
\sum_{J_2<\dots <J_{n-p}}\left(
\left({}^{\S}_{\vec{u}}H\right)_{J_2\dots J_{n-p}}\right)^2\right\}\, .
\e*
Due to (\ref{ba}), the factor in square brackets satisfies
\b*
\left[c_2-\frac{b^2}{4a^2}c_1\left(1+\frac{c_2}{c_1}\right)^2\right]
\geq \left[c_2-\frac{c_1}{4}\left(1+\frac{c_2}{c_1}\right)^2\right]=
-\frac{c_1}{4}\left(1-\frac{c_2}{c_1}\right)^2
\e*
and thus 
\b*
T_{\lambda\mu}\left\{\S_{[p]};c_1,c_2\right\}u^{\lambda}v^{\mu}\geq
a\left\{c_1\sum_{i_3<\dots <i_p}\left(
\left({}^{\S}_{\vec{u}}E\right)_{1i_3\dots i_p}\right)^2
+c_2\sum_{j_3<\dots <j_{n-p}}\left(
\left({}^{\S}_{\vec{u}}H\right)_{1j_3\dots j_{n-p}}\right)^2\right.\\
+c_1\sum_{I_2<\dots <I_p, J_2<\dots <J_{n-p}}^{I_2,\dots ,I_p\neq  
J_2,\dots ,J_{n-p}}\left(\left({}^{\S}_{\vec{u}}E\right)_{I_2\dots I_p}+
(-1)^{n+|\sigma |}\frac{b}{2a}\left(1+\frac{c_2}{c_1}\right)
\left({}^{\S}_{\vec{u}}H\right)_{J_2\dots J_{n-p}}\right)^2-\\
-\left.\frac{c_1}{4}\left(1-\frac{c_2}{c_1}\right)^2
\sum_{J_2<\dots <J_{n-p}}\left(
\left({}^{\S}_{\vec{u}}H\right)_{J_2\dots J_{n-p}}\right)^2\right\}\, .
\e*
From here, keeping in mind that the result must hold for generic
$\bm{\S}$, if $c_1\neq c_2$ the last term is strictly negative,
and given that all the other terms can vanish for some $\bm{\S}$ and that
the equality can be achieved for $b=a$, it follows that
$T_{\lambda\mu}\left\{\S_{[p]};c_1,c_2\right\}u^{\lambda}v^{\mu}$ is
certainly negative for some $\bm{\S}$ if $c_1\neq c_2$.
On the other hand, one immediately sees that if $c_1=c_2$ the last term
vanishes and the righthand side is a sum of squares, so that
$T_{\lambda\mu}\left\{\S_{[p]};c_1,c_1\right\}u^{\lambda}v^{\mu}\geq 0$.
This proves the result except for the minor point that $\vec{u}$ was assumed to
be timelike from the beginning. However, as this is valid for arbitrary
$\vec{u}$, the result also follows by continuity for null vectors.\N

An interesting corollary of this result is
\begin{coro}
For all timelike vectors $\vec{u}$ and $\vec{w}$ with the same time
orientation and any simple $p$-form $\bm{\S}$ one has
$T_{\lambda\mu}\left\{\S_{[p]}\right\}u^{\lambda}w^{\mu}>0$, that is
\b*
\frac{1}{(p-1)!}\left({}^{\S}_{\vec{u}}E\right)_{\rho_2\dots\rho_p}
\left({}^{\S}_{\vec{w}}E\right)^{\rho_2\dots\rho_p}+
\frac{1}{(n-p-1)!}\left({}^{\S}_{\vec{u}}H\right)_{\rho_{p+2}\dots\rho_n}
\left({}^{\S}_{\vec{w}}H\right)^{\rho_{p+2}\dots\rho_n}>0\, .
\e*
\label{strict}
\end{coro}
\P
This follows at once from the proof of Lemma \ref{r=1} by using
$c_1=c_2=1$ together with $b^2<a^2$ (strict inequality). Then, as long
as $\bm{\S}$ is not zero,
$T_{\lambda\mu}\left\{\S_{[p]}\right\}u^{\lambda}w^{\mu}$ is proved to be
a sum of squares which cannot vanish simultaneously.\N

Notice that, in general, neither $\left({}^{\S}_{\vec{u}}E
\right)_{\rho_2\dots\rho_p}
\left({}^{\S}_{\vec{w}}E\right)^{\rho_2\dots\rho_p}$ nor
$\left({}^{\S}_{\vec{u}}H\right)_{\rho_{p+2}\dots\rho_n}
\left({}^{\S}_{\vec{w}}H\right)^{\rho_{p+2}\dots\rho_n}$ are positive
by themselves, nor any other combination of them with weights {\it different}
from those of the Corollary \ref{strict}.

\begin{theo}
The s-e tensors (\ref{set}) satisfy the DSEP.
\label{th:dsep}
\end{theo}
For General Relativity ($n=4$), the proof of this result has been given in
an elegant manner by Bergqvist \cite{Ber3} using spinors. Apparently, the
use of Clifford algebra techniques may also provide a proof of this
theorem \cite{PP}.

\P
The proof is by mathematical induction on the natural number $r$ of the
$r$-fold forms. From Lemma \ref{r=1} the result holds for $r=1$. Thus, let us
make the induction hypothesis that the result holds for the basic s-e tensor
of $(r-1)$-fold forms and try to prove the result for $r$-fold forms.
Same notation as in Lemma \ref{r=1} will be used so that $\vec{u}$
is any unit future-pointing timelike vector, $\vec{v}$ is any future-pointing
causal vector and relations (\ref{v}) and (\ref{ba}) hold. Furthermore,
the orthonormal basis $\left\{\vec{e}_{\mu}\right\}$ with $\vec{u}=\vec{e}_0$
and $\vec{b}=b\vec{e}_1$ with $b\geq 0$ is also used as well as the notation
for Latin small and capital letters.

Let $t_{[n_1],\dots,[n_r]}$ be any $r$-fold $(n_1,\dots,n_r)$-form and
$T_{\lambda_1\mu_1\dots\lambda_r\mu_r}\left\{t\right\}$ its corresponding
basic s-e tensor (\ref{set}). The idea is to prove that 
$T_{\lambda_1\mu_1\dots\lambda_{r-1}\mu_{r-1}\lambda_r\mu_r}\left\{t\right\}
u^{\lambda_r}v^{\mu_r}$ is a linear combination with non-negative coefficients
of basic s-e tensors of type (\ref{set}) for some $(r-1)$-fold forms. To that
end, define the following set of $(r-1)$-fold $(n_1,\dots,n_{r-1})$-forms
\bea
Y_{\mu_1\dots\mu_{n_1},\dots,\nu_1\dots\nu_{n_{r-1}}}
\left(i_2,\dots,i_{n_r}\right)\equiv 
\tilde{t}_{\mu_1\dots\mu_{n_1},\dots,
\nu_1\dots\nu_{n_{r-1}},\rho_1\rho_2\dots\rho_{n_r}}
u^{\rho_1}e^{\rho_2}_{i_2}\dots e^{\rho_{n_r}}_{i_{n_r}}
\label{Y}\\
Z_{\mu_1\dots\mu_{n_1},\dots,\nu_1\dots\nu_{n_{r-1}}}
\left(i_2,\dots,i_{n-n_r}\right)\equiv 
\tilde{t}_{\mu_1\dots\mu_{n_1},\dots,
\nu_1\dots\nu_{n_{r-1}},\stackrel{*}{\rho_1\rho_2\dots\rho_{n-n_r}}}
u^{\rho_1}e^{\rho_2}_{i_2}\dots e^{\rho_{n-n_r}}_{i_{n-n_r}}
\label{Z}
\eea
which contains $\left(\begin{array}{c}
n-1 \\ n_r -1 \end{array}\right)+\left(\begin{array}{c}
n-1 \\ n_r \end{array}\right)=\left(\begin{array}{c}
n \\ n_r \end{array}\right)$ different $(r-1)$-fold forms.
From (\ref{set}) we have
\b*
T_{\lambda_1\mu_1\dots\lambda_r\mu_r}\left\{t\right\}u^{\lambda_r}v^{\mu_r}
=\frac{1}{2}\left\{\begin{array}{l} \\ \\ \end{array}\hspace{-4mm}
\left(t_{[n_1],\dots,[n_r]}\times t_{[n_1],\dots,
[n_r]}\right)_{\lambda_1\mu_1\dots\lambda_r\mu_r}+\right.  \\
+\left(t_{\stackrel{*}{[n-n_1]},\dots,[n_r]}\times
t_{\stackrel{*}{[n-n_1]},\dots,[n_r]}\right)_{\lambda_1\mu_1\dots\lambda_r\mu_r}
+\dots + \\
+\dots + \left(t_{[n_1],\dots,\stackrel{*}{[n-n_r]}}\times
t_{[n_1],\dots,\stackrel{*}{[n-n_r]}}\right)_{\lambda_1\mu_1\dots\lambda_r\mu_r}
+ \dots \\
+\dots +\left(t_{\stackrel{*}{[n-n_1]},\stackrel{*}{[n-n_2]},\dots,[n_r]}\times
t_{\stackrel{*}{[n-n_1]},\stackrel{*}{[n-n_2]},\dots,[n_r]}
\right)_{\lambda_1\mu_1\dots\lambda_r\mu_r}+\dots +\\
\left. +\dots \, \dots +
\left(t_{\stackrel{*}{[n-n_1]},\dots,\stackrel{*}{[n-n_r]}}\times
t_{\stackrel{*}{[n-n_1]},\dots,
\stackrel{*}{[n-n_r]}}\right)_{\lambda_1\mu_1\dots\lambda_r\mu_r}\right\}
u^{\lambda_r}v^{\mu_r}\, .
\e*
Let us concentrate on the first term of the righthand side. By using the
notation introduced in (\ref{Y}-\ref{Z}) and definition (\ref{semi-s})
we get due to (\ref{v})
\b*
\left(t_{[n_1],\dots,[n_r]}\times t_{[n_1],\dots,
[n_r]}\right)_{\lambda_1\mu_1\dots\lambda_{r-1}\mu_{r-1}\lambda_r\mu_r}
u^{\lambda_r}v^{\mu_r}=\left(\prod_{\U=1}^{r}\frac{1}{(n_\U-1)!}\right)
\cdot \,\\
\left\{
a\sum_{i_2,\dots ,i_{n_r}}Y_{\lambda_1\rho_2\dots\rho_{n_1},\dots,
\lambda_{r-1}\sigma_2\dots\sigma_{n_{r-1}}}
\left(i_2,\dots,i_{n_r}\right)
Y_{\mu_1}{}^{\rho_2\dots\rho_{n_1},}{}_{\dots,
\mu_{r-1}}{}^{\sigma_2\dots\sigma_{n_{r-1}}}
\left(i_2,\dots,i_{n_r}\right)+\right.\\
+\left.\sum_{i_2,\dots ,i_{n_r}}Y_{\lambda_1\rho_2\dots\rho_{n_1},\dots,
\lambda_{r-1}\sigma_2\dots\sigma_{n_{r-1}}}
\left(i_2,\dots,i_{n_r}\right)\tilde{t}_{\mu_1}{}^{\rho_2\dots\rho_{n_1},}
{}_{\dots,\mu_{r-1}}{}^{\sigma_2\dots\sigma_{n_{r-1}},}
{}_{\mu_r i_2\dots i_{n_r}}b^{\mu_r}\right\}
\e*
and taking into account that, in the second summatory, only the values
of $i_2,\dots,i_{n_r}$ different from 1 are relevant and that
\b*
\tilde{t}_{\mu_1}{}^{\rho_2\dots\rho_{n_1},}
{}_{\dots,\mu_{r-1}}{}^{\sigma_2\dots\sigma_{n_{r-1}},}
{}_{\mu_r I_2\dots I_{n_r}}b^{\mu_r}=
b\,\tilde{t}_{\mu_1}{}^{\rho_2\dots\rho_{n_1},}
{}_{\dots,\mu_{r-1}}{}^{\sigma_2\dots\sigma_{n_{r-1}},}
{}_{1 I_2\dots I_{n_r}}=\\
(-1)^{n+|\sigma|}b\, Z_{\mu_1}{}^{\rho_2\dots\rho_{n_1},}
{}_{\dots,\mu_{r-1}}{}^{\sigma_2\dots\sigma_{n_{r-1}}}
\left(J_2,\dots,J_{n-n_r}\right) \hspace{7mm} \mbox{for}
\hspace{7mm} J_2,\dots,J_{n-n_r}\neq I_2\dots I_{n_r}
\e*
where $(-1)^{|\sigma|}$ is the sign of the permutation
\b*
\left(\begin{array}{lccccr}
I_2 &\dots &I_{n_r} & J_2 &\dots & J_{n-n_r}\\
2 & . & . & . & . & (n-1)
\end{array}\right)
\e*
we arrive at
\b*
\left(t_{[n_1],\dots,[n_r]}\times t_{[n_1],\dots,
[n_r]}\right)_{\lambda_1\mu_1\dots\lambda_{r-1}\mu_{r-1}\lambda_r\mu_r}
u^{\lambda_r}v^{\mu_r}=\hspace{2cm}\\
=a\sum_{i_2<\dots <i_{n_r}}
\left(Y_{[n_1],\dots,[n_{r-1}]}\left(i_2,\dots,i_{n_r}\right)\times
Y_{[n_1],\dots,[n_{r-1}]}\left(i_2,\dots,
i_{n_r}\right)\right)_{\lambda_1\mu_1\dots\lambda_{r-1}\mu_{r-1}}+\\
+\sum_{I_2<\dots <I_{n_r},\, J_2<\dots <J_{n-n_r}}^{I_2,\dots ,I_{n_r}\neq  
J_2,\dots ,J_{n-n_r}}(-1)^{n+|\sigma|}b
\left(Y_{[n_1],\dots,[n_{r-1}]}\left(I_2,\dots,I_{n_r}\right)\times
\right.\hspace{25mm}\\
\left.Z_{[n_1],\dots,[n_{r-1}]}\left(J_2,\dots,
J_{n-n_r}\right)\right)_{\lambda_1\mu_1\dots\lambda_{r-1}\mu_{r-1}}
\hspace{15mm}
\e*
or equivalently
\b*
\left(t_{[n_1],\dots,[n_r]}\times t_{[n_1],\dots,
[n_r]}\right)_{\lambda_1\mu_1\dots\lambda_{r-1}\mu_{r-1}\lambda_r\mu_r}
u^{\lambda_r}v^{\mu_r}=\hspace{2cm}\\
=a\left\{\sum_{I_3<\dots <I_{n_r}}
\left(Y_{[n_1],\dots,[n_{r-1}]}\left(1,I_3,\dots,I_{n_r}\right)\times
Y_{[n_1],\dots,[n_{r-1}]}\left(1,I_3,\dots,
I_{n_r}\right)\right)_{\lambda_1\mu_1\dots\lambda_{r-1}\mu_{r-1}}+\right.\\
+\sum_{I_2<\dots <I_{n_r},\, J_2<\dots <J_{n-n_r}}^{I_2,\dots ,I_{n_r}\neq  
J_2,\dots ,J_{n-n_r}}
\left(\frac{}{}Y_{[n_1],\dots,[n_{r-1}]}\left(I_2,\dots,I_{n_r}\right)\times
\right.\hspace{35mm}\\
\left.\left.\left[Y_{[n_1],\dots,[n_{r-1}]}\left(I_2,\dots,I_{n_r}\right)+
\frac{b}{a}(-1)^{n+|\sigma|}Z_{[n_1],\dots,[n_{r-1}]}\left(J_2,\dots,
J_{n-n_r}\right)\right]\right)_{\lambda_1\mu_1\dots\lambda_{r-1}\mu_{r-1}}
\right\}.
\e*
By doing the same with all the terms and using (\ref{set}) one can write
\b*
T_{\lambda_1\mu_1\dots\lambda_r\mu_r}\left\{t\right\}u^{\lambda_r}v^{\mu_r}=
a\left\{\sum_{I_3<\dots <I_{n_r}}
T_{\lambda_1\mu_1\dots\lambda_{r-1}\mu_{r-1}}
\left\{Y_{[n_1],\dots,[n_{r-1}]}\left(1,I_3,\dots,I_{n_r}\right)\right\}+
\right.\\
+\sum_{J_3<\dots <J_{n-n_r}}T_{\lambda_1\mu_1\dots\lambda_{r-1}\mu_{r-1}}
\left\{Z_{[n_1],\dots,[n_{r-1}]}\left(1,J_3,\dots,J_{n-n_r}\right)\right\}+\\
+\frac{1}{2}
\sum_{I_2<\dots <I_{n_r},\, J_2<\dots <J_{n-n_r}}^{I_2,\dots ,I_{n_r}
\neq  J_2,\dots ,J_{n-n_r}}\left[
\left(\frac{}{}Y_{[n_1],\dots,[n_{r-1}]}\left(I_2,\dots,I_{n_r}\right)\times
\right.\right.\hspace{35mm}\\
\left.\left[Y_{[n_1],\dots,[n_{r-1}]}\left(I_2,\dots,I_{n_r}\right)+
\frac{b}{a}(-1)^{n+|\sigma|}Z_{[n_1],\dots,[n_{r-1}]}\left(J_2,\dots,
J_{n-n_r}\right)\right]\right)_{\lambda_1\mu_1\dots\lambda_{r-1}\mu_{r-1}}+\\
+\left(\frac{}{}Y_{\stackrel{*}{[n_1]},\dots,[n_{r-1}]}
\left(I_2,\dots,I_{n_r}\right)\times
\right.\hspace{45mm}\\
\left.\left[Y_{\stackrel{*}{[n_1]},\dots,[n_{r-1}]}
\left(I_2,\dots,I_{n_r}\right)+
\frac{b}{a}(-1)^{n+|\sigma|}Z_{\stackrel{*}{[n_1]},\dots,[n_{r-1}]}
\left(J_2,\dots,J_{n-n_r}\right)\right]
\right)_{\lambda_1\mu_1\dots\lambda_{r-1}\mu_{r-1}}+\\
\\
+\dots \,\, \dots + \mbox{all duals} \, \, + \hspace{35mm}\\
+\left(\frac{}{}Z_{[n_1],\dots,[n_{r-1}]}\left(J_2,\dots,J_{n-n_r}\right)\times
\right.\hspace{35mm}\\
\left.\left[Z_{[n_1],\dots,[n_{r-1}]}\left(J_2,\dots,J_{n-n_r}\right)+
\frac{b}{a}(-1)^{n+|\sigma|}Y_{[n_1],\dots,[n_{r-1}]}\left(I_2,\dots,
I_{n_r}\right)\right]\right)_{\lambda_1\mu_1\dots\lambda_{r-1}\mu_{r-1}}+\\
+\left(\frac{}{}Z_{\stackrel{*}{[n_1]},\dots,[n_{r-1}]}
\left(J_2,\dots,J_{n-n_r}\right)\times \right.\hspace{35mm}\\
\left.\left[Z_{\stackrel{*}{[n_1]},\dots,[n_{r-1}]}
\left(J_2,\dots,J_{n-n_r}\right)+
\frac{b}{a}(-1)^{n+|\sigma|}Y_{\stackrel{*}{[n_1]},\dots,[n_{r-1}]}
\left(I_2,\dots,I_{n_r}\right)\right]
\right)_{\lambda_1\mu_1\dots\lambda_{r-1}\mu_{r-1}}+\\
\\
\left.\left.+\dots \,\, \dots + \mbox{all duals} \, \, 
\frac{}{}\right]\begin{array}{l} \\ \\ \\ \end{array}\hspace{-4mm}
\right\}\hspace{35mm}
\e*
which after reorganizing a little bit becomes finally
\b*
T_{\lambda_1\mu_1\dots\lambda_r\mu_r}\left\{t\right\}u^{\lambda_r}v^{\mu_r}=
a\left\{\sum_{I_3<\dots <I_{n_r}}
T_{\lambda_1\mu_1\dots\lambda_{r-1}\mu_{r-1}}
\left\{Y_{[n_1],\dots,[n_{r-1}]}\left(1,I_3,\dots,I_{n_r}\right)\right\}+
\right.\\
+\sum_{J_3<\dots <J_{n-n_r}}T_{\lambda_1\mu_1\dots\lambda_{r-1}\mu_{r-1}}
\left\{Z_{[n_1],\dots,[n_{r-1}]}\left(1,J_3,\dots,J_{n-n_r}\right)\right\}+\\
+\sum_{I_2<\dots <I_{n_r},\, J_2<\dots <J_{n-n_r}}^{I_2,\dots ,I_{n_r}
\neq  J_2,\dots ,J_{n-n_r}}
T_{\lambda_1\mu_1\dots\lambda_{r-1}\mu_{r-1}}
\left\{\begin{array}{l} \\ \\ \end{array} \hspace{-4mm}
\left[\frac{}{}Y\left(I_2,\dots,I_{n_r}\right)+
\right.\right.\hspace{3cm}\\
\left.\left.+(-1)^{n+|\sigma|}\frac{b}{a}
Z\left(J_2,\dots,J_{n-n_r}\right)\right]_{[n_1],\dots,[n_{r-1}]}\right\}+
\hspace{1cm}\\
+\left.\left(1-\frac{b^2}{a^2}\right)\sum_{J_2<\dots <J_{n-n_r}}
T_{\lambda_1\mu_1\dots\lambda_{r-1}\mu_{r-1}}
\left\{Z_{[n_1],\dots,[n_{r-1}]}\left(J_2,\dots,J_{n-n_r}\right)\right\}
\right\}.
\e*
This is a sum of basic s-e tensors for $(r-1)$-fold forms with non-negative
coefficients (due to (\ref{v}-\ref{ba})), and thus by the induction hypothesis
it follows
\b*
T_{\lambda_1\mu_1\dots\lambda_{r-1}\mu_{r-1}\lambda_r\mu_r}
\left\{t\right\}k^{\lambda_1}_1k^{\mu_1}_2\dots
k^{\lambda_{r-1}}_{2r-3}k^{\mu_{r-1}}_{2r-2}u^{\lambda_r}v^{\mu_r}\geq 0
\e*
for all future pointing causal vectors
$\vec{k}_1,\vec{k}_2,\dots,\vec{k}_{2r-2}$. As $\vec{u}$ and $\vec{v}$ are also
arbitrary, the theorem follows.\N

With this result at hand, the proof of Property \ref{square} is very simple.
\begin{coro}
\label{square3}
Property \ref{square} holds.
\end{coro}
\P From Property \ref{square2}, the result holds for $r=1$. Let us use
again the mathematical induction on $r$, and assume that the result is true
for $(r-1)$-fold forms. From the proof of Theorem \ref{th:dsep}, but with
$\vec{v}=\vec{u}$ (which is equivalent to setting $a=1$ and $b=0$), it follows
that
\b*
T_{\lambda_1\mu_1\dots\lambda_r\mu_r}\left\{t\right\}u^{\lambda_r}u^{\mu_r}=
\sum_{i_2<\dots <i_{n_r}}
T_{\lambda_1\mu_1\dots\lambda_{r-1}\mu_{r-1}}
\left\{Y_{[n_1],\dots,[n_{r-1}]}\left(i_2,\dots,i_{n_r}\right)\right\}+\\
+\sum_{j_2<\dots <j_{n-n_r}}T_{\lambda_1\mu_1\dots\lambda_{r-1}\mu_{r-1}}
\left\{Z_{[n_1],\dots,[n_{r-1}]}\left(j_2,\dots,j_{n-n_r}\right)\right\}
\hspace{1cm}
\e*
so that 
\b*
W_{t}\left(\vec{u}\right)=\sum_{i_2<\dots <i_{n_r}}
W_{Y(i_2,\dots ,i_{n_r})}\left(\vec{u}\right)+
\sum_{j_2<\dots <j_{n-n_r}}W_{Z(j_2,\dots ,j_{n-n_r})}\left(\vec{u}\right)
\e*
and using here the induction hypothesis one can write
\b*
W_{t}\left(\vec{u}\right)=\frac{1}{2}
\left(\prod_{\U=1}^{r-1}\frac{1}{n_{\U}!}\right)
h^{\mu_1\nu_1}\dots h^{\mu_{n_1}\nu_{n_1}}\dots
h^{\rho_1\sigma_1}\dots h^{\rho_{n_{r-1}}\sigma_{n_{r-1}}} \hspace{2cm}\\
\cdot \left(\sum_{i_2<\dots <i_{n_r}}
\tilde{Y}_{\mu_1\dots\mu_{n_1},\dots ,\rho_1\dots\rho_{n_{r-1}}}
\left(i_2,\dots ,i_{n_r}\right)
\tilde{Y}_{\nu_1\dots\nu_{n_1},\dots ,\sigma_1\dots\sigma_{n_{r-1}}}
\left(i_2,\dots ,i_{n_r}\right)+\right.\\
\left.+\sum_{j_2<\dots <j_{n-n_r}}
\tilde{Z}_{\mu_1\dots\mu_{n_1},\dots ,\rho_1\dots\rho_{n_{r-1}}}
\left(j_2,\dots ,j_{n-n_r}\right)
\tilde{Z}_{\nu_1\dots\nu_{n_1},\dots ,\sigma_1\dots\sigma_{n_{r-1}}}
\left(j_2,\dots ,j_{n-n_r}\right)\right).
\e*
A calculation similar to that leading to (\ref{square4}) and the use of
the definitions (\ref{Y}-\ref{Z}) proves then the desired result.\N

A combination of the proofs and results of Lemma \ref{r=1} and
Theorem \ref{th:dsep} leads easily to the following
\begin{coro}
\label{uniqueness}
Among all the tensors that one can form similar to (\ref{set}) by adding the
different terms of the righthand side in (\ref{set}) multiplied by arbitrary
constants, the only ones which satisfy
the DSEP are precisely those proportional to the basic s-e tensor
(\ref{set}).\N
\end{coro}
\begin{coro}
For all future-pointing {\em timelike} vectors
$\vec{w}_1,\vec{w}_2,\dots,\vec{w}_{2r}$, the strict inequality
\b*
T_{\lambda_1\mu_1\dots\lambda_{r-1}\mu_{r-1}\lambda_r\mu_r}
\left\{t\right\}w^{\lambda_1}_1w^{\mu_1}_2\dots
w^{\lambda_r}_{2r-1}w^{\mu_r}_{2r} >0
\e*
holds as long as $t_{[n_1],\dots,[n_r]}$ does not vanish. 
\label{strict2}
\end{coro}
\P
This is similar to Corollary \ref{strict}.\N

Now, it is very easy to prove the first result of Property \ref{pr:2}.
\begin{coro}
The s-e flux vectors (\ref{sev}) are causal with the same time
orientation than $\vec{u}$.
\label{Pt}
\end{coro}
\P
By Corollary \ref{strict2} and definition (\ref{sev}),
${}^{\U}\vec{P}_t\left(\vec{u}\right)$ satisfy 
\b*
{}^{\U}P^{\rho}_t\left(\vec{u}\right)w_{\rho}<0
\e*
for all possible future-pointing timelike vectors $\vec{u}$ and $\vec{w}$.\N

This result seems to indicate a {\it causal propagation} of the corresponding
super-energy if one interprets ${}^{\U}\vec{P}_t\left(\vec{u}\right)$ as flux
vectors. However, as have been proved, a s-e tensor with the DSEP (and therefore
with causal flux s-e vectors) can {\it always} be constructed for any
given field $t_{\mu_1\dots\mu_m}$. This cannot mean that causal propagation
is `universal', because one can very easily construct tensor fields
$t_{\mu_1\dots\mu_m}$ by hand which do not propagate causally. What happens
here is that the super-energy does propagate causally in the sense
that ${}^{\U}\vec{P}_t\left(\vec{u}\right)$ are causal, but the real field
$t_{\mu_1\dots\mu_m}$ does not necessarily propagate causally. In order to
link these two propagations one needs a further condition which is related
to the divergence of the s-e tensor (\ref{set}) and thereby to the 
derivatives (`field equations') of the field $t_{\mu_1\dots\mu_m}$.
For instance, in \cite{BerS} we have been able to prove the following
interesting theorem. Let $\z$ be any closed achronal set in $V_n$ and
$D(\z)$ its total Cauchy development (see \cite{HE,S0,Wald} for definitions
and notation).
\begin{theo}
If the s-e tensor (\ref{set}) satisfies the following divergence condition
\be
\nabla_{\rho}T^{\rho\mu_1\dots\lambda_r\mu_r}\left\{t\right\}
w_{\mu_1}\dots w_{\lambda_r}w_{\mu_r}\leq f\,
T^{\lambda_1\mu_1\dots\lambda_r\mu_r}\left\{t\right\}
w_{\lambda_1}w_{\mu_1}\dots w_{\lambda_r}w_{\mu_r}\label{divcon}
\ee
where $f$ is a continuous function and $\bm{w}=-d\tau$ is any timelike
1-form foliating $D(\z)$ with hypersurfaces $\tau=$const., then
\b*
\left.t_{\mu_1\dots\mu_m}\right|_{\z}=0 \hspace{1cm} \Longrightarrow
\hspace{1cm} \left.t_{\mu_1\dots\mu_m}\right|_{\overline{D(\z)}}=0.
\e*
\label{th:causal}
\end{theo}
\P
See \cite{BerS}.\N

\noindent
In fact, condition (\ref{divcon}) can be relaxed substantially by allowing
`powers of $T$' greater than one on the righthand side. This theorem allows
to prove the causal propagation of massive and massless fields in General
Relativity, as well as in higher dimensional theories, including that of the
gravitational field, see \cite{BerS,BS}. Furthermore, the uniqueness of the
solution to the field equations in $\overline{D(\z)}$ also follows from
theorem \ref{th:causal}, see \cite{BerS}.
Let us remark that a key point in the proof of Theorem \ref{th:causal}
is the DSEP for the s-e tensor, so that the universality of the
construction of the s-e tensor (\ref{set}) for arbitrary fields may lead
to very general results.

\section{The general super-energy tensor}
\label{sec:general}
As remarked after the definition of the basic s-e tensor (\ref{set}), the
word `basic' was used because more general tensors constructed by
permutting indices in $T^{\lambda_1\mu_1\dots\lambda_r\mu_r}$ are also
good s-e tensors. This freedom will be studied in this section.

The first thing to note is that every tensor of type
\b*
\hat{T}_{\mu_1\mu_2\dots\mu_{2r-1}\mu_{2r}}\left\{t\right\}\equiv
T_{\mu_{\sigma (1)}\mu_{\sigma (2)}\dots\mu_{\sigma (2r-1)}
\mu_{\sigma (2r)}}\left\{t\right\}
\e*
where $\sigma (1),\dots ,\sigma(2r)$ denotes any possible permutation of
$1,\dots ,2r$, satisfies the good properties shown so far for (\ref{set})
including the DSEP. Thus, let us make the following
\begin{defi}
The general s-e tensor of $t$ is defined by
\be
\bbb{T}_{\mu_1\mu_2\dots\mu_{2r-1}\mu_{2r}}\left\{t\right\}\equiv
\sum_{\sigma} c_{\sigma}T_{\mu_{\sigma (1)}\mu_{\sigma (2)}\dots
\mu_{\sigma (2r-1)}\mu_{\sigma (2r)}} \label{gset}
\ee
with non-negative coefficients $c_{\sigma}$.
\end{defi}
Notice that, in general, due to the Property \ref{pr:sym}
not all the permutations above will give rise to a new tensor. Thus, it
is very simple to check that there are only $(2r)!/2^r$ meaningful
independent constants $c_{\sigma}$ or, in other words, 
$\bbb{T}$ is a $(2r)!/2^r$-parameter family of s-e tensors for $t$.
If the Property \ref{pr:sympair} also holds for some blocks $[n_{\U}]$ and
$[n_{\U'}]$, then the number $(2r)!/2^r$ is further reduced. Thus, if
there are $\hat{r}\leq r$ blocks $[n_{\U}]$ which can be interchanged,
$\bbb{T}$ reduces to a $\displaystyle{\frac{(2r)!}{2^r\hat{r}!}}$-parameter
family.

Concerning the properties of $\bbb{T}$, let us first note that, in general,
it does not have any symmetry property, nor any particular traceless
property. The only thing that can be said is that, if the basic s-e
tensor is symmetric with respect to {\it all} $(\lambda_{\U}\mu_{\U})$-pairs,
then the general s-e tensor $\bbb{T}$ keeps this property. In this
particular case, $\bbb{T}$ is a $\displaystyle{\frac{(2r)!}{2^r r!}}=
(2r-1)!!$-parameter family of s-e tensors. With respect to the general
super-energy density $\bbb{W}_t(\vec{u})$ defined as the total timelike
component of $\bbb{T}$ with respect to $\vec{u}$, we obviously have
\b*
\bbb{W}_t\left(\vec{u}\right)=\left(\sum_{\sigma} c_{\sigma}\right)
W_t\left(\vec{u}\right)
\e*
so that the original function $W_t(\vec{u})$ is enough to capture this
concept. In this sense, Properties \ref{pr:1}, \ref{pr:1'} are valid for
$\bbb{W}_t(\vec{u})$ (with the proportionality factor
$\sum_{\sigma} c_{\sigma}$ in the second case). 

With regard to the s-e flux vectors, all of them are combined into a single
$r$-parameter family of s-e flux vectors which can always be written as
\b*
\bbb{P}^{\nu}_t\left(\vec{u}\right)\equiv
-\bbb{T}^{\nu}{}_{\mu_1\dots\lambda_r\mu_r}\{t\}
u^{\mu_1}\dots u^{\lambda_r}u^{\mu_r}=
\sum_{\U =1}^{r} C_{\U} \left({}^{\U}P^{\nu}_t\left(\vec{u}\right)\right)
\e*
where the non-negative constants $C_{\U}$ are appropriate linear
combinations of the $c_{\sigma}$. From this one sees that
$\vec{\bbb{P}}_t(\vec{u})$ is causal and with the same time-orientation than
$\vec{u}$, so that Property \ref{pr:2} is also kept.

In fact, from Property \ref{pr:alg}, Theorem \ref{th:dsep} and definition
(\ref{gset}) we have
\begin{coro}
The DSEP holds for the general s-e tensor $\bbb{T}$.\N
\end{coro}
In other words, a $(2r)!/2^r$-parameter family of s-e tensors satisfying the
good mathematical properties mentioned above and including the fundamental DSEP
has been constructed.

A possible way to avoid this freedom (if this is desired) is to take the
completely symmetric part of the general s-e tensor, which coincides
with the fully symmetric part of the basic s-e tensor (\ref{set}):
$\bbb{T}_{(\mu_1\mu_2\dots\mu_{2r-1}\mu_{2r})}\left\{t\right\}\propto
T_{(\mu_1\mu_2\dots\mu_{2r-1}\mu_{2r})}\left\{t\right\}$. This has no
incidence on the super-energy density, in which definition only the
completely symmetric part
$T_{(\mu_1\mu_2\dots\mu_{2r-1}\mu_{2r})}\left\{t\right\}$ is relevant, and
selects a unique s-e flux vector. In fact, in some occasions this
completely symmetric part arises also in relation with the existence
of conserved currents constructed from s-e tensors, see Sections
\ref{sec:physics} and \ref{sec:cons}. Nevertheless, I have preferred to
maintain the generality and the general s-e tensor (\ref{gset}) will be used
too.

To illustrate these points, let us consider the simplest cases explicitly.
For $r=1$ (simple $p$-forms $\bm{\S}$), the basic tensor
(\ref{set1}-\ref{set1'}) coincides obviously with the general s-e
tensor $\bbb{T}_{\lambda\mu}\left\{\S_{[p]}\right\}=c
T_{\lambda\mu}\left\{\S_{[p]}\right\}$ so that this case is trivial.
When $r=2$ (double $(p,q)$-forms $K_{[p],[q]}$), the general s-e tensor
$\bbb{T}_{\alpha\beta\lambda\mu}\left\{K_{[p],[q]}\right\}$ is constructed
from the basic s-e tensor (\ref{set11}-\ref{set11'}) as follows
\bea
\bbb{T}_{\alpha\beta\lambda\mu}\left\{K_{[p],[q]}\right\}=
c_1T_{\alpha\beta\lambda\mu}\left\{K_{[p],[q]}\right\}+
c_2 T_{\alpha\lambda\beta\mu}\left\{K_{[p],[q]}\right\}+
c_3T_{\alpha\mu\lambda\beta}\left\{K_{[p],[q]}\right\}+\nonumber\\
+c_4T_{\lambda\beta\alpha\mu}\left\{K_{[p],[q]}\right\}+
c_5T_{\mu\beta\lambda\alpha}\left\{K_{[p],[q]}\right\}+
c_6T_{\lambda\mu\alpha\beta}\left\{K_{[p],[q]}\right\}.\label{c16}
\eea
Here $c_1,\dots ,c_6$ are assumed to be non-negative. This general s-e
tensor does not have any symmetry in general. However, it is automatically
symmetric in the interchange of $(\alpha\beta)\leftrightarrow (\lambda\mu)$
whenever the original $T_{\alpha\beta\lambda\mu}\left\{K_{[p],[q]}\right\}$
has this property, in which case the following notation will be
used: $\hat{c}_1=c_1+c_6$, $\hat{c}_2=c_2+c_5$,
$\hat{c}_3=c_3+c_4$, $\hat{c}_4=\hat{c}_5=\hat{c}_6=0$. Thus, there appears a
six-parameter ($4!/2^2=6$)
family (reduced to a three-parameter one when the interchange
between pairs holds for $T_{\alpha\beta\lambda\mu}\left\{K_{[p],[q]}\right\}$)
of s-e tensors satisfying the fundamental DSEP. Furthermore,
$\bbb{T}_{\alpha\beta\lambda\mu}\left\{K_{[p],[q]}\right\}$ is symmetric in
$\alpha\beta$ iff $c_2=c_4$ and $c_3=c_5$ (or $\hat{c}_2=\hat{c}_3$ in the
special case), and symmetric in $\lambda\mu$ if and only if $c_2=c_3$ and
$c_4=c_5$ (respectively $\hat{c}_2=\hat{c}_3$). This provides a
three-parameter (resp.\ two-parameter) family of s-e tensors with the same
symmetries as the basic one. Concerning the interchange of pairs,
$\bbb{T}_{\alpha\beta\lambda\mu}\left\{K_{[p],[q]}\right\}$ is symmetric
with respect to the exchange $(\alpha\beta)\leftrightarrow (\lambda\mu)$ if
and only if $c_1=c_6$ and $c_3=c_4$, or in general if the basic one had
already this property.

For triple $(p,q,s)$-forms $A_{[p],[q],[s]}$, the general s-e tensor
(\ref{gset}) becomes a $6!/2^3=90$-parameter family of s-e tensors, and
thus the formula analogous to (\ref{c16}) is very large and will
be skipped here.

\section{Application to physical fields. (Super)$^k$-energy tensors.}
\label{sec:physics}
The applications of the above mathematical constructions to the real
physical fields is considered in detail in this section, containing three
different subsections devoted to the gravitational field, to typical 
massless fields (scalar and electromagnetic ones), and
to the corresponding massive fields (scalar and Proca fields),
respectively. Some particular cases of interest will be remarked, such
as that of General Relativity. The traditional s-e tensors of Bel
\cite{B4,Beltesis}, Bel-Robinson \cite{B1,B2,Rob}, and Chevreton \cite{C}
will be rederived as particular
cases of the general construction.

\subsection{The gravitational field}
\label{subsec:grav}
Let us assume that the gravitational field is described by the curvature tensor
$R^{\alpha}{}_{\beta\lambda\mu}$ of the Lorentzian manifold $(V_n,g)$. Given
that $R_{\alpha\beta,\lambda\mu}$ is a double symmetric (2,2)-form,
the basic s-e tensor for the gravitational field is simply the appropriate
restriction of (\ref{set11}):
\b*
T_{\alpha}{}^{\beta}{}_{\lambda}{}^{\mu}\left\{R_{[2],[2]}\right\}=
\frac{1}{2}\left(R_{\alpha\rho,\lambda\sigma}
R^{\beta\rho,\mu\sigma}
+\frac{1}{(n-3)!}\,R_{\alpha\rho,
\stackrel{*}{\lambda\sigma_{4}\dots\sigma_n}}
R^{\beta\rho,\stackrel{*}{\mu\sigma_{4}\dots\sigma_n}}
+\right.\\
+\left.\frac{1}{(n-3)!}\,
R_{\stackrel{*}{\alpha\rho_{4}\dots\rho_n},\lambda\sigma}
R^{\stackrel{*}{\beta\rho_{4}\dots\rho_n},\mu\sigma}
+\frac{1}{\left[(n-3)!\right]^2}
R_{\stackrel{*}{\alpha\rho_{4}\dots\rho_n},
\stackrel{*}{\lambda\sigma_{4}\dots\sigma_n}}
R^{\stackrel{*}{\beta\rho_{4}\dots\rho_n},
\stackrel{*}{\mu\sigma_{4}\dots\sigma_n}}\right) 
\e*
which is a straightforward generalization of the original definition
given by Bel \cite{B4,Beltesis}. As remarked at the end of section
\ref{sec:set}, after expanding the duals the expression for this generalized
Bel tensor does not depend on the
dimension $n$, and its form (\ref{set11'}) reads (see \cite{C}
for $n=4$, and also \cite{Gal})
\bea
B_{\alpha\beta\lambda\mu}\equiv 
T_{\alpha\beta\lambda\mu}\left\{R_{[2],[2]}\right\}=
R_{\alpha\rho,\lambda\sigma}
R_{\beta}{}^{\rho,}{}_{\mu}{}^{\sigma}
+R_{\alpha\rho,\mu\sigma}
R_{\beta}{}^{\rho,}{}_{\lambda}{}^{\sigma}-\nonumber\\
-\frac{1}{2}g_{\alpha\beta}
R_{\rho\tau,\lambda\sigma}R^{\rho\tau,}{}_{\mu}{}^{\sigma}
-\frac{1}{2}g_{\lambda\mu}
R_{\alpha\rho,\sigma\tau}R_{\beta}{}^{\rho,\sigma\tau}+
\frac{1}{8}g_{\alpha\beta}g_{\lambda\mu}
R_{\rho\tau,\sigma\nu}
R^{\rho\tau,\sigma\nu} \label{Bel}
\eea
from where properties (\ref{sym11}) are manifest and also
\be
B_{\alpha\beta\lambda\mu}=B_{(\alpha\beta)(\lambda\mu)}=
B_{\lambda\mu\alpha\beta}.\label{belsym}
\ee
Furthermore, using the first Bianchi identity
$R^{\alpha}{}_{[\beta\lambda\mu]}=0$ if necessary
\bea
B^{\rho}{}_{\rho\lambda\mu}=B_{\lambda\mu}{}^{\rho}{}_{\rho}=
\frac{(4-n)}{2}\left(R_{\rho\tau,\lambda\sigma}
R^{\rho\tau,}{}_{\mu}{}^{\sigma}
-\frac{1}{4}g_{\lambda\mu}R_{\rho\tau,\sigma\nu}R^{\rho\tau,\sigma\nu}
\right), \label{beltr} \\
B^{\rho}{}_{\rho}{}^{\sigma}{}_{\sigma}=
\frac{(4-n)^2}{8}R_{\rho\tau,\sigma\nu}R^{\rho\tau,\sigma\nu},
\hspace{15mm} \label{beltrtr} \\
B^{\rho}{}_{\beta\rho\mu}=R^{\rho\sigma}R_{\beta\rho,\mu\sigma}-
\frac{1}{2}R_{\beta\rho,\tau\sigma}R_{\mu}{}^{\rho,\tau\sigma}+
\frac{1}{8}g_{\beta\mu}R_{\rho\tau,\sigma\nu}R^{\rho\tau,\sigma\nu},
\label{beltrr}\\
B^{\rho\sigma}{}_{\rho\sigma}=R^{\rho\sigma}R_{\rho\sigma}+
\frac{n-4}{8}R_{\rho\tau,\sigma\nu}R^{\rho\tau,\sigma\nu}.\hspace{1cm}
\label{beltrrtrr}
\eea

Due to the general results proved in the previous sections, the generalized
Bel tensor has the DSEP and the generalized Bel s-e density is defined by
\b*
W_B\left(\vec{u}\right)\equiv
B_{\alpha\beta\lambda\mu}u^{\alpha}u^{\beta}u^{\lambda}u^{\mu}
\e*
so that
\b*
W_B\left(\vec{u}\right)= \frac{1}{2}\left([{}^{R}_{\vec{u}}EE]^2+
[{}^{R}_{\vec{u}}EH]^2+[{}^{R}_{\vec{u}}HE]^2+[{}^{R}_{\vec{u}}HH]^2
\right)\geq 0
\e*
where $\left({}^{R}_{\vec{u}}EE\right)_{[1],[1]}$,
$\left({}^{R}_{\vec{u}}EH\right)_{[1],[n-3]}$,
$\left({}^{R}_{\vec{u}}HE\right)_{[n-3],[1]}$ and
$\left({}^{R}_{\vec{u}}HH\right)_{[n-3],[n-3]}$ are the electric-electric,
electric-magnetic, magnetic-electric and magnetic-magnetic parts of the
Riemann tensor according to the definition of Section \ref{sec:EH}.
For General Relativity ($n=4$) they were first introduced by Bel in
\cite{XYZZ}, see also \cite{BS2}. These E-H parts satisfy in general
\b*
\left({}^{R}_{\vec{u}}EE\right)_{\mu\nu}=
\left({}^{R}_{\vec{u}}EE\right)_{\nu\mu}, \hspace{3mm}
\left({}^{R}_{\vec{u}}EH\right)_{\mu,\nu_1\dots\nu_{n-3}}=
\left({}^{R}_{\vec{u}}HE\right)_{\nu_1\dots\nu_{n-3},\mu},\\
\left({}^{R}_{\vec{u}}HH\right)_{\mu_1\dots\mu_{n-3},\nu_1\dots\nu_{n-3}}=
\left({}^{R}_{\vec{u}}HH\right)_{\nu_1\dots\nu_{n-3},\mu_1\dots\mu_{n-3}}.
\e*
Again, $W_B$ vanishes if and only if the whole Riemann tensor vanishes too
\b*
\left\{\exists \vec{u}\hspace{3mm} \mbox{such that} \hspace{2mm}
W_{B}\left(\vec{u}\right)=0 \right\} \Longleftrightarrow
B_{\alpha\beta\lambda\mu}=0 \Longleftrightarrow
R_{\alpha\beta\lambda\mu}=0 .
\e*
Concerning the generalized Bel s-e flux, from properties (\ref{belsym}) and
the results of Section \ref{sec:general} it follows that there is only one
independent such vector, given by
\b*
P^{\alpha}_B\left(\vec{u}\right)\equiv -B^{\alpha}{}_{\beta\mu\nu}u^{\beta}
u^{\mu}u^{\nu}
\e*
which is a causal vector with the same time orientation than $\vec{u}$.

In general, from the analysis presented in Section \ref{sec:general}, the
general s-e tensor for the gravitational field is the following 3-parameter
family of tensors
\be
\bbb{B}_{\alpha\beta\lambda\mu}\equiv \hat{c}_1 B_{\alpha\beta\lambda\mu}+
\hat{c}_2 B_{\alpha\lambda\beta\mu}+\hat{c}_3 B_{\alpha\mu\lambda\beta}
\label{belfam}
\ee
whose symmetry properties are
\b*
\bbb{B}_{\alpha\beta\lambda\mu}=\bbb{B}_{\beta\alpha\mu\lambda}=
\bbb{B}_{\lambda\mu\alpha\beta}.
\e*
However, we also have
\b*
\mbox{If} \hspace{5mm} \hat{c}_2=\hat{c}_3 \hspace{1cm} \Longrightarrow
\hspace{1cm} \bbb{B}_{\alpha\beta\lambda\mu}=
\bbb{B}_{(\alpha\beta)(\lambda\mu)},\\
\mbox{If} \hspace{5mm} \hat{c}_1=\hat{c}_2=\hat{c}_3 \hspace{1cm}
\Longrightarrow \hspace{1cm}
\bbb{B}_{\alpha\beta\lambda\mu}=\bbb{B}_{(\alpha\beta\lambda\mu)}.
\e*
Both cases have been recently analyzed in \cite{Tey}, and the second one
leads to a tensor considered also in \cite{Rob}.

One of the important properties of the generalized Bel tensor is the
expression for its divergence, which does not vanish in general. By
using the second Bianchi identity $\nabla_{[\nu}R_{\alpha\beta]\lambda\mu}=0$,
one can deduce from (\ref{Bel})
\be
\nabla_{\alpha}B^{\alpha\beta\lambda\mu}=
R^{\beta\hspace{1mm}\lambda}_{\hspace{1mm}\rho\hspace{2mm}\sigma}
J^{\mu\sigma\rho}+R^{\beta\hspace{1mm}\mu}_{\hspace{1mm}\rho\hspace{2mm}\sigma}
J^{\lambda\sigma\rho}-\frac{1}{2}g^{\lambda\mu}
R^{\beta}_{\hspace{1mm}\rho\sigma\gamma}J^{\sigma\gamma\rho}\label{divbel}
\ee
where $J_{\lambda\mu\beta}=-J_{\mu\lambda\beta}\equiv
\nabla_{\lambda}R_{\mu\beta}-\nabla_{\mu}R_{\lambda\beta}$. Thus, the
fundamental result that $B$ is divergence-free when the `current' of matter
$J_{\lambda\mu\beta}$ vanishes has been proved. More precisely
\begin{theo}
\label{th:divbel}
If $J_{\lambda\mu\beta}=0$ then the generalized Bel tensor (\ref{Bel}) as well
as the 3-parameter family of tensors (\ref{belfam}) generated from it are 
divergence-free. This includes all Einstein spaces (i.e. with
$R_{\mu\nu}=\Lambda g_{\mu\nu}$).\N
\end{theo}

As is well-known, the Riemann tensor can be always decomposed into the
Ricci tensor $R_{\beta\mu}\equiv R^{\rho}{}_{\beta\rho\mu}$ and a traceless
part denoted by $C_{\alpha\beta\lambda\mu}$ and called the Weyl tensor
\cite{Ei,HE,Lan2}.
This tensor is conformally invariant \cite{Ei,HE} and possesses the same
symmetry properties as the Riemann tensor but is also traceless
\b*
C_{\alpha\beta,\lambda\mu}=C_{[\alpha\beta],[\lambda\mu]}, \hspace{3mm}
C_{\alpha[\beta\lambda\mu]}=0, \hspace{3mm} C^{\rho}{}_{\beta,\rho\mu}=0
\e*
so that $C$ is another double symmetric (2,2)-form. The explicit formula
relating these curvature tensors is \cite{Ei,HE,Lan2}
\be
R_{\alpha\beta,\lambda\mu}=C_{\alpha\beta,\lambda\mu}+\frac{2}{n-2}\left(
R_{\alpha[\lambda}g_{\mu]\beta}-R_{\beta[\lambda}g_{\mu]\alpha}\right)-
\frac{R}{(n-1)(n-2)}\left(g_{\alpha\lambda}g_{\beta\mu}-
g_{\alpha\mu}g_{\beta\lambda}\right)\label{weyl}
\ee
where $R$ stands for the scalar curvature $R\equiv R^{\rho}{}_{\rho}$. Due to
the fact that the gravitational field equations usually involve the Ricci
tensor in terms of the energy-momentum tensor (such as in the Einstein
field equations $R_{\mu\nu}-(1/2)g_{\mu\nu}R=\kappa T_{\mu\nu}$), one
thinks of the Ricci tensor as the part of the curvature {\it directly}
related to the matter fields, and of the Weyl tensor as the {\it free}
gravitational field induced by the sources. Of course, one can construct
the basic s-e tensor for the Weyl tensor, which generalizes the classical
Bel-Robinson tensor \cite{B1,P,PR,Wald} constructed in General Relativity
($n=4$):
\b*
T_{\alpha}{}^{\beta}{}_{\lambda}{}^{\mu}\left\{C_{[2],[2]}\right\}=
\frac{1}{2}\left(C_{\alpha\rho,\lambda\sigma}
C^{\beta\rho,\mu\sigma}
+\frac{1}{(n-3)!}\,C_{\alpha\rho,
\stackrel{*}{\lambda\sigma_{4}\dots\sigma_n}}
C^{\beta\rho,\stackrel{*}{\mu\sigma_{4}\dots\sigma_n}}
+\right.\\
+\left.\frac{1}{(n-3)!}\,
C_{\stackrel{*}{\alpha\rho_{4}\dots\rho_n},\lambda\sigma}
C^{\stackrel{*}{\beta\rho_{4}\dots\rho_n},\mu\sigma}
+\frac{1}{\left[(n-3)!\right]^2}
C_{\stackrel{*}{\alpha\rho_{4}\dots\rho_n},
\stackrel{*}{\lambda\sigma_{4}\dots\sigma_n}}
C^{\stackrel{*}{\beta\rho_{4}\dots\rho_n},
\stackrel{*}{\mu\sigma_{4}\dots\sigma_n}}\right) .
\e*
This can be re-written, analogously to (\ref{Bel}), independently of the
dimension $n$ as (compare with \cite{Des,Gal})
\bea
{\cal T}_{\alpha\beta\lambda\mu}\equiv 
T_{\alpha\beta\lambda\mu}\left\{C_{[2],[2]}\right\}=
C_{\alpha\rho,\lambda\sigma}
C_{\beta}{}^{\rho,}{}_{\mu}{}^{\sigma}
+C_{\alpha\rho,\mu\sigma}
C_{\beta}{}^{\rho,}{}_{\lambda}{}^{\sigma}-\nonumber\\
-\frac{1}{2}g_{\alpha\beta}
C_{\rho\tau,\lambda\sigma}C^{\rho\tau,}{}_{\mu}{}^{\sigma}
-\frac{1}{2}g_{\lambda\mu}
C_{\alpha\rho,\sigma\tau}C_{\beta}{}^{\rho,\sigma\tau}+
\frac{1}{8}g_{\alpha\beta}g_{\lambda\mu}
C_{\rho\tau,\sigma\nu}
C^{\rho\tau,\sigma\nu} \label{BR}
\eea
form where one deduces the symmetry properties
\be
{\cal T}_{\alpha\beta\lambda\mu}={\cal T}_{(\alpha\beta)(\lambda\mu)}=
{\cal T}_{\lambda\mu\alpha\beta}.\label{BRsym}
\ee
Taking into acount that the Weyl tensor is traceless, the formulae
analogous to (\ref{beltr}-\ref{beltrrtrr}) are
\bea
{\cal T}^{\rho}{}_{\rho\lambda\mu}={\cal T}_{\lambda\mu}{}^{\rho}{}_{\rho}=
\frac{(4-n)}{2}\left(C_{\rho\tau,\lambda\sigma}
C^{\rho\tau,}{}_{\mu}{}^{\sigma}
-\frac{1}{4}g_{\lambda\mu}C_{\rho\tau,\sigma\nu}C^{\rho\tau,\sigma\nu}
\right), \label{BRtr} \\
{\cal T}^{\rho}{}_{\rho}{}^{\sigma}{}_{\sigma}=
\frac{(4-n)^2}{8}C_{\rho\tau,\sigma\nu}C^{\rho\tau,\sigma\nu},
\hspace{15mm} \label{BRtrtr} \\
{\cal T}^{\rho}{}_{\beta\rho\mu}=-\frac{1}{2}\left(
C_{\beta\rho,\tau\sigma}C_{\mu}{}^{\rho,\tau\sigma}-
\frac{1}{4}g_{\beta\mu}C_{\rho\tau,\sigma\nu}C^{\rho\tau,\sigma\nu}\right),
\label{BRtrr}\\
{\cal T}^{\rho\sigma}{}_{\rho\sigma}=
\frac{n-4}{8}C_{\rho\tau,\sigma\nu}C^{\rho\tau,\sigma\nu}.\hspace{1cm}
\label{BRtrrtrr}
\eea
The generalized Bel-Robinson tensor (\ref{BR}) satisfies the DSEP, and one
can define the Bel-Robinson s-e density $W_{{\cal T}}\left(\vec{u}\right)$,
the flux vector $\vec{P}_{{\cal T}}\left(\vec{u}\right)$ and the corresponding
E-H parts of the Weyl tensor. All these objects have the same good properties
as those considered in the case of the generalized Bel tensor, and their
definitions can be obtained by simply substituting the Weyl tensor for the
Riemann tensor everywhere (and ${\cal T}$ for $B$). Similarly, there is a
3-parameter family of s-e tensors for the Weyl tensor with properties
analogous to those of $\bbb{B}$. Nevertheless, all these general properties
degenerate in the case of $n=4$ (including General Relativity), where
the Bel-Robinson tensor has special properties to be shown presently.

Concerning the divergence of (\ref{BR}), by using (\ref{weyl}), the
formula for $\nabla_{[\nu}C_{\alpha\beta]\lambda\mu}$ deduced from it and
\b*
\nabla_{\rho}C^{\rho}{}_{\beta\lambda\mu}=2\frac{n-3}{n-2}\left(
\nabla_{[\lambda}R_{\mu]\beta}+
\frac{1}{2(n-1)}g_{\beta[\lambda}\nabla_{\mu]}R\right)
\e*
one arrives to
\begin{prop}
If $R_{\mu\nu}=\Lambda g_{\mu\nu}$, then the generalized Bel-Robinson tensor
(\ref{BR}) as well as the 3-parameter family of tensors generated from it are 
divergence-free.\N
\end{prop}

Obviously, the generalized Bel-Robinsor tensor (\ref{BR}) is always a part of
the generalized Bel tensor (\ref{Bel}), as follows from (\ref{weyl}). Actually,
by introducing (\ref{weyl}) into (\ref{Bel}) one gets the canonical
decomposition of $B$ as
\b*
B_{\alpha\beta\lambda\mu}={\cal T}_{\alpha\beta\lambda\mu}+
{\cal M}_{\alpha\beta\lambda\mu}+{\cal Q}_{\alpha\beta\lambda\mu}
\e*
where ${\cal M}_{\alpha\beta\lambda\mu}$ is called the {\it pure matter}
gravitational s-e tensor (it depends exclusively on the Ricci tensor) and
is given by
\be
{\cal M}_{\alpha\beta\lambda\mu}\equiv
T_{\alpha\beta\lambda\mu}\left\{R_{[2],[2]}-C_{[2],[2]}\right\},
\hspace{1cm} {\cal M}_{\alpha\beta\lambda\mu}=
{\cal M}_{(\alpha\beta)(\lambda\mu)}=
{\cal M}_{\lambda\mu\alpha\beta}\label{M}
\ee
and ${\cal Q}_{\alpha\beta\lambda\mu}$ is the matter-gravity coupling tensor,
containing all the terms with products of the Weyl and Ricci components and
whose explicit expression reads
\b*
{\cal Q}_{\alpha\beta\lambda\mu}\equiv \frac{1}{(n-2)}\left[
4C^{\sigma}{}_{(\lambda\mu)(\alpha}\bar{R}_{\beta)\sigma}+
4C^{\sigma}{}_{(\alpha\beta)(\lambda}\bar{R}_{\mu)\sigma}+
2\bar{R}_{\rho\sigma}\left(C_{\alpha}{}^{\rho}{}_{(\lambda}{}^{\sigma}
g_{\mu)\beta}-C_{\lambda}{}^{\rho}{}_{\mu}{}^{\sigma}g_{\alpha\beta}+
\right.\right.\\
\left.\left.+C_{\beta}{}^{\rho}{}_{(\lambda}{}^{\sigma}g_{\mu)\alpha}-
C_{\alpha}{}^{\rho}{}_{\beta}{}^{\sigma}g_{\lambda\mu}\right)\right],
\hspace{2cm}
\bar{R}_{\alpha\beta}=\bar{R}_{\beta\alpha}\equiv R_{\alpha\beta}-
\frac{R}{2(n-1)}g_{\alpha\beta} \hspace{1cm}
\e*
from where it is easily derived that
\be
{\cal Q}_{\alpha\beta\lambda\mu}={\cal Q}_{(\alpha\beta)(\lambda\mu)}=
{\cal Q}_{\lambda\mu\alpha\beta},\hspace{1cm}
{\cal Q}_{\alpha(\beta\lambda\mu)}=0,\hspace{1cm}
{\cal Q}^{\rho}{}_{\rho\lambda\mu}=
\frac{2(4-n)}{(n-2)}\bar{R}^{\rho\sigma}C_{\rho\lambda\sigma\mu}.\label{Q}
\ee
The form of ${\cal Q}$ is in fact simpler in the case of $n=4$, where
$6\,{\cal Q}_{\alpha\beta\lambda\mu}=R(C_{\alpha\lambda\beta\mu}+
C_{\alpha\mu\beta\lambda})$, as can be proved using spinors \cite{Zu} or
traditional methods \cite{BS2}. Using (\ref{Q}), in general $n$ we have
\be
B_{\alpha(\beta\lambda\mu)}={\cal T}_{\alpha(\beta\lambda\mu)}+
{\cal M}_{\alpha(\beta\lambda\mu)}\label{decom2}
\ee
so that the completely symmetric part of the generalized Bel tensor decomposes
simply as the sum of the completely symmetric parts of the generalized
Bel-Robinson tensor plus that of the pure-matter gravitational s-e tensor.

The pure-matter gravitational s-e tensor (\ref{M}) is interesting because
only the Ricci tensor of the spacetime is involved, which in most cases
is equivalent to saying, due to typical field equations, that only the
energy-matter content is involved. Given that 
${\cal M}$ is the basic s-e tensor (\ref{set}) for the double symmetrical
(2,2)-form $\left(R_{[2],[2]}-C_{[2],[2]}\right)$, it follows from the results
of Sections \ref{sec:set} and \ref{sec:dsep} that the pure-matter gravitational
s-e density $W_{{\cal M}}\left(\vec{u}\right)$ is positive definite, that
the pure-matter gravitational s-e flux vector
$\vec{P}_{{\cal M}}\left(\vec{u}\right)$ is causal and that in general
${\cal M}_{\alpha\beta\lambda\mu}$ satisfies the DSEP. Moreover, one has
from Property \ref{pr:1} and (\ref{cero})
\be
\left\{\exists \vec{u}\hspace{3mm} \mbox{such that} \hspace{2mm}
W_{{\cal M}}\left(\vec{u}\right)=0 \right\} \Longleftrightarrow
{\cal M}_{\alpha\beta\lambda\mu}=0 \Longleftrightarrow
R_{\alpha\beta,\lambda\mu}=C_{\alpha\beta,\lambda\mu} \Longleftrightarrow
R_{\mu\nu}=0. \label{ceroM}
\ee
Even though these properties of ${\cal M}$ indicate that it is a good
s-e tensor for the Ricci tensor\footnote{Notice that another similar basic
s-e tensor can be constructed using that the Ricci tensor $R_{[1][1]}$ is
a double symmetrical (1,1)-form, i.e.,
$T_{\alpha\beta\lambda\mu}\left\{R_{[1][1]}\right\}$. This tensor has all the
above good properties too, and in fact it can be related to
${\cal M}_{\alpha\beta\lambda\mu}$, see for instance \cite{S}.}, one should
bear in mind that this is {\it not} a s-e tensor for the particular field 
(or type of matter) which generates the gravitational field through the field
equations. A good s-e tensor for a matter field must have a well-defined
sense in the absence of gravitation (for instance in Special Relativity),
and this is not true for ${\cal M}$. Somehow, the term {\it gravitational}
pure-matter s-e tensor has been systematically used to remark that this is
a {\it gravitational super-energy}, even though it appears due to
the matter contents of the spacetime {\it exclusively}. The good s-e tensors
for physical fields will be constructed in the next subsections.

From the second property shown in (\ref{Q}), or directly from (\ref{decom2}),
it follows that the matter-gravity coupling tensor ${\cal Q}$ does {\it not}
contribute to the generalized Bel s-e flux vector. More precisely
\begin{prop}
The generalized Bel s-e flux vector is the direct sum of the generalized
Bel-Robinson s-e flux vector and the pure-matter gravitational s-e flux vector.
\end{prop}
\P
Contracting (\ref{decom2}) with the unit timelike $\vec{u}$ thrice one gets
\b*
\vec{P}_{B}\left(\vec{u}\right)=
\vec{P}_{{\cal T}}\left(\vec{u}\right)+\vec{P}_{{\cal M}}\left(\vec{u}\right)
\e*
which proves the result. \N
\begin{coro}
The generalized Bel s-e density decomposes as the simple sum of the generalized
Bel-Robinson s-e density plus the pure-matter gravitational s-e density:
\be
W_{B}\left(\vec{u}\right)=
W_{{\cal T}}\left(\vec{u}\right)+W_{{\cal M}}\left(\vec{u}\right). \label{decom3}
\ee
\N
\end{coro}
\begin{coro}
In general $n$, one has
\b*
B_{\alpha\beta\lambda\mu}={\cal T}_{\alpha\beta\lambda\mu}
\hspace{1cm} \Longleftrightarrow \hspace{1cm} R_{\mu\nu}=0
\e*
\end{coro}
\P
If $R_{\mu\nu}=0$ the result is trivial. Conversely, if
$B_{\alpha\beta\lambda\mu}={\cal T}_{\alpha\beta\lambda\mu}$, then
$W_{B}\left(\vec{u}\right)=W_{{\cal T}}\left(\vec{u}\right)$ so that from
(\ref{decom3}) we arrive at $W_{{\cal M}}\left(\vec{u}\right)=0$. Using
(\ref{ceroM}) the corollary follows.\N

Therefore, in Ricci-flat spacetimes there is no need to distinguish between
the generalized Bel and Bel-Robinson tensors.

The properties of the different s-e densities allow to compare the relative
strength of the Riemman, Weyl and Ricci tensors in a given $(V_n,g)$. Thus,
for instance, one can define the three positive scalars
\b*
q_1\equiv \frac{W_{{\cal M}}\left(\vec{u}\right)}{W_{B}\left(\vec{u}\right)},
\hspace{1cm}
q_2\equiv
\frac{W_{{\cal T}}\left(\vec{u}\right)}{W_{{\cal M}}\left(\vec{u}\right)},
\hspace{1cm}
q_3\equiv \frac{W_{{\cal T}}\left(\vec{u}\right)}{W_{B}\left(\vec{u}\right)}
=q_1q_2
\e*
relative to any observer $\vec{u}$, which are not independent in general and 
such that, due to (\ref{decom3})
\b*
0\leq q_1\leq 1, \hspace{1cm} 0\leq q_2\leq \infty , \hspace{1cm}
0\leq q_3\leq 1 .
\e*
These may serve to measure the strength of
the Ricci tensor in front of the pure Weyl tensor or of the whole Riemann
tensor. Thus for example, $q_2$ can be used to check the so-called
Penrose's Weyl tensor hypothesis \cite{P6,W3} and similar conjectures concerning
the entropy of the gravitational field \cite{Bon}. On the other hand,
$q_1$ may be a `quality factor' for approximate solutions of some field
equations as explained in \cite{calidad,BS2}. In this sense, notice that
$q_1=0$ if and only if $R_{\mu\nu}=0$, which is usually equivalent to the
absence of matter-energy content. Thus, the closest the adimensional number
$q_1$ is to zero for any given metric and some $\vec{u}$, the best resemblance
it has to an exact `vacuum' solution for that observer $\vec{u}$. Observe
that one does not need to know the explicit exact solution one wishes to
approximate in order to compute $q_1$, so that the goodness of
approximate solutions may be measured without ever knowing the form of the
exact solution. Finally, $q_3$ gives the relative strength of the Weyl and
the Riemann tensors or, in other words, of the free gravitational field
with respect to the total one. The number $q_3$ vanishes if and only if 
the metric is conformally flat, so that $q_3$ measures the departure from
this condition somehow. This can have relevance in, for instance, General
Relativity and other typical theories, where the standard cosmological models
are conformally flat.

In General Relativity ($n=4$), the Bel-Robinson tensor has been widely
studied and used, and it is well-known that it is a completely symmetric
and traceless tensor
\cite{Beltesis,B2,Ber,BS2,BFI,Lord,HS,MB,MCQ,MTW,P,PR,Rob,S,S2,Wald,Z,Zu}.
These properties do not hold in general, and in fact (see also \cite{Gal}):
\begin{prop}
The generalized Bel-Robinson tensor is completely symmetric if and only if
$n=4,5$.
\end{prop}
\P
From (\ref{BRtrtr}) and (\ref{BRtrrtrr}) one gets
\b*
{\cal T}^{\rho}{}_{\rho}{}^{\sigma}{}_{\sigma}-
{\cal T}^{\rho\sigma}{}_{\rho\sigma}=
\frac{(4-n)(5-n)}{8}C_{\rho\tau,\sigma\nu}C^{\rho\tau,\sigma\nu}
\e*
and this must be zero if the tensor ${\cal T}$ is completely symmetric.
Thus, a necessary condition for the complete symmetry of the generalized
Bel-Robinson tensor is that $n$ be either 4 or 5. To see that it is
also sufficient, let us consider both cases separately.

\underline{Case $n=5$}.
From formula (\ref{ap:3}) in the Appendix, if $n=5$
\bea
C_{\stackrel{*}{\alpha\rho\sigma}}{}^{\stackrel{*}{\lambda\tau\nu}}
C^{\stackrel{*}{\beta\rho\sigma}}{}_{\stackrel{*}{\mu\tau\nu}}=
4C_{\alpha\rho}{}^{\lambda\sigma}C^{\beta\rho}{}_{\mu\sigma}-\hspace{5mm}
\nonumber\\
-2\delta^{\lambda}_{\mu}C_{\alpha\rho\tau\nu}C^{\beta\rho\tau\nu}-
2\delta^{\beta}_{\alpha}C^{\lambda\rho\tau\nu}C_{\mu\rho\tau\nu}+
\delta^{\beta}_{\alpha}\delta^{\lambda}_{\mu}
C_{\rho\tau,\sigma\nu}C^{\rho\tau,\sigma\nu}\label{1}
\eea
but from the identity (\ref{**1}) or directly from (\ref{ap:slanczos})
we also have
\b*
C_{\stackrel{*}{\alpha\rho\sigma}}{}^{\stackrel{*}{\lambda\tau\nu}}=
-9C^{[\lambda\tau}{}_{[\alpha\rho}\delta^{\nu]}_{\sigma]}
\e*
so that using repeatedly the trace-free property of the Weyl tensor, a
straightforward calcualtion leads to
\bea
C_{\stackrel{*}{\alpha\rho\sigma}}{}^{\stackrel{*}{\lambda\tau\nu}}
C^{\stackrel{*}{\beta\rho\sigma}}{}_{\stackrel{*}{\mu\tau\nu}}=
-4C_{\alpha\rho}{}^{\lambda\sigma}C^{\beta\rho}{}_{\mu\sigma}+
8C_{\alpha\sigma}{}^{\beta\rho}C^{\lambda\sigma}{}_{\mu\rho}+
4C_{\alpha\mu}{}^{\rho\sigma}C^{\lambda\beta}{}_{\rho\sigma}+
2\delta^{\beta}_{\alpha}C^{\lambda\rho\tau\nu}C_{\mu\rho\tau\nu}+\nonumber\\
+2\delta^{\lambda}_{\mu}C_{\alpha\rho\tau\nu}C^{\beta\rho\tau\nu}-
4\delta^{\beta}_{\mu}C_{\alpha\rho\tau\nu}C^{\lambda\rho\tau\nu}
-4\delta^{\lambda}_{\alpha}C^{\beta\rho\tau\nu}C_{\mu\rho\tau\nu}+
\delta^{\lambda}_{\alpha}\delta^{\beta}_{\mu}
C_{\rho\tau,\sigma\nu}C^{\rho\tau,\sigma\nu}.\hspace{5mm}\label{2}
\eea
As (\ref{1}) and (\ref{2}) must be equal, the following identity follows
(see also \cite{Lov})
\b*
\mbox{If $n=5$} \hspace{5mm} \Longrightarrow \hspace{5mm}
2\left(C_{\alpha\rho}{}^{\lambda\sigma}C^{\beta\rho}{}_{\mu\sigma}-
C_{\alpha\rho}{}^{\beta\sigma}C^{\lambda\rho}{}_{\mu\sigma}\right)-
C_{\alpha\mu}{}^{\rho\sigma}C^{\lambda\beta}{}_{\rho\sigma}
-\delta^{\lambda}_{\mu}C_{\alpha\rho\tau\nu}C^{\beta\rho\tau\nu}-\\
-\delta^{\beta}_{\alpha}C^{\lambda\rho\tau\nu}C_{\mu\rho\tau\nu}+
\delta^{\beta}_{\mu}C_{\alpha\rho\tau\nu}C^{\lambda\rho\tau\nu}
+\delta^{\lambda}_{\alpha}C^{\beta\rho\tau\nu}C_{\mu\rho\tau\nu}+
\frac{1}{4}\left(\delta^{\beta}_{\alpha}\delta^{\lambda}_{\mu}-
\delta^{\lambda}_{\alpha}\delta^{\beta}_{\mu}\right)
C_{\rho\tau,\sigma\nu}C^{\rho\tau,\sigma\nu}=0.
\e*
But due to $C_{\alpha[\beta\mu\nu]}=0$, 
\b*
C_{\alpha\mu}{}^{\rho\sigma}C^{\lambda\beta}{}_{\rho\sigma}=
2C_{\alpha\rho\mu\sigma}\left(C^{\beta\sigma\lambda\rho}-
C^{\beta\rho\lambda\sigma}\right)
\e*
so that the previous identity, using (\ref{BR}), can be simply written as
\b*
{\cal T}_{\alpha\beta\lambda\mu}-{\cal T}_{\alpha\lambda\beta\mu}=0
\hspace{1cm} \mbox{if $n=5$},
\e*
which together with (\ref{BRsym}) proves that
${\cal T}_{\alpha\beta\lambda\mu}$ is completely symmetric in $n=5$.

\underline{Case $n=4$} (see \cite{Beltesis}).
From (\ref{ap:3}) in the Appendix, if $n=4$ 
\b*
C_{\stackrel{*}{\alpha\rho}}{}^{\stackrel{*}{\lambda\sigma}}
C^{\stackrel{*}{\beta\rho}}{}_{\stackrel{*}{\mu\sigma}}=
C_{\alpha\rho}{}^{\lambda\sigma}C^{\beta\rho}{}_{\mu\sigma}
-\frac{1}{2}\delta^{\lambda}_{\mu}C_{\alpha\rho\tau\nu}C^{\beta\rho\tau\nu}-
\frac{1}{2}\delta^{\beta}_{\alpha}C^{\lambda\rho\tau\nu}C_{\mu\rho\tau\nu}+
\frac{1}{4}\delta^{\beta}_{\alpha}\delta^{\lambda}_{\mu}
C_{\rho\tau,\sigma\nu}C^{\rho\tau,\sigma\nu}
\e*
but the Lanczos identity (\ref{ap:lanczos}) implies
\be
C_{\stackrel{*}{\alpha\rho}}{}^{\stackrel{*}{\lambda\tau}}=-
C^{\lambda\tau}{}_{\alpha\rho}\label{3}
\ee
so that we obtain \cite{Lov}
\b*
\delta^{\lambda}_{\mu}C_{\alpha\rho\tau\nu}C^{\beta\rho\tau\nu}+
\delta^{\beta}_{\alpha}C^{\lambda\rho\tau\nu}C_{\mu\rho\tau\nu}-
\frac{1}{2}\delta^{\beta}_{\alpha}\delta^{\lambda}_{\mu}
C_{\rho\tau,\sigma\nu}C^{\rho\tau,\sigma\nu}=0
\e*
or equivalently
\b*
C_{\alpha\rho\tau\nu}C^{\beta\rho\tau\nu}-
\frac{1}{4}\delta^{\beta}_{\alpha}
C_{\rho\tau,\sigma\nu}C^{\rho\tau,\sigma\nu}=0
\hspace{1cm} \mbox{if $n=4$}.
\e*
Observe that this implies the complete tracelessness of the Bel-Robinson
tensor in $n=4$ due to (\ref{BRtrr}). From (\ref{3}) it follows that
\be
C_{\stackrel{*}{\alpha\beta},\lambda\mu}=
C_{\alpha\beta,\stackrel{*}{\lambda\mu}}\equiv
\stackrel{*}{C}_{\alpha\beta\lambda\mu}\hspace{1cm} \mbox{if $n=4$}\label{c*}
\ee
so that only one independent dual can be constructed in $n=4$. Thus, the
traditional formulas for the Bel-Robinson tensor in $n=4$ can now be
written \cite{B1,B2,P,PR,Wald,Z}:
\bea
{\cal T}_{\alpha}{}^{\beta}{}_{\lambda}{}^{\mu}=
C_{\alpha\rho\lambda\sigma}C^{\beta\rho\mu\sigma}+
\stackrel{*}{C}_{\alpha\rho\lambda\sigma}
\stackrel{*}{C}{}^{\beta\rho\mu\sigma} , \hspace{1cm} \mbox{if $n=4$,}
\label{BR*}\\
{\cal T}_{\alpha\beta\lambda\mu}=C_{\alpha\rho\lambda\sigma}
C_{\beta}{}^{\rho}{}_{\mu}{}^{\sigma}+C_{\alpha\rho\mu\sigma}
C_{\beta}{}^{\rho}{}_{\lambda}{}^{\sigma}-\frac{1}{8}g_{\alpha\beta}
g_{\lambda\mu}C_{\rho\tau,\sigma\nu}C^{\rho\tau,\sigma\nu} , \, \, \,
\mbox{if $n=4$} .\nonumber
\eea
Using $C_{\alpha[\beta\mu\nu]}=0$ one can write
\b*
C_{\alpha\rho\lambda\sigma}=\frac{1}{2}\left(C_{\alpha\rho\lambda\sigma}+
C_{\lambda\rho\alpha\sigma}+C_{\alpha\lambda\rho\sigma}\right)
\e*
and analogously
\b*
\stackrel{*}{C}_{\alpha\rho\lambda\sigma}=\frac{1}{2}
\left(\stackrel{*}{C}_{\alpha\rho\lambda\sigma}+
\stackrel{*}{C}_{\lambda\rho\alpha\sigma}+
\stackrel{*}{C}_{\alpha\lambda\rho\sigma}\right)
\e*
so that formula (\ref{BR*}) becomes
\b*
{\cal T}_{\alpha\beta\lambda\mu}=\frac{1}{4}
\left(C_{\alpha\rho\lambda\sigma}+
C_{\lambda\rho\alpha\sigma}\right)
\left(C_{\beta}{}^{\rho}{}_{\mu}{}^{\sigma}+
C_{\mu}{}^{\rho}{}_{\beta}{}^{\sigma}\right)+\frac{1}{4}
C_{\alpha\lambda\rho\sigma}C_{\beta\mu}{}^{\rho\sigma}+\hspace{1cm} \\
\frac{1}{4}
\left(\stackrel{*}{C}_{\alpha\rho\lambda\sigma}+
\stackrel{*}{C}_{\lambda\rho\alpha\sigma}\right)
\left(\stackrel{*}{C}_{\beta}{}^{\rho}{}_{\mu}{}^{\sigma}+
\stackrel{*}{C}_{\mu}{}^{\rho}{}_{\beta}{}^{\sigma}\right)+\frac{1}{4}
\stackrel{*}{C}_{\alpha\lambda\rho\sigma}
\stackrel{*}{C}_{\beta\mu}{}^{\rho\sigma} \hspace{1cm} \mbox{if $n=4$}
\e*
and using here that, from (\ref{ap:form2}) or directly from (\ref{**1}),
\b*
C_{\alpha\lambda\rho\sigma}C_{\beta\mu}{}^{\rho\sigma}+
\stackrel{*}{C}_{\alpha\lambda\rho\sigma}
\stackrel{*}{C}_{\beta\mu}{}^{\rho\sigma}=0 \hspace{1cm} \mbox{if $n=4$}
\e*
one obtains an expression for ${\cal T}_{\alpha\beta\lambda\mu}$ which
has manifestly the property
\b*
{\cal T}_{\alpha\beta\lambda\mu}={\cal T}_{\lambda\beta\alpha\mu}
\hspace{1cm} \mbox{if $n=4$}
\e*
which together with (\ref{BRsym}) proves the result. \N

From the above proof, only one independent dual of the Weyl tensor is
possible in $n=4$, see (\ref{c*}). Thus, the E-H decomposition of the Weyl
tensor in $n=4$ is very simple, because from (\ref{3}) and (\ref{c*})
it follows
\b*
\left({}^{C}_{\vec{u}}EE\right)_{\mu\nu}=
-\left({}^{C}_{\vec{u}}HH\right)_{\mu\nu},\hspace{5mm}
\left({}^{C}_{\vec{u}}EH\right)_{\mu\nu}=
\left({}^{C}_{\vec{u}}HE\right)_{\mu\nu}\hspace{1cm} \mbox{if $n=4$}
\e*
which are called the electric and magnetic parts of the Weyl tensor
relative to $\vec{u}$ if $n=4$ \cite{XYZZ,Beltesis,B2,MB,Mat,MAWH}. These
tensors are spatial relative to $\vec{u}$, symmetric and traceless; they were
first introduced by Matte \cite{Mat}. An analysis of the positivity properties
derived from the DSEP for (\ref{BR*}) appears in \cite{Ber}. Such an analysis,
and its consequences, may be translated to the general (\ref{set}) in 
principle.

A traditional question concerning the Bel-Robinson tensor is what type
of physical quantity, if any, is described by it. In this sense, one must
notice that the physical dimensions of ${\cal T}$ or $B$ are $L^{-4}$, where
$L$ means length. This has been usually separated into two $L^{-2}$'s,
and led to two different classical interpretations: either both $L^{-2}$'s
are on the same footing describing an energy density each so that
$W_{B}\left(\vec{u}\right)$ is an energy density ``square''
(\cite{Ber2,BS2,Ro} and references therein), or one of the $L^{-2}$'s is
a `true' $L^{-2}$ and $W_{B}\left(\vec{u}\right)$ represents energy density per
unit surface (or second derivatives of energy density).
The first interpretation was supported because, for instance,
in Einstein-Maxwell spacetimes and $n=4$, the pure-matter gravitational
tensor takes the simple form ${\cal M}_{\alpha\beta\lambda\mu}=
R_{\alpha\beta}R_{\lambda\mu}$, which through Einstein's field equations
can be rewritten as the square of the energy-momentum tensor \cite{BS2}.
This led to the possibility of finding a `square root' of the Bel-Robinson
tensor, which would be a good two-index tensor with appropriate physical
dimensions. Unfortunately, this square root exists only in some particular
cases \cite{BS2,Ber2}. A generalization of the square root to two different
factors is always possible \cite{BS2,Ber2}, but in this case the factors
are not unique in general \cite{BS2,Ber2}. Thus, the first interpretation
seems dubious. On the other hand, the second interpretation seems correct
as can be deduced from several independent results \cite{Ga,HS,Tey}. Notably,
the relationship \cite{Ber0,BerL,BY,BLY,Dou,DM,HS,K,LV,Sza,Sza4}
\b*
E_a = \mbox{const.\,} W_{{\cal T}}\left(\vec{u}\right)\mid_{r=0} a^5 + O(a^6)
\e*
between any of the quasilocal energies $E_a$ of small 2-spheres of radius $a$
in vacuum and the Bel-Robinson s-e density $W_{{\cal T}}\left(\vec{u}\right)$
supports clearly this view (${\vec u}$ orthogonal to the 2-sphere and $n=4$).
Similarly, the results in \cite{Ga} point into this direction and, more
importantly, Teyssandier \cite{Tey} has recently found a formula for the
`super-Hamiltonian' associated to the super-energy of the scalar field
(see next subsections) in Special Relativity from where one checks that
this super-energy is interchanged in quanta of $\hbar \omega_k^3/c^2$, where
$\omega_k$ is the frequency of the $k$-mode. All in all, due to the
independence of these three results, it seems that the second interpretation
is valid, at least in $n=4$, as also suggested in \cite{Des2}.

To end this section, let us finally make some comments about the higher
order super-energy tensors for the gravitational field. Instead of
considering the Riemann tensor as fundamental field, one can also take
its covariant derivative and repeat all the construction above. Specifically,
one can consider the tensor $\nabla_{\nu}R_{\alpha\beta\lambda\mu}$ as
a triple (1,2,2)-form and construct its corresponding s-e tensor (\ref{set}),
$T_{\alpha\beta\lambda\mu\tau\nu}\left\{\nabla_{[1]}R_{[2],[2]}\right\}$,
and analogously for the Weyl tensor, etcetera. This provides an
infinite set of (super)$^k$-energy tensors for the gravitational field,
one for each natural number $k\geq 1$, according to the following
\begin{defi}
The (super)$^k$-energy tensor of the gravitational field ($k\geq 1$)
is the tensor (\ref{set}) for the $(k+1)$-fold $(1,\dots ,1,2,2)$-form
describing the $(k-1)$-th covariant derivative of the Riemann tensor, that
is
\b*
T_{\lambda_1\mu_1\dots\lambda_k\mu_k\lambda_{k+1}\mu_{k+1}}
\left\{\underbrace{\nabla_{[1]}\dots\nabla_{[1]}}_{k-1} R_{[2],[2]}\right\}.
\e*
\end{defi}
A similar definition can be given using the Weyl tensor. From (\ref{set111'}),
the explicit formula for the (super)$^2$-energy tensor of the gravitational
field is
\bea
T_{\alpha\beta\lambda\mu\tau\nu}\left\{\nabla_{[1]}R_{[2],[2]}\right\}=
\nabla_{\alpha}R_{\lambda\rho\tau\sigma}
\nabla_{\beta}R_{\mu}{}^{\rho}{}_{\nu}{}^{\sigma}+
\nabla_{\beta}R_{\lambda\rho\tau\sigma}
\nabla_{\alpha}R_{\mu}{}^{\rho}{}_{\nu}{}^{\sigma}+
\nabla_{\alpha}R_{\mu\rho,\tau\sigma}
\nabla_{\beta}R_{\lambda}{}^{\rho}{}_{\nu}{}^{\sigma}+\nonumber\\
+\nabla_{\alpha}R_{\lambda\rho\nu\sigma}
\nabla_{\beta}R_{\mu}{}^{\rho}{}_{\tau}{}^{\sigma}-g_{\alpha\beta}
\left(\nabla_{\z}R_{\lambda\rho\tau\sigma}
\nabla^{\z}R_{\mu}{}^{\rho}{}_{\nu}{}^{\sigma}+
\nabla_{\z}R_{\lambda\rho\nu\sigma}
\nabla^{\z}R_{\mu}{}^{\rho}{}_{\tau}{}^{\sigma}\right)-\hspace{1cm}\nonumber\\
-\frac{1}{2}g_{\lambda\mu}\left(\nabla_{\alpha}R_{\z\rho\tau\sigma}
\nabla_{\beta}R^{\z\rho}{}_{\nu}{}^{\sigma}+
\nabla_{\alpha}R_{\z\rho\nu\sigma}
\nabla_{\beta}R^{\z\rho}{}_{\tau}{}^{\sigma}\right)-\hspace{3cm}\nonumber\\
-\frac{1}{2}g_{\tau\nu}\left(
\nabla_{\alpha}R_{\lambda\rho\z\sigma}
\nabla_{\beta}R_{\mu}{}^{\rho}{}^{\z\sigma}+
\nabla_{\alpha}R_{\mu\rho\z\sigma}
\nabla_{\beta}R_{\lambda}{}^{\rho}{}^{\z\sigma}\right)+\hspace{2cm}\nonumber\\
+\frac{1}{2}g_{\alpha\beta}g_{\lambda\mu}
\nabla_{\z}R_{\gamma\rho\tau\sigma}
\nabla^{\z}R^{\gamma\rho}{}_{\nu}{}^{\sigma}+
\frac{1}{2}g_{\alpha\beta}g_{\tau\nu}
\nabla_{\z}R_{\lambda\rho\gamma\sigma}
\nabla^{\z}R_{\mu}{}^{\rho}{}^{\gamma\sigma}+\nonumber\\
+\frac{1}{4}g_{\lambda\mu}g_{\tau\nu}
\nabla_{\alpha}R_{\z\rho\gamma\sigma}
\nabla_{\beta}R^{\z\rho\gamma\sigma}-
\frac{1}{8}g_{\alpha\beta}g_{\lambda\mu}g_{\tau\nu}
\nabla_{\delta}R_{\z\rho\gamma\sigma}
\nabla^{\delta}R^{\z\rho\gamma\sigma}.\label{sseg}
\eea
These (super)$^k$-energy tensors may have some relevance at points of $V_n$
where the Riemann tensor vanishes but such that some derivative of Riemann is
non-zero there (so that every neighbourhood of the point has gravitational
field). From the general results of sections \ref{sec:set} and \ref{sec:dsep}
it follows that all the (super)$^k$-energy tensors satisfy the corresponding
dominant (super)$^k$-energy property, and one can define the
(super)$^k$-energy densities relative to $\vec{u}$. Moreover, we have
\begin{prop}
The gravitational (super)$^k$-energy (tensor) vanishes in a domain $D\subseteq
V_n$ if and only if the $(k-1)$-th covariant derivative of the Riemann tensor is
zero in $D$. In particular, {\em all} gravitational (super)$^k$-energy tensors
vanish in flat regions of $(V_n,g)$. If the gravitational (super)$^k$-energy
(tensor) vanishes in $D$, so do all the (super)$^{\tilde k}$-energy tensors
with $\tilde{k}>k$. The $(V_n,g)$ of constant curvature are characterized by
the vanishing of the gravitational (super)$^2$-energy (tensor).\N
\end{prop}
Thus, for example, in Special Relativity the gravitational
(super)$^k$-energies (and therefore the corresponding (super)$^k$-energy
tensors) vanish. Besides, the previous Proposition clarifies somehow the
reasons why the Bel-Robinson and Bel tensors arise {\it naturally} in
General Relativity: from the equivalence principle, the energy density of the
gravitational field can be always set to zero at any point $x\in V_n$ by
appropriate choice of the observer, and thus the natural concept at the given
point is the super-energy density.
Further related considerations regarding these tensors will
be made in the next subsections.

\subsection{Massless fields}
General massless fields can be considered from the point of view of
super-energy tensors and, sometimes, the use of spinors simplifies things
substantially \cite{Ber,Ber2,Ber3,BerS,C,Lord,PR}. However, the spinor
simplification is usually restricted to dimension $n=4$ and thus, in this
subsection, only the outstanding cases of the electromagnetic and the massless
scalar fields are going to be treated in general dimension $n$.

\subsubsection{The massless scalar field}
Consider a massless scalar field $\phi$, that is, any function $\p$ satisfying
the $n$-dimensional massless Klein-Gordon equation
\be
\nabla_{\rho}\nabla^{\rho}\p=0. \label{KG}
\ee
This is usually referred to as a minimally coupled massless scalar field,
and is not conformally invariant in general \cite{PR}. As the basic object
is $\p$, the simplest s-e tensor (\ref{set}) for $\phi$ would simply be
$T\{\phi\}=\phi^2/2$. This object is obviously positive but it seems to have
no physical relevance. However, one can also take the first covariant
derivative $\nabla _{\mu}\phi$ as starting object so that a case of physical
interest arises by constructing the tensor (\ref{set}) associated to
$\nabla_{[1]}\p$. Using (\ref{set1'}) one gets
\be
T_{\lambda\mu}\{\nabla_{[1]}\phi\}=\nabla_{\lambda}\phi\nabla_{\mu}\phi
-\frac{1}{2}g_{\lambda\mu}\nabla_{\rho}\phi\nabla^{\rho}\phi ,
\hspace{1cm} T_{\mu}{}^{\mu}\{\nabla_{[1]}\phi\}=\left(1-\frac{n}{2}\right)
\nabla_{\rho}\phi\nabla^{\rho}\phi \label{ems}
\ee
which is, in fact, the standard energy-momentum tensor of a {\it massless}
scalar field. The tensor (\ref{ems}) is symmetric and identically
divergence-free if the field equation (\ref{KG}) holds. These two properties
lead to the existence of conserved currents for the scalar
field in any spacetime with a Killing vector field $\vec{\xi}$, that is, with
a vector field satisfying \cite{Ei,HE}
\be
\nabla_{\mu}\xi_{\nu}+\nabla_{\nu}\xi_{\mu}=0.\label{kil}
\ee
The well-known idea is to construct the current
\b*
j^{\mu}\left(\nabla_{[1]} \p;\vec{\xi}\right)\equiv
\xi_{\rho}T^{\rho\mu}\{\nabla_{[1]}\phi\}
\e*
which due to (\ref{kil}) and the properties of
$T_{\lambda\mu}\{\nabla_{[1]}\phi\}$ satisfies
\be
\nabla_{\mu}j^{\mu}\left(\nabla_{[1]} \p;\vec{\xi}\right)=
T^{\rho\mu}\{\nabla_{[1]}\phi\}\nabla_{(\mu}\xi_{\rho)}=0.\label{divems}
\ee
Notice that in this simple calculation {\it both} the divergence-free
as well as the symmetry properties of $T^{\rho\mu}\{\nabla_{[1]}\phi\}$ are
needed. Once that we have a divergence-free current such as
$j^{\mu}\left(\nabla_{[1]} \p;\vec{\xi}\right)$, conserved quantities are
readily constructed by means of the Gauss theorem \cite{HE} and the
integration over appropriate domains of the manifold $V_n$ (sometimes called
Gaussian flux integrals, \cite{MTW}). In flat $(V_n,g)$, they provide
$n(n+1)/2$ independent conserved quantities for the scalar field
(one for each independent Killing vector) related to the energy, the momentum
and the angular momentum of the scalar field. In general non-flat $(V_n,g)$,
they give some conserved quantities related to the intrinsic symmetries
of the spacetime, and to the corresponding combinations of the pertinent
quantity for the scalar field {\it together} with the gravitational field
(described by the curvature of the $(V_n,g)$). Observe finally that only
a conformal Killing vector $\vec{\z}$, that is
\b*
\nabla_{\mu}\z_{\nu}+\nabla_{\nu}\z_{\mu}=\psi g_{\mu\nu},
\e*
is necessary in case that the tensor $T_{\lambda\mu}\{\nabla_{[1]}\phi\}$
be traceless, because then $j^{\mu}\left(\nabla_{[1]} \p;\vec{\z}\right)$ is
also divergence-free in general. In the case of the scalar field, this only
happens for $n=2$ as can be seen from (\ref{ems}).

One can go on with the above construction and use now the double symmetric
(1,1)-form $\nabla_{\alpha}\nabla_{\beta}\phi$ as the starting object, so that
the corresponding tensor (\ref{set}) becomes by using (\ref{set11'})
\bea
S_{\alpha\beta\lambda\mu}\equiv
T_{\alpha\beta\lambda\mu}\{\nabla_{[1]}\nabla_{[1]}\phi\}=
\nabla_{\alpha}\nabla_{\lambda}\phi \nabla_{\mu}\nabla_{\beta}\phi 
+\nabla_{\alpha}\nabla_{\mu}\phi \nabla_{\lambda}\nabla_{\beta}\phi - 
\hspace{1cm} \nonumber \\
-g_{\alpha\beta}\nabla_{\lambda}\nabla^{\rho}\phi\nabla_{\mu}\nabla_{\rho}\phi 
- g_{\lambda\mu}\nabla_{\alpha}\nabla^{\rho}\phi\nabla_{\beta}\nabla_{\rho}\phi 
+\frac{1}{2} g_{\alpha\beta}g_{\lambda\mu}
\nabla_{\sigma}\nabla_{\rho}\phi\nabla^{\sigma}\nabla^{\rho}\phi \, .
\label{ses}
\eea
In fact this tensor was previously found by Bel \cite{B5} and Teyssandier
\cite{Tey} in Special Relativity ($n=4$). In general,
it will be called the basic s-e tensor of the massless scalar field. Its
symmetry properties are
\b*
S_{\alpha\beta\lambda\mu}=S_{(\alpha\beta)(\lambda\mu)}=
S_{\lambda\mu\alpha\beta}
\e*
so that, following the considerations of section \ref{sec:general}, the
general s-e tensor for the massless scalar field is a three-parameter
family given by
\b*
\bbb{S}_{\alpha\beta\lambda\mu}\equiv
\hat{b}_1 S_{\alpha\beta\lambda\mu}+
\hat{b}_2 S_{\alpha\lambda\beta\mu}+\hat{b}_3 S_{\alpha\mu\lambda\beta}
\e*
for arbitrary non-negative constants $\hat{b}_1,\hat{b}_2,\hat{b}_3$.
The symmetry properties of $\bbb{S}$ are the same as that of $\bbb{B}$
\b*
\bbb{S}_{\alpha\beta\lambda\mu}=\bbb{S}_{\beta\alpha\mu\lambda}=
\bbb{S}_{\lambda\mu\alpha\beta}\hspace{2cm}\\
\mbox{If} \hspace{5mm} \hat{b}_2=\hat{b}_3 \hspace{1cm} \Longrightarrow
\hspace{1cm} \bbb{S}_{\alpha\beta\lambda\mu}=
\bbb{S}_{(\alpha\beta)(\lambda\mu)},\\
\mbox{If} \hspace{5mm} \hat{b}_1=\hat{b}_2=\hat{b}_3 \hspace{1cm}
\Longrightarrow \hspace{1cm}
\bbb{S}_{\alpha\beta\lambda\mu}=\bbb{S}_{(\alpha\beta\lambda\mu)}.
\e*
The last two cases have been also found in \cite{Tey} for the
case of Special Relativity ($n=4$).

All in all, the tensors $S_{\alpha\beta\lambda\mu}$ and
$\bbb{S}_{\alpha\beta\lambda\mu}$ have mathematical properties analogous
to those of the generalized Bel tensor and the general s-e tensor of the
gravitational field. This analogy can also be confirmed on physical grounds
by several independent methods, such as a) the analysis performed in
\cite{Tey}, b) the interplay between the propagation of discontinuities of
the Riemann tensor and that of the second derivatives of $\p$ leading to mixed
conserved quantities along null hypersurfaces \cite{S}, or c) the existence
of general conserved currents involving necessarily both the generalized Bel
tensor and the tensor (\ref{ses}), see \cite{S2} and
section \ref{sec:cons}. Therefore, it seems that the tensor
$S_{\alpha\beta\lambda\mu}$ represents for the scalar field the same
type of physical quantities as the traditional Bel and Bel-Robinson tensors
do for the gravitational field. Thus, the super-energy density and the s-e
flux vector of the massless scalar field can be defined using the tensor
(\ref{ses}).

Concerning the divergence of $S_{\alpha\beta\lambda\mu}$, a straightforward
computation using the Ricci identity \cite{Ei,HE,MTW} leads to
\bea
\nabla_{\alpha}S^{\alpha}{}_{\beta\lambda\mu}=
2\nabla_{\beta}\nabla_{(\lambda}\phi
R_{\mu)\rho}\nabla^{\rho}\phi -g_{\lambda\mu} R^{\sigma\rho}
\nabla_{\beta}\nabla_{\rho}\phi\nabla_{\sigma}\phi -\nonumber\\
-\nabla_{\sigma}\phi \left(2\nabla^{\rho}\nabla_{(\lambda}\phi \,
R^{\sigma}_{\mu)\rho\beta} +g_{\lambda\mu}
R^{\sigma}_{\rho\beta\tau}\nabla^{\rho}\nabla^{\tau}\phi\right) \label{divses}
\eea
so that, as we can see, all the terms on the righthand side involve
components of the Riemann tensor. Notice that, due to the symmetry
properties of $S_{\alpha\beta\lambda\mu}$, only one independent divergence
can be computed. Thus, one has the important result:
\begin{theo}
\label{th:cons}
In any flat region of $(V_n,g)$ (that is, with vanishing Riemann tensor),
the super-energy tensor of the massless scalar field (\ref{ses})
as well as the general 3-parameter family of tensors generated from it are 
divergence-free. \N
\end{theo}
In other words, in absence of gravitational field (represented by the Riemann
tensor), the general super-energy tensor of the scalar field is
divergence-free. In fact, this leads to conserved currents and quantities
for the scalar field in flat $(V_n,g)$ in the same way as was shown before for
the energy-momentum tensor of $\p$, given that any flat $(V_n,g)$ always
has $n(n+1)/2$ independent Killing vectors \cite{Ei}. Thus, define
\b*
j^{\mu}\left(\nabla_{[1]}\nabla_{[1]}\p;\vec{\xi}\right)\equiv
S^{\mu\rho\sigma\tau}\xi_{\rho}\xi_{\sigma}\xi_{\tau}=
S^{(\mu\rho\sigma\tau)}\xi_{\rho}\xi_{\sigma}\xi_{\tau}
\e*
for any given Killing vector $\vec{\xi}$, where the symmetry properties of
(\ref{ses}) have been used to write the last expression. This $\vec{j}$ is
divergence-free in flat $(V_n,g)$ because then
\b*
\nabla_{\mu}j^{\mu}\left(\nabla_{[1]}\nabla_{[1]}\p;\vec{\xi}\right)=
S^{(\mu\rho\sigma\tau)}\nabla_{\mu}\left(\xi_{\rho}\xi_{\sigma}\xi_{\tau}
\right)=0 \hspace{1cm} \mbox{in flat $(V_n,g)$.}
\e*
Actually, one can put different Killing vectors in the definition of
$\vec{j}$ if at the same time only the completely symmetric part of
(\ref{ses}) is used. Thus, the vector fields
\be
j_{\mu}\left(\nabla_{[1]}\nabla_{[1]}\p;
\vec{\xi}_1,\vec{\xi}_2,\vec{\xi}_3\right)\equiv
S_{(\mu\rho\sigma\tau)}\xi_1{}^{\rho}\xi_2{}^{\sigma}\xi_3{}^{\tau}
\label{jses}
\ee
are divergence-free in flat manifolds for arbitrary Killing vectors
$\vec{\xi}_1,\vec{\xi}_2,\vec{\xi}_3$. This provides a total of 
\b*
\left(\begin{array}{c}
N+2\\N-1\end{array}\right)
\e*
conserved currents (\ref{jses}) in flat spacetimes, where $N$ stands for the
number of independent Killing vectors: $n(n+1)/2$. These are currents
quadratic in the second derivatives of $\p$. Then, one can find the
corresponding conserved quantities using the Gauss theorem, as always.
In the particular case of $n=4$ there are 220 currents
in the set (\ref{jses}), to be compared with the results found by
Teyssandier \cite{Tey}.

Continuing the process, one builds the (super)$^2$-energy tensor
$T_{\alpha\beta\lambda\mu\tau\nu}\{\nabla_{[1]}\nabla_{[1]}\nabla_{[1]}\phi\}$
associated with the triple (1,1,1)-form
$\nabla_{\alpha}\nabla_{\beta}\nabla_{\mu}\phi$, which using (\ref{set111'})
is given by
\bea
T_{\alpha\beta\lambda\mu\tau\nu}\{\nabla_{[1]}\nabla_{[1]}\nabla_{[1]}\phi\}=
\nabla_{\alpha}\nabla_{\lambda}\nabla_{\tau}\p\,
\nabla_{\beta}\nabla_{\mu}\nabla_{\nu}\p+
\nabla_{\beta}\nabla_{\lambda}\nabla_{\tau}\p\,
\nabla_{\alpha}\nabla_{\mu}\nabla_{\nu}\p+\nonumber\\
+\nabla_{\alpha}\nabla_{\mu}\nabla_{\tau}\p\,
\nabla_{\beta}\nabla_{\lambda}\nabla_{\nu}\p+
\nabla_{\alpha}\nabla_{\lambda}\nabla_{\nu}\p\,
\nabla_{\beta}\nabla_{\mu}\nabla_{\tau}\p -\hspace{1cm}\nonumber\\
-g_{\alpha\beta}\left(\nabla_{\rho}\nabla_{\lambda}\nabla_{\tau}\p\,
\nabla^{\rho}\nabla_{\mu}\nabla_{\nu}\p+
\nabla_{\rho}\nabla_{\lambda}\nabla_{\nu}\p\,
\nabla^{\rho}\nabla_{\mu}\nabla_{\tau}\p\right)-\nonumber\\
-g_{\lambda\mu}\left(\nabla_{\alpha}\nabla_{\rho}\nabla_{\tau}\p\,
\nabla_{\beta}\nabla^{\rho}\nabla_{\nu}\p+
\nabla_{\alpha}\nabla_{\rho}\nabla_{\nu}\p\,
\nabla_{\beta}\nabla^{\rho}\nabla_{\tau}\p\right)-\nonumber\\
-g_{\tau\nu}\left(\nabla_{\alpha}\nabla_{\lambda}\nabla_{\rho}\p\,
\nabla_{\beta}\nabla_{\mu}\nabla^{\rho}\p+
\nabla_{\alpha}\nabla_{\mu}\nabla_{\rho}\p\,
\nabla_{\beta}\nabla_{\lambda}\nabla^{\rho}\p
\right)+\nonumber\\
+g_{\alpha\beta}g_{\lambda\mu}\nabla_{\rho}\nabla_{\sigma}\nabla_{\tau}\p\,
\nabla^{\rho}\nabla^{\sigma}\nabla_{\nu}\p+
g_{\alpha\beta}g_{\tau\nu}\nabla_{\rho}\nabla_{\lambda}\nabla_{\sigma}\p\,
\nabla^{\rho}\nabla_{\mu}\nabla^{\sigma}\p+\nonumber\\
+g_{\lambda\mu}g_{\tau\nu}\nabla_{\alpha}\nabla_{\rho}\nabla_{\sigma}\p\,
\nabla_{\beta}\nabla^{\rho}\nabla^{\sigma}\p-
\frac{1}{2}g_{\alpha\beta}g_{\lambda\mu}g_{\tau\nu}
\nabla_{\gamma}\nabla_{\rho}\nabla_{\sigma}\p\,
\nabla^{\gamma}\nabla^{\rho}\nabla^{\sigma}\p  \label{sses}
\eea
and so on. This produces an infinite set of basic (super)$^k$-energy tensors,
one for each natural number $k$.
\begin{defi}
The (super)$^k$-energy tensor ($k\geq 0$) of the massless scalar field
is the tensor (\ref{set}) for the $(k+1)$-fold $(1,\dots ,1)$-form
describing the $(k+1)$-th covariant derivative of $\p$, that is
\b*
T_{\lambda_1\mu_1\dots\lambda_k\mu_k\lambda_{k+1}\mu_{k+1}}
\left\{\underbrace{\nabla_{[1]}\dots\nabla_{[1]}}_{k+1} \p\right\}.
\e*
\end{defi}
Each of these tensors describes the same physical characteristics of the
massless scalar field as the corresponding gravitational (super)$^k$-energy
tensor does for the gravitational field. They may again have some relevance at
points of $V_n$ where the derivatives of a given order, but not of all orders,
of the scalar field vanish. From the general results of sections \ref{sec:set}
and \ref{sec:dsep} the dominant (super)$^k$-energy properties hold, and one
can define the (super)$^k$-energy densities relative to $\vec{u}$. Moreover,
the following fundamental result holds:
\begin{prop}
\label{prop:ceros}
The (super)$^k$-energy (tensor) of the massless scalar field vanishes at a
point $x\in V_n$ if and only if the $(k+1)$-th covariant derivative of $\p$ is
zero at $x$. In particular, {\em all} (super)$^k$-energy tensors of the massless
scalar field vanish in a domain $D\subseteq V_n$ if $\p$ is constant in $D$.
If the (super)$^k$-energy (tensor) of $\p$ vanishes in $D$, so do all the
(super)$^{\tilde k}$-energy tensors with $\tilde{k}>k$. The manifolds
containing a constant $\nabla \p$ are characterized by
the vanishing of the super-energy tensor of $\p$.\N
\end{prop}

Notice that the (super)$^k$-energy tensors of the massless scalar field
have a precise and definite meaning in flat manifolds $(V_n,g)$, and one
wonders whether these tensors give also rise to conserved currents similar
to (\ref{jses}). In general $(V_n,g)$, one can compute the divergence of
the (super)$^k$-energy tensors to obtain very long formulae involving
the Riemann tensor. Observe that now there may appear, in fact,
several different divergences for these tensors, depending on the index
used to contract with the covariant derivative. Nevertheless, in flat $(V_n,g)$
it is obvious that $\nabla_{[1]}\dots\nabla_{[1]}\p$ is completely
symmetric, and thus all the $(\lambda_{\U}\mu_{\U})$-pairs of the
(super)$^k$-energy tensors can be interchanged in this case. In other
words, in the case of flat $(V_n,g)$ only one independent divergence of the
(super)$^k$-energy tensors exists. Computing it in the special case of
flat manifolds one straightforwardly gets using (\ref{KG})
\b*
\nabla_{\lambda_1}T^{\lambda_1\mu_1\dots\lambda_k\mu_k\lambda_{k+1}\mu_{k+1}}
\left\{\underbrace{\nabla_{[1]}\dots\nabla_{[1]}}_{k+1} \p\right\}=0
\hspace{2cm} \mbox{in flat $(V_n,g)$.}
\e*
\begin{theo}
\label{th:cons2}
In any flat region of $(V_n,g)$, all the basic (super)$^k$-energy tensors of the
massless scalar field, as well as the general $(2k+1)!!$-parameter family of
tensors generated from it according to (\ref{gset}), are 
divergence-free. \N
\end{theo}
From this theorem one is led to construct the following currents
\b*
j_{\lambda}\left(\underbrace{\nabla_{[1]}\dots\nabla_{[1]}}_{k+1} \p;
\vec{\xi_2},\dots,\vec{\xi}_{2k+2}\right)\equiv
T_{(\lambda\mu_1\dots\lambda_{k+1}\mu_{k+1})}
\left\{\underbrace{\nabla_{[1]}\dots\nabla_{[1]}}_{k+1} \p\right\}
\xi_2^{\mu_1}\dots \xi_{2k+1}^{\lambda_{k+1}}\xi_{2k+2}^{\mu_{k+1}}
\e*
which are divergence-free in flat manifolds $(V_n,g)$ for arbitrary Killing
vectors $\vec{\xi}_2,\dots ,\vec{\xi}_{2k+2}$. This provides an
{\it infinite set} of conserved quantities in flat $(V_n,g)$ for the massless
scalar field. This result is not surprising, because if $\p$ satisfies
the Klein-Gordon equation (\ref{KG}) in flat $(V_n,g)$, then the {\it scalars}
$\nabla_{|\alpha_1|}\dots\nabla_{|\alpha_{k+1}|}\p$ for {\it fixed} values
of $\alpha_1,\dots,\alpha_{k+1}$ so do, and their respective 
``energy-momentum'' tensors give rise to an infinite set of conserved
quantities. For $n=4$ a clarifying discussion can be
found in \cite{Tey}.

\subsubsection{The electromagnetic field}
Let $F_{\mu\nu}=F_{[\mu\nu]}$ be an electromagnetic field satisfying the
source-free Maxwell equations
\be
F_{\mu\nu}=\nabla_{\mu}A_{\nu}-\nabla_{\nu}A_{\mu} \hspace{3mm} (\,
\Longrightarrow \hspace{2mm} \nabla_{[\tau}F_{\mu\nu]}=0\,), \hspace{1cm}
\nabla_{\rho}F^{\rho\nu}=0 \label{Max}
\ee
where $A_{\mu}$ is the electromagnetic potential, which is defined up to
the gauge transformation $A_{\mu}\longrightarrow A_{\mu}+\nabla_{\mu}\varphi$
for any arbitrary function $\varphi$. One could first consider the
electromagnetic potential $\bm{A}$ as starting tensor, and build up the
tensor (\ref{set}) associated to $A_{[1]}$, given by
\be
T_{\lambda\mu}\left\{A_{[1]}\right\}=A_{\lambda}A_{\mu}-\frac{1}{2}
g_{\lambda\mu}A_{\rho}A^{\rho}. \label{emA}
\ee
However, this tensor is not invariant under gauge transformations and thus
is not well-defined. Nevertheless, this tensor will have some role to play
in the case of the massive Proca field, to be studied in the next
subsection.

One thus must take $\bm{F}$, which is gauge-invariant, as the tensor
describing the electromagnetic field. Constructing the expression (\ref{set})
associated to $\bm{F}$, or using directly (\ref{set1'}), one gets
\be
T_{\lambda\mu}\left\{F_{[2]}\right\}=F_{\lambda\rho}F_{\mu}{}^{\rho}-
\frac{1}{4}g_{\lambda\mu}F_{\rho\sigma}F^{\rho\sigma} ,
\hspace{1cm} T_{\mu}{}^{\mu}\{F_{[2]}\}=\left(1-\frac{n}{4}\right)
F_{\rho\sigma}F^{\rho\sigma} \label{emF}
\ee 
which is the standard energy-momentum tensor of the electromagnetic
field. This tensor is symmetric and identically divergence-free if the
Maxwell equations (\ref{Max}) hold, but it is traceless {\it only} in $n=4$,
as seen from (\ref{emF}). As in the case of the massless scalar field, this
leads to divergence-free currents whenever there is a Killing vector
$\vec{\xi}$, defined by
\b*
j^{\mu}\left(F_{[2]};\vec{\xi}\right)\equiv \xi_{\rho}
T^{\rho\mu}\left\{F_{[2]}\right\} \hspace{1cm} \Longrightarrow\hspace{1cm}
\nabla_{\mu}j^{\mu}\left(F_{[2]};\vec{\xi}\right)=0.
\e*
In $(V_n,g)$ of constant curvature, there are $n(n+1)/2$ independent such
currents, which are related to the fluxes of energy, momentum and angular
momentum of the electromagnetic field. In general non-flat $(V_n,g)$, they
provide conserved quantities arising because of the intrinsic symmetries of the
spacetime, and related to combinations of the pertinent quantity for the
electromagnetic field {\it together} with the gravitational one. Notice also
that a conformal Killing vector $\vec{\z}$ will give rise to a divergence-free
current $j^{\mu}\left(F_{[2]};\vec{\z}\right)$ in $n=4$, when the
energy-momentum tensor is traceless. Thus, in conformally flat $V_4$ manifolds
there appear 15 such conserved currents.

In order to get the basic super-energy tensor of the electromagnetic field,
the double (1,2)-form $\nabla_{\alpha}F_{\mu\nu}$ can be used as the starting
object, so that the tensor (\ref{set}) becomes by using (\ref{set11'})
\bea
E_{\alpha\beta\lambda\mu}\equiv
T_{\alpha\beta\lambda\mu}\{\nabla_{[1]}F_{[2]}\}=
\nabla_{\alpha}F_{\lambda\rho}\nabla_{\beta}F_{\mu}{}^{\rho}+
\nabla_{\alpha}F_{\mu\rho}\nabla_{\beta}F_{\lambda}{}^{\rho}-
g_{\alpha\beta}\nabla_{\sigma}F_{\lambda\rho}
\nabla^{\sigma}F_{\mu}{}^{\rho}- \nonumber \\
-\frac{1}{2}g_{\lambda\mu}
\nabla_{\alpha}F_{\sigma\rho}\nabla_{\beta}F^{\sigma\rho}+
\frac{1}{4}g_{\alpha\beta}g_{\lambda\mu}
\nabla_{\tau}F_{\sigma\rho}\nabla^{\tau}F^{\sigma\rho} \, .
\label{seF}
\eea
Its symmetry properties are
\b*
E_{\alpha\beta\lambda\mu}=E_{(\alpha\beta)(\lambda\mu)}
\e*
but it is not symmetric in the exchange of $\alpha\beta$ with $\lambda\mu$.
Thus, following the considerations of section \ref{sec:general}, the
general s-e tensor for the electromagnetic field is the six-parameter
family (\ref{c16}) given by
\be
\bbb{E}_{\alpha\beta\lambda\mu}\equiv
c_1 E_{\alpha\beta\lambda\mu}+c_2 E_{\alpha\lambda\beta\mu}+
c_3E_{\alpha\mu\lambda\beta}+c_4E_{\lambda\beta\alpha\mu}+
c_5E_{\mu\beta\lambda\alpha}+c_6E_{\lambda\mu\alpha\beta} \label{c16F}
\ee
for arbitrary non-negative constants $c_1,c_2,c_3,c_4,c_5,c_6$.
The tensor $\bbb{E}$ has no index symmetries in general. However, there
are the following particular cases: 
\b*
\mbox{If} \hspace{5mm} c_2=c_4, \, \, c_3=c_5 \hspace{1cm} \Longrightarrow
\hspace{1cm} \bbb{E}_{\alpha\beta\lambda\mu}=
\bbb{E}_{(\alpha\beta)\lambda\mu},\\
\mbox{If} \hspace{5mm} c_2=c_3, \, \, c_4=c_5 \hspace{1cm} \Longrightarrow
\hspace{1cm} \bbb{E}_{\alpha\beta\lambda\mu}=
\bbb{E}_{\alpha\beta(\lambda\mu)},\\
\mbox{If} \hspace{5mm} c_1=c_6, \, \, c_3=c_4 \hspace{1cm} \Longrightarrow
\hspace{1cm} \bbb{E}_{\alpha\beta\lambda\mu}=
\bbb{E}_{\lambda\mu\alpha\beta},\\
\mbox{If} \hspace{5mm} c_2=c_3=c_4=c_5 \hspace{1cm} \Longrightarrow
\hspace{1cm} \bbb{E}_{\alpha\beta\lambda\mu}=
\bbb{E}_{(\alpha\beta)(\lambda\mu)},\\
\mbox{If} \hspace{5mm} c_2=c_3=c_4=c_5, \, \, c_1=c_6 \hspace{1cm}
\Longrightarrow \hspace{1cm} \bbb{E}_{\alpha\beta\lambda\mu}=
\bbb{E}_{(\alpha\beta)(\lambda\mu)}=\bbb{E}_{\lambda\mu\alpha\beta},\\
\mbox{If} \hspace{5mm} c_1=c_2=c_3=c_4=c_5=c_6 \hspace{1cm}
\Longrightarrow \hspace{1cm}
\bbb{E}_{\alpha\beta\lambda\mu}=\bbb{E}_{(\alpha\beta\lambda\mu)}.
\e*
The last but one case was considered many years ago by Chevreton \cite{C}
in $n=4$, see also \cite{WZ}, and this together with the last case have
been recently considered in \cite{Tey} in Special Relativity.

The same considerations as those made for the case of the massless scalar
field apply now, and thus the tensors (\ref{seF}) or (\ref{c16F}) seem
to describe the same physical quantities for the electromagnetic field as
the traditional Bel and Bel-Robinson tensors
do for the gravitational field, or the tensor (\ref{ses}) does for the
massless scalar field. Thus, the super-energy density and the s-e
flux vectors of the electromagnetic field can be defined using the tensors
(\ref{seF}) or (\ref{c16F}).

Due to the symmetry properties of $E_{\alpha\beta\lambda\mu}$, one can
compute two independent divergences, which after a straightforward calculation
using (\ref{Max}) and the Ricci identity become
\b*
\nabla_{\rho}E^{\rho\beta\lambda\mu}=2\left(R_{\rho\tau}{}^{\beta\sigma}
F^{\tau(\mu}+R^{\beta\sigma\tau(\mu}F_{\tau\rho}\right)\nabla_{\sigma}
F^{\lambda)\rho}+\\
+2\nabla^{\beta}F^{(\mu}{}_{\rho}\left(R^{\lambda)\sigma}F_{\sigma}{}^{\rho}
+F^{\lambda)}{}_{\sigma}R^{\sigma\rho}-R^{\lambda)\rho\sigma\nu}F_{\sigma\nu}
\right)+\\
+g^{\lambda\mu}\left(R^{\sigma\nu\beta\tau}F_{\nu\rho}
\nabla_{\tau}F_{\sigma}{}^{\rho}-R_{\tau\sigma}F^{\tau\rho}
\nabla^{\beta}F^{\sigma}{}_{\rho}-\frac{1}{2}R_{\tau\nu\sigma\rho}F^{\tau\nu}
\nabla^{\beta}F^{\sigma\rho}\right)
\e*
and
\b*
\nabla_{\rho}E^{\alpha\beta\rho\mu}=2\left(R^{\sigma(\alpha}F_{\sigma\rho}
-\frac{1}{2}F^{\sigma\tau}R^{(\alpha}{}_{\rho\sigma\tau}\right)
\nabla^{\beta)}F^{\mu\rho}+
3\nabla^{(\alpha}F_{\sigma\rho}R^{\beta)\tau[\mu\sigma}F^{\rho]}{}_{\tau}+\\
-g^{\alpha\beta}\left[\left(R^{\sigma\nu}F_{\sigma\rho}-
\frac{1}{2}F^{\sigma\tau}R^{\nu}{}_{\rho\sigma\tau}\right)
\nabla^{\nu}F^{\mu\rho}+\frac{3}{4}
\nabla_{\nu}F_{\sigma\rho}R^{\nu\tau[\mu\sigma}F^{\rho]}{}_{\tau}\right].
\e*
As is explicit, these righthand sides vanish if the Riemann tensor is zero.
\begin{theo}
\label{th:conF}
In any flat region of $(V_n,g)$, the basic super-energy tensor of the
electromagnetic field (\ref{seF}) as well as the general 6-parameter family 
(\ref{c16F}) generated from it are divergence-free. \N
\end{theo}
Thus, in absence of curvature, the general super-energy tensor of the
electromagnetic field is divergence-free. This again leads to conserved
currents for the electromagnetic field in flat $(V_n,g)$ similarly to those
shown before for the scalar field. Thus, the vector fields
\b*
j_{\mu}\left(\nabla_{[1]}F_{[2]};\vec{\xi}_1,\vec{\xi}_2,\vec{\xi}_3\right)\equiv
E_{(\mu\rho\sigma\tau)}\xi_1{}^{\rho}\xi_2{}^{\sigma}\xi_3{}^{\tau}
\e*
are divergence-free in flat manifolds for arbitrary Killing vectors
$\vec{\xi}_1,\vec{\xi}_2,\vec{\xi}_3$. The total number of independent
such conserved currents is the same as that given for the scalar field.
Again, these are currents quadratic in the first derivatives of $\bm{F}$ and
one can find the corresponding conserved quantities using the Gauss
theorem.

As in previous cases, higher order s-e tensors can be constructed. The basic
(super)$^2$-energy tensor of the electromagnetic field is the expression
(\ref{set}) associated to the triple (1,1,2)-form
$\nabla_{\rho}\nabla_{\sigma}F_{\mu\nu}$. Its explicit expression is
\bea
T_{\alpha\beta\lambda\mu\tau\nu}\left\{\nabla_{[1]}\nabla_{[1]}F_{[2]}\right\}
=\nabla_{\alpha}\nabla_{\lambda}F_{\tau\rho}
\nabla_{\beta}\nabla_{\mu}F_{\nu}{}^{\rho}+
\nabla_{\beta}\nabla_{\lambda}F_{\tau\rho}
\nabla_{\alpha}\nabla_{\mu}F_{\nu}{}^{\rho}+\nonumber\\
+\nabla_{\alpha}\nabla_{\mu}F_{\tau\rho}
\nabla_{\beta}\nabla_{\lambda}F_{\nu}{}^{\rho}
+\nabla_{\alpha}\nabla_{\lambda}F_{\nu\rho}
\nabla_{\beta}\nabla_{\mu}F_{\tau}{}^{\rho}-\nonumber\\-g_{\alpha\beta}\left(
\nabla_{\sigma}\nabla_{\lambda}F_{\tau\rho}
\nabla^{\sigma}\nabla_{\mu}F_{\nu}{}^{\rho}+
\nabla_{\sigma}\nabla_{\lambda}F_{\nu\rho}
\nabla^{\sigma}\nabla_{\mu}F_{\tau}{}^{\rho}\right)-\nonumber\\
-g_{\lambda\mu}\left(\nabla_{\alpha}\nabla_{\sigma}F_{\tau\rho}
\nabla_{\beta}\nabla^{\sigma}F_{\nu}{}^{\rho}+
\nabla_{\alpha}\nabla_{\sigma}F_{\nu\rho}
\nabla_{\beta}\nabla^{\sigma}F_{\tau}{}^{\rho}\right)-\nonumber\\
-\frac{1}{2}g_{\tau\nu}
\left(\nabla_{\alpha}\nabla_{\lambda}F_{\sigma\rho}
\nabla_{\beta}\nabla_{\mu}F^{\sigma\rho}+
\nabla_{\alpha}\nabla_{\mu}F_{\sigma\rho}
\nabla_{\beta}\nabla_{\lambda}F^{\sigma\rho}\right)+\nonumber\\
+g_{\alpha\beta}g_{\lambda\mu}\nabla_{\sigma}\nabla_{\gamma}F_{\tau\rho}
\nabla^{\sigma}\nabla^{\gamma}F_{\nu}{}^{\rho}+\frac{1}{2}
g_{\alpha\beta}g_{\tau\nu}\nabla_{\gamma}\nabla_{\lambda}F_{\sigma\rho}
\nabla^{\gamma}\nabla_{\mu}F^{\sigma\rho}+\nonumber\\
+\frac{1}{2}g_{\lambda\mu}g_{\tau\nu}
\nabla_{\alpha}\nabla_{\gamma}F_{\sigma\rho}
\nabla_{\beta}\nabla^{\gamma}F^{\sigma\rho}-
\frac{1}{4}g_{\alpha\beta}g_{\lambda\mu}g_{\tau\nu}
\nabla_{\delta}\nabla_{\gamma}F_{\sigma\rho}
\nabla^{\delta}\nabla^{\gamma}F^{\sigma\rho}.\label{sseF}
\eea
In general we have
\begin{defi}
The (super)$^k$-energy tensor ($k\geq 0$) of the electromagnetic field
is the tensor (\ref{set}) for the $(k+1)$-fold $(1,\dots ,1,2)$-form
describing the $k$-th covariant derivative of $\bm{F}$, that is
\b*
T_{\lambda_1\mu_1\dots\lambda_k\mu_k\lambda_{k+1}\mu_{k+1}}
\left\{\underbrace{\nabla_{[1]}\dots\nabla_{[1]}}_{k} F_{[2]}\right\}.
\e*
\end{defi}
Each of these tensors describes the same physical properties of the
electromagnetic field as the corresponding (super)$^k$-energy tensors do for
the gravitational and massless scalar fields. They may arise at points of
$V_n$ where the derivatives up to a given order of the electromagnetic field
vanish. From the discussion in sections \ref{sec:set} and \ref{sec:dsep}
the dominant (super)$^k$-energy properties hold for the electromagnetic
field, and one can define its (super)$^k$-energy densities and flux vectors
relative to $\vec{u}$. As in previous cases, we have
\begin{prop}
\label{prop:ceroF}
The (super)$^k$-energy (tensor) of the electromagnetic field vanishes at a
point $x\in V_n$ if and only if the $k$-th covariant derivative of $\bm{F}$ is
zero at $x$. In particular, {\em all} (super)$^k$-energy tensors of the 
electromagnetic field vanish in a domain $D\subseteq V_n$ if $\bm{F}$ vanishes
in $D$. If the (super)$^k$-energy (tensor) of $\bm{F}$ vanishes in $D$, so do
all the (super)$^{\tilde k}$-energy tensors with $\tilde{k}>k$. The manifolds
containing a constant $\bm{F}$ are characterized by
the vanishing of the super-energy (tensor) of $\bm{F}$.\N
\end{prop}

In general $(V_n,g)$, one can compute the divergences of the (super)$^k$-energy
tensors to obtain very long formulae involving the Riemann tensor. 
However, in flat $(V_n,g)$ it is obvious that
$\nabla_{[1]}\dots\nabla_{[1]}F_{[2]}$ is completely
symmetric in all the $[1]$-blocks, and thus all the corresponding
$(\lambda_{\U}\mu_{\U})$-pairs of the
(super)$^k$-energy tensors can be interchanged in this case. This implies
that, in a flat $(V_n,g)$, only two independent divergences of the
(super)$^k$-energy tensors exist in principle. Computing them in this special
case of flat manifolds one straightforwardly can prove using (\ref{Max}) that
in fact they are equal and vanishing, so that
\b*
\nabla_{\lambda_1}
\bbb{T}^{\lambda_1\mu_1\dots\lambda_k\mu_k\lambda_{k+1}\mu_{k+1}}
\left\{\underbrace{\nabla_{[1]}\dots\nabla_{[1]}}_{k} F_{[2]}\right\}=0
\hspace{2cm} \mbox{in flat $(V_n,g)$.}
\e*
\begin{theo}
\label{th:conF2}
In any flat region of $(V_n,g)$, all the basic (super)$^k$-energy tensors of the
electromagnetic field, as well as the general $(k+1) (2k+1)!!$-parameter
family of tensors generated from it according to (\ref{gset}), are 
divergence-free. \N
\end{theo}
From this theorem one can define the following currents
\b*
j_{\lambda}\left(\underbrace{\nabla_{[1]}\dots\nabla_{[1]}}_{k} F_{[2]};
\vec{\xi_2},\dots,\vec{\xi}_{2k+2}\right)\equiv
T_{(\lambda\mu_1\dots\lambda_{k+1}\mu_{k+1})}
\left\{\underbrace{\nabla_{[1]}\dots\nabla_{[1]}}_{k} F_{[2]}\right\}
\xi_2^{\mu_1}\dots \xi_{2k+1}^{\lambda_{k+1}}\xi_{2k+2}^{\mu_{k+1}}
\e*
which are divergence-free in flat manifolds $(V_n,g)$ for arbitrary Killing
vectors $\vec{\xi}_2,\dots ,\vec{\xi}_{2k+2}$, leading to 
{\it infinitely many} conserved quantities in flat $(V_n,g)$ for the
electromagnetic field. This result is again not surprising, as explained
for the massless scalar field, because if $\bm{F}$ satisfies
the Maxwell equations (\ref{Max}) in flat $(V_n,g)$, then the {\it 2-forms}
$\nabla_{|\alpha_1|}\dots\nabla_{|\alpha_{k+1}|}\bm{F}$ for {\it fixed} values
of $\alpha_1,\dots,\alpha_{k+1}$ are also solutions of the same equations,
and their respective ``energy-momentum'' tensors of type (\ref{emF})
give rise to infinitely many conserved quantities.

\subsection{Massive fields}
\label{subsec:massive}
The previous s-e tensors have been defined for the cases where the physical
fields are massless. However, sometimes these fields carry mass, denoted
generically by $m$, and then the field equations as well as the 
energy-momentum properties change. In this subsection the treatment of
the massive cases is explained in detail and the corresponding
(super)$^k$-energy tensors are derived. Only the prominent cases of the
scalar and Proca fields will be considered explicitly, but the general
rules will be valid for any other fields.

In general, the existence of the mass amounts to having another quantity
of physical relevance denoted by $m$ which can be assumed to have (in
natural units) dimensions of the inverse of length, $L^{-1}$. Thus, at
first sight one realizes that the derivative of the field (say) has the
same physical dimensions as the field multiplied by $m$, so that they are
on the same footing as per how they can contribute to the energy or the
higher order super-energies. Nevertheless, the derivative of the field has
one more index block than the product of the field by $m$, so that one has
to use the metric in order to combine them appropriately. Fortunately,
the metric satisfies (up to sign) the DSEP, and thus one can construct
new s-e tensors containing the mass which keep this fundamental property.
As an illustrative example, before giving the general definition, consider
the energy-momentum tensor of a massive scalar field $\p$,
\be
T_{\lambda\mu}=\nabla_{\lambda}\phi\nabla_{\mu}\phi
-\frac{1}{2}g_{\lambda\mu}\nabla_{\rho}\phi\nabla^{\rho}\phi 
-\frac{1}{2}g_{\lambda\mu}m^2\p^2 \label{emsm}
\ee
which is symmetric and identically divergence-free if the Klein-Gordon
equation
\be
\nabla_{\rho}\nabla^{\rho}\p=m^2\p \label{KGm}
\ee
holds. Let us observe that this tensor can be written as
\b*
T_{\lambda\mu}=T_{\lambda\mu}\left\{\nabla_{[1]}\p\right\}+
T\left\{m \p\right\}(-g_{\lambda\mu}).
\e*
The key idea here is that both $\nabla_{[1]}\p$ and $m \p$ have the same
physical dimensions, but of course the first is a 1-form and the second
is a scalar. This mathematical difference is corrected by use of the
fundamental metric tensor $g$. Other important points are: first, due to
Property \ref{pr:alg} the above tensor satisfies the DSEP because both
$T_{\lambda\mu}\left\{\nabla_{[1]}\p\right\}$ and $T\left\{m \p\right\}$
so do; and second, the divergence of the tensor $T_{\lambda\mu}$ is zero
due to (\ref{KGm}).

Bearing all this in mind, the generalization to arbitrary fields can be
given in an inductive manner as follows. Assume that some physical massive
field is described by an $r$-fold form $t_{[n_1],\dots,[n_r]}$ (then, the
`smallest' s-e tensor of type (\ref{set}) that can be formed will have $2r$
indices, so that this will be at the (super)$^{(r-1)}$-energy level).
Imagine then that we know the (super)$^{(k+r-1)}$-energy tensor of
this field which, by consideration of the
physical dimensions, will involve basic s-e tensors of type (\ref{set})
for i) the derivatives of the field of order $k$; ii) products of $m$ by
derivatives of the field of order $k-1$; iii) products of $m^2$ by derivatives
of the field of order $k-2$; and so on. In other words, there will appear
the s-e tensors of type (\ref{set}) for a {\it definite and precise} set of
tensors of type
\be
m^{k-\ell}\nabla_{[1]}^{\ell}t_{[n_1],\dots,[n_r]} \label{list}
\ee
where $\ell$ can take values $\ell=0,\dots,k$. Then, the basic
(super)$^{(k+r)}$-energy tensor of the massive $t_{[n_1],\dots,[n_r]}$ is the
one constructed adding all the s-e tensors (\ref{set}) for the derivatives
\b*
\nabla_{[1]}\left(m^{k-\ell}\nabla_{[1]}^{\ell}t_{[n_1],\dots,[n_r]}\right)
\e*
of precisely the tensors appearing in the set (\ref{list}), plus all the
s-e tensors of type (\ref{set}) for the products of $m$ with exactly
those in the list (\ref{list})
\b*
m\, \left(m^{k-\ell}\nabla_{[1]}^{\ell}t_{[n_1],\dots,[n_r]}\right).
\e*
It is important to remark that not {\it all} possible tensors of the
type $m^{k-\ell}\nabla_{[1]}^{\ell}t_{[n_1],\dots,[n_r]}$ will necessarily
appear in the list (\ref{list}), and thus, in the next step, one only has to
consider those arising from the ones really appearing in the list by
taking the covariant derivative and multiplying by $m$. Observe further that
the position of the $m$ and of the derivative is important, and for instance
the tensors $m\nabla_{[1]}t_{[n_1],\dots,[n_r]}$ and
$\nabla_{[1]}(m\,t_{[n_1],\dots,[n_r]})$
will produce s-e tensors with different index order. 

Some explicit examples will make all this clear. Starting with the
scalar field $\p$, at the energy level one can use the derivative of
$\p$ as well as the product of $\p$ with $m$, or schematically
\be
\nabla_{[1]}\p \hspace{5mm} \oplus \hspace{5mm} m\, \p , \label{lists1}
\ee
as tensors to construct the corresponding expressions (\ref{set}). As we have
seen in (\ref{emsm}), this gives the correct energy-momentum tensor. Notice
that in the first case the index in $\nabla_{[1]}$ gives rise to the pair of
indices in the energy-momentum tensor, while in the second case it is $m$ (so
to speak) that gives rise to these indices by means of the corresponding
$-g_{\lambda\mu}$. Now, in order to get the super-energy tensor of the massive
scalar field, and following the general rule explained before, one must
use the derivatives of the tensors in the list (\ref{lists1}) plus the
products of the mass $m$ with the tensors in that list, that is
\b*
\nabla_{[1]}\left(\nabla_{[1]}\p\, \oplus \, m \p\right) \hspace{2mm} \oplus
 \hspace{2mm} m\, \left(\nabla_{[1]}\p\, \oplus \, m \p\right)
\e*
which produces the new explicit list
\be
\nabla_{[1]}\nabla_{[1]}\p \hspace{2mm} \oplus \hspace{2mm}
\nabla_{[1]}\left(m\p\right) \hspace{2mm} \oplus 
 \hspace{2mm} m\, \left(\nabla_{[1]}\p\right) \hspace{2mm} \oplus
m\, m\, \p .\label{lists2}
\ee
Observe that the second and third terms are different because they produce
different indices in the super-energy tensor. Explicitly, with the four
tensors in the list (\ref{lists2}), the following s-e tensors are built
(all of them at the same level from the {\it physical dimension} viewpoint)
\b*
T_{\alpha\beta\lambda\mu}\left\{\nabla_{[1]}\nabla_{[1]}\p\right\},
\hspace{5mm}
T_{\alpha\beta}\left\{\nabla_{[1]}\left(m\p\right)\right\}, \hspace{5mm}
T_{\lambda\mu}\left\{ m\, \left(\nabla_{[1]}\p\right)\right\}, \hspace{5mm}
T\left\{m\, m\, \p \right\}
\e*
and the sum of all of them, including a $-g$ for each $m$, produces the basic
s-e tensor of the massive scalar field \cite{S2}:
\bea
{\cal S}_{\alpha\beta\lambda\mu}\equiv
T_{\alpha\beta\lambda\mu}\left\{\nabla_{[1]}\nabla_{[1]}\p\right\}+
T_{\alpha\beta}\left\{\nabla_{[1]}\left(m\p\right)\right\}(-g_{\lambda\mu})+
T_{\lambda\mu}\left\{m\left(\nabla_{[1]}\p\right)\right\} (-g_{\alpha\beta})+
\nonumber\\
+T\left\{m\, m\, \p \right\}(-g_{\alpha\beta})(-g_{\lambda\mu})=
2\nabla_{\alpha}\nabla_{(\lambda}\phi \nabla_{\mu)}\nabla_{\beta}\phi
-g_{\alpha\beta}\left(\nabla_{\lambda}\nabla^{\rho}\phi
\nabla_{\mu}\nabla_{\rho}\phi+m^2\nabla_{\lambda}\phi\nabla_{\mu}\phi \right)
\nonumber \\
- g_{\lambda\mu}\left(\nabla_{\alpha}\nabla^{\rho}\phi
\nabla_{\beta}\nabla_{\rho}\phi
+m^2\nabla_{\alpha}\phi\nabla_{\beta}\phi\right)
+\frac{1}{2} g_{\alpha\beta}g_{\lambda\mu}\left(
\nabla_{\sigma}\nabla_{\rho}\phi\nabla^{\sigma}\nabla^{\rho}\phi +
2m^2\nabla_{\rho}\phi\nabla^{\rho}\phi +m^4\phi^2\right) \label{sesm}
\eea
which has the same symmetry properties as (\ref{ses}). This tensor has been
also found in \cite{Tey} for the case of Special Relativity ($n=4$).

Similarly, the (super)$^2$-energy tensor for the massive $\p$ is obtained by
using the list produced by taking the derivative and multiplying by $m$ the
tensors in (\ref{lists2}), that is
\b*
\nabla_{[1]}\left[\nabla_{[1]}\nabla_{[1]}\p  \oplus
\nabla_{[1]}\left(m\p\right) \oplus m\, \left(\nabla_{[1]}\p\right)\oplus
m\, m\, \p \right] \hspace{2mm} \oplus \\
\oplus \hspace{2mm}
m\,\left[\nabla_{[1]}\nabla_{[1]}\p  \oplus
\nabla_{[1]}\left(m\p\right) \oplus m\, \left(\nabla_{[1]}\p\right)\oplus
m\, m\, \p \right].
\e*
Therefore, the basic (super)$^2$-energy tensor of the massive scalar field
is given by
\b*
T_{\alpha\beta\lambda\mu\tau\nu}\left\{\nabla_{[1]}\nabla_{[1]}\nabla_{[1]}\p
\right\}+
T_{\alpha\beta\lambda\mu}\left\{\nabla_{[1]}\nabla_{[1]}(m\p)\right\}
(-g_{\tau\nu})+\nonumber\\
+T_{\alpha\beta\tau\nu}\left\{\nabla_{[1]}\left(m\nabla_{[1]}\p\right)\right\}
(-g_{\lambda\mu})+
T_{\alpha\beta}\left\{\nabla_{[1]}(m\, m\p)\right\}(-g_{\lambda\mu})
(-g_{\tau\nu})+\nonumber\\
+T_{\lambda\mu\tau\nu}\left\{m\nabla_{[1]}\nabla_{[1]}\p\right\}
(-g_{\alpha\beta})+
T_{\lambda\mu}\left\{m\nabla_{[1]}(m\p)\right\}
(-g_{\alpha\beta})(-g_{\tau\nu})+\nonumber \\
+T_{\tau\nu}\left\{m\, m\nabla_{[1]}\p\right\}
(-g_{\alpha\beta})(-g_{\lambda\mu})+
T\left\{m\, m\, m\p\right\}(-g_{\alpha\beta})(-g_{\lambda\mu})(-g_{\tau\nu})
\e*
whose explicit expression can be read using (\ref{sses}), (\ref{ses}) and
(\ref{ems}). The (super)$^k$-energy tensors for the massive scalar field 
can be thus constructed for any $k$ in this way. 

Consider now a Proca field, also considerable as a massive electromagnetic
field \cite{BLP,IZ}. This is a field satisfying the Proca field equations
\be
F_{\mu\nu}=\nabla_{\mu}A_{\nu}-\nabla_{\nu}A_{\mu} \hspace{3mm} (\,
\Longrightarrow \hspace{2mm} \nabla_{[\tau}F_{\mu\nu]}=0\,), \hspace{1cm}
\nabla_{\rho}F^{\rho\nu}=m^2\, A^{\nu} \label{Pro}
\ee
from where one immediately deduces
\b*
m\neq 0 \hspace{5mm} \Longrightarrow \hspace{5mm} \nabla_{\nu}A^{\nu}=0.
\e*
Therefore, in this case there is no gauge freedom and the `potential'
$\vec{A}$ is well-defined. In this sense, the tensor (\ref{emA}) has
a precise meaning, but unfortunately it does not have the proper physical
dimensions comparable to those of (\ref{emF}). However, the existence of
mass makes it possible to construct the corresponding tensor (\ref{emA})
but using $m\, \vec{A}$ instead of $\vec{A}$. This has the proper physical
dimensions. Combining the two tensors we obtain
\b*
T_{\lambda\mu}\equiv T_{\lambda\mu}\left\{F_{[2]}\right\}+
T_{\lambda\mu}\left\{m\, A_{[1]}\right\}=\nonumber\\
=F_{\lambda\rho}F_{\mu}{}^{\rho}-
\frac{1}{4}g_{\lambda\mu}F_{\rho\sigma}F^{\rho\sigma}+
m^2\, A_{\lambda}A_{\mu}-\frac{1}{2}m^2\,
g_{\lambda\mu}A_{\rho}A^{\rho}
\e*
which is the standard symmetric energy-momentum tensor of the Proca field
\cite{BLP}. This tensor is divergence-free due to the Proca equations
(\ref{Pro}).

Hence, the list of tensors used at the energy level has been found, namely
\b*
F_{[2]} \hspace{5mm} \oplus \hspace{5mm} m\, A_{[1]}
\e*
and thereby, the list for the higher-order (super)$^k$-energy tensors can
be deduced step by step. At the super-energy level, the list is given by
\b*
\nabla_{[1]}\left(F_{[2]} \, \oplus \, m\, A_{[1]}\right) \hspace{3mm} \oplus
\hspace{3mm} m\, \left(F_{[2]} \, \oplus \, m\, A_{[1]}\right)
\e*
or equivalently
\b*
\nabla_{[1]}F_{[2]} \hspace{2mm} \oplus \hspace{2mm}
\nabla_{[1]}\left(m\, A_{[1]}\right) \hspace{2mm}  \oplus \hspace{2mm} 
m\, F_{[2]} \hspace{2mm} \oplus \hspace{2mm} m\, m\, A_{[1]}.
\e*
This example clarifies also that terms of type
$F_{[2]}m$ or $mA_{[1]}m$ (order is important because of the 
places of indices) will never appear according to our general rule.
If one included those type of terms, the s-e tensor would not be
divergence-free in flat $(V_n,g)$, see below. In summary, the basic s-e tensor
of the Proca field becomes, by using (\ref{seF}), (\ref{emF}) and (\ref{emA})
\bea
{\cal E}_{\alpha\beta\lambda\mu}\equiv 
T_{\alpha\beta\lambda\mu}\left\{\nabla_{[1]}F_{[2]}\right\}+
T_{\alpha\beta\lambda\mu}\left\{\nabla_{[1]}\left(mA_{[1]}\right)\right\}+
T_{\lambda\mu}\left\{m\, F_{[2]}\right\}(-g_{\alpha\beta})+\nonumber\\
+T_{\lambda\mu}\left\{m\, m\, A_{[1]}\right\}(-g_{\alpha\beta})=
\nabla_{\alpha}F_{\lambda\rho}\nabla_{\beta}F_{\mu}{}^{\rho}+
\nabla_{\alpha}F_{\mu\rho}\nabla_{\beta}F_{\lambda}{}^{\rho}-
g_{\alpha\beta}\nabla_{\sigma}F_{\lambda\rho}
\nabla^{\sigma}F_{\mu}{}^{\rho}- \nonumber \\
-\frac{1}{2}g_{\lambda\mu}
\nabla_{\alpha}F_{\sigma\rho}\nabla_{\beta}F^{\sigma\rho}+
\frac{1}{4}g_{\alpha\beta}g_{\lambda\mu}
\nabla_{\tau}F_{\sigma\rho}\nabla^{\tau}F^{\sigma\rho}+
m^2\left(\frac{}{}\nabla_{\alpha}A_{\lambda}\nabla_{\beta}A_{\mu}+
\nabla_{\alpha}A_{\mu}\nabla_{\beta}A_{\lambda}-\right.\nonumber\\
\left.-g_{\alpha\beta}\nabla_{\rho}A_{\lambda}\nabla^{\rho}A_{\mu}-
g_{\lambda\mu}\nabla_{\alpha}A_{\rho}\nabla_{\beta}A^{\rho}+
\frac{1}{2}g_{\alpha\beta}g_{\lambda\mu}
\nabla_{\sigma}A_{\rho}\nabla^{\sigma}A^{\rho}\right)-\nonumber\\
-m^2g_{\alpha\beta}\left(F_{\lambda\rho}F_{\mu}{}^{\rho}-
\frac{1}{4}g_{\lambda\mu}F_{\rho\sigma}F^{\rho\sigma}\right)-
m^4g_{\alpha\beta}\left(A_{\lambda}A_{\mu}-\frac{1}{2}
g_{\lambda\mu}A_{\rho}A^{\rho}\right).\label{seP}
\eea
The next step is the (super)$^2$-energy tensor of the Proca field, 
constructed using the list obtained by doing
\b*
\nabla_{[1]}\left(\nabla_{[1]}F_{[2]} \, \oplus \,
\nabla_{[1]}\left(m\, A_{[1]}\right) \,  \oplus \,
m\, F_{[2]} \, \oplus \,\, m\, m\, A_{[1]}\right) \, \oplus\\
\oplus \, \, m\,\left(\nabla_{[1]}F_{[2]} \, \oplus \,
\nabla_{[1]}\left(m\, A_{[1]}\right) \,  \oplus \,
m\, F_{[2]} \, \oplus \,\, m\, m\, A_{[1]}\right)
\e*
which can be written as
\b*
T_{\alpha\beta\lambda\mu\tau\nu}\left\{\nabla_{[1]}\nabla_{[1]}F_{[2]}\right\}+
T_{\alpha\beta\lambda\mu\tau\nu}
\left\{\nabla_{[1]}\nabla_{[1]}\left(mA_{[1]}\right)\right\}+\\
+T_{\alpha\beta\tau\nu}\left\{\nabla_{[1]}\left(mF_{[2]}\right)\right\}
(-g_{\lambda\mu})+T_{\alpha\beta\tau\nu}\left\{\nabla_{[1]}\left(
m\,mA_{[1]}\right)\right\}
(-g_{\lambda\mu})-m^2g_{\alpha\beta}{\cal E}_{\lambda\mu\tau\nu}
\e*
with an explicit expression easily written by using (\ref{sseF}), (\ref{seP})
and (\ref{set}). All (super)$^k$-energy tensors are built similarly.

Due to Property \ref{pr:alg}, all the (super)$^k$-energy tensors constructed
in this manner for the massive fields satisfy the DSEP, and obviously
one can define the positive (super)$^k$-energy density, the corresponding
flux vectors, etcetera. Moreover, the study of section \ref{sec:general}
applies and one can construct, for instance, a 3-parameter family of
s-e tensors for the massive scalar field starting from (\ref{sesm}), and
a 6-parameter family of s-e tensors for the Proca field starting from
(\ref{seP}). All this is obvious and will be skipped here.

More important is the question of the divergence of the above tensors.
For instance, for the s-e tensor (\ref{sesm}) one gets exactly the same
formula (\ref{divses})
\bea
\nabla_{\alpha}{\cal S}^{\alpha}{}_{\beta\lambda\mu}=
2\nabla_{\beta}\nabla_{(\lambda}\phi
R_{\mu)\rho}\nabla^{\rho}\phi -g_{\lambda\mu} R^{\sigma\rho}
\nabla_{\beta}\nabla_{\rho}\phi\nabla_{\sigma}\phi -\nonumber\\
-\nabla_{\sigma}\phi \left(2\nabla^{\rho}\nabla_{(\lambda}\phi \,
R^{\sigma}_{\mu)\rho\beta} +g_{\lambda\mu}
R^{\sigma}_{\rho\beta\tau}\nabla^{\rho}\nabla^{\tau}\phi\right)
\label{divsesm}
\eea
and similarly for other cases. Thus, theorems \ref{th:cons}, \ref{th:cons2},
\ref{th:conF} and \ref{th:conF2} remain true for the massive fields.
\begin{theo}
In any flat region of $(V_n,g)$, all the basic (super)$^k$-energy tensors of the
massive fields, as well as the general families of
tensors generated from it according to (\ref{gset}), are 
divergence-free.
\end{theo}
This leads to the conserved currents in flat $(V_n,g)$ using the same
definitions as in the previous subsection. Details again are trivial and will
be omitted here.

Finally, there is a small change in the results proven in Propositions
\ref{prop:ceros} and \ref{prop:ceroF}, because the existence of mass
implies that, at the (super)$^k$-energy level, {\it all} derivatives of
the given field {\it up to} a given order (and not only those of this order)
are involved, and thus the vanishing of the corresponding (super)$^k$-energy
density implies the vanishing of all the derivatives up to the given order.
More precisely
\begin{prop}
The (super)$^k$-energy (tensor) of the massive scalar field
vanishes at a point $x\in V_n$ if and only if $\p$ and all its derivatives up
to the $(k+1)$-th order are zero at $x$. Analogously, the (super)$^k$-energy
(tensor) of the Proca field vanishes at $x$ if and only if the $k$-th
covariant derivative of $\bm{F}$ as well as $\bm{A}$ and all its derivatives
up to the $k$-th order vanish at $x$. In particular, {\em all}
(super)$^k$-energy tensors of the massive scalar field, or of the Proca field,
vanish in a domain $D\subseteq V_n$ if $\p$, or respectively $\bm{A}$,
vanish in $D$.\N
\end{prop}

\section{Exchange of super-energy quantities: conserved currents}
\label{sec:cons}
We finally arrive at the fundamental question of whether the s-e tensors
have any physical importance or describe some kind of relevant physical
quantity. Obviously, the s-e tensors generalize the typical energy-momentum
tensors, which are assumed to describe the energy-momentum and stress
properties of a physical system. Hence, a naive guessing
would lead to consider the s-e tensors as describing some kind of
generalized energy-momenta or stresses of the same
physical system. As far as mathematics are concerned, all the good properties
of the energy-momentum tensors have been kept for the s-e tensors, namely,
the positivity of the s-e density, the causality of the s-e flux vectors,
and in general the DSEP which, as mentioned in section \ref{sec:dsep}, allows
to prove very general results on the causal propagation of fields and its
uniqueness \cite{BerS,BS}. Let us also mention that the DSEP may serve to prove
stability results such as those in \cite{CK} for the global stability of flat
spacetime, but generalizing them to cases with matter or other asymptotic
conditions. In general, the DSEP of the (super)$^k$-energy tensors provides
estimates and a priori bounds on all the components of a given tensor and its
derivatives of any order.
Thus, from the mathematical point of view there is little doubt
that the s-e construction herein presented is relevant.

But, what about physics? Energy-momentum tensors are not only important from
the mathematical point of view, but also from the physical one. The main
role they play is in the definition of conserved quantities and its
interchange between {\it different} physical systems. Thus, the main question
arises: can the s-e tensors give rise to conserved currents? And more
importantly, can s-e quantities be exchanged between different physical
systems?

The first question has been already answered in this paper. As has been seen
in section \ref{sec:physics}, the (super)$^k$-energy tensors of the scalar,
electromagnetic or Proca fields provide infinitely many divergence-free
currents {\it in flat spacetime}. Similarly, the generalized Bel tensor 
was proved to be divergence-free {\it in Einstein spaces} ($R_{\mu\nu}=\Lambda
g_{\mu\nu}$), see Theorem \ref{th:divbel}. This last property leads to
conserved currents if there exists at least a Killing vector,
for due to the symmetries (\ref{belsym}) of the generalized Bel tensor only one
independent divergence can be computed, and moreover, one has for arbitrary 
Killing vectors $\vec{\xi}_1,\vec{\xi}_2,\vec{\xi}_3$
\be
j_{\mu}\left(R_{[2],[2]};\vec{\xi}_1,\vec{\xi}_2,\vec{\xi}_3\right)\equiv
B_{(\alpha\beta\lambda)\mu}\,\xi^{\alpha}_1\xi^{\beta}_2\xi^{\lambda}_3=
B_{(\alpha\beta\lambda\mu)}\,\xi^{\alpha}_1\xi^{\beta}_2\xi^{\lambda}_3
\label{gc}
\ee
so that from Theorem \ref{th:divbel} it follows
\b*
\nabla_{\mu}j^{\mu}\left(R_{[2],[2]};
\vec{\xi}_1,\vec{\xi}_2,\vec{\xi}_3\right)=0 \hspace{1cm}
\mbox{in Einstein spaces: $R_{\mu\nu}=\Lambda g_{\mu\nu}$.}
\e*
As already remarked after formula (\ref{divems}), not {\it only} the
divergence-free property of $B$ is needed, but {\it also} its symmetry
properties which allow to form a divergence-free {\it vector} with any
given Killing vector. It is noteworthy that sometimes (\ref{gc}) is
divergence-free under much more general conditions (such as for instance
when the Killings are hypersurface-orthogonal, and many others,
{\it independently} of the form of the Ricci tensor), but these special cases
fall out of the general treatment considered in this paper and will be
presented elsewhere.

Recapitulating, the generalized Bel tensor provides
conserved currents if $R_{\mu\nu}=\Lambda g_{\mu\nu}$ (i.e., {\it in absence
of matter}), and the s-e tensors of the physical fields give conserved
currents in flat $(V_n,g)$ (that is, {\it in the absence of gravitational field}
if this is described by the curvature). This situation is, to say the
least, encouraging, and one is led to the natural question of whether or not
these respective currents can be combined to produce new {\it conserved vectors}
involving several fields, such as gravitation and others. Let us remark
that this situation is similar to that appearing in classical electrodynamics,
where the energy-momentum tensor of the electromagnetic field is
divergence-free in the absence of charges, but it
is not if they are present. As is well-known, see for instance \cite{LL},
the divergence-free property (and a fortiori the conserved currents) is
restored by adding the energy-momentum tensor of the existing charges,
and the combined {\it total} energy-momentum tensor is the
one without divergence, showing that the conserved quantities are the
{\it total} energy-momentum and angular momentum of the charges {\it plus}
the electromagnetic field. Alternatively, this can be seen as
direct proof that part of the energy-momentum of the charges can be
transferred to the field, and viceversa. This is what I call 
``exchange'' in the second question asked above and the title of this
section.

Let us state, from the beginning, that the answer to this second question is
positive, as is going to be shown explicitly in several cases. It is important
to remark that, in order to find such mixed conserved currents, the
corresponding field equations must be used. But, of course, the field equations
will depend on the particular theory one is using (higher-order Lagrangians,
scalar-tensor, string, dilaton, General Relativity,....), and thus the seeking
of the conserved currents must be performed in each of these theories separately.
In this paper, only the traditional case of Einstein's equations (in arbitrary
dimension, units chosen such that $8\pi G=c=1$) will be considered from now on.
In particular, under this hypothesis, the interchange of super-energy between
the scalar and the gravitational fields will be proved in full generality by
constructing explicitly a divergence-free current from the sum of the
generalized Bel tensor (\ref{Bel}) and the s-e tensor (\ref{sesm}) of the
scalar field. Similarly, the exchange of super-energy between the
electromagnetic and the gravitational fields will be shown by considering the
propagation of discontinuities of Einstein-Maxwell fields. Nevertheless, before 
entering into this, let us make some remarks about other traditional
treatments of the problem and the reasons of their failure.

\subsection{Non-validity of general divergence-free tensors}
The typical way to attack the problem has usually been the seeking of
divergence-free tensors generalizing the Bel and Bel-Robinson tensors.
Thus, soon after the publication of the Bel tensor, Sachs \cite{Sa} found a
four-index tensor $T'{}^{\alpha\beta\lambda\mu}$ in General Relativity ($n=4$)
which is identically divergence-free. This tensor was later generalized
by Collinson \cite{Col}, who found the most general divergence-free tensor
quadratic in the Riemann tensor and linear in its second derivatives if $n=4$ .
These results can be generalized to arbitrary $n$, see \cite{Rob} (see also
\cite{BL} for related results). Unfortunately, these tensors do not satisfy
any of the good mathematical properties of the s-e tensors (\ref{set})
herein presented, and their completely timelike component is not 
positive definite in general, nor they satisfy the DSEP. Furthermore, despite
being divergence-free, all these tensors do {\it not} have any symmetry
property, and this
invalidates them as candidates to produce conserved quantities. In order
to have a conserved quantity one needs a divergence-free {\it vector}, so
that Gauss' theorem is applicable. When one has a divergence-free
tensor, in principle no conserved current can be built, unless there
exists an auxiliary object which permits to find a true vector without
divergence. Of course, this auxiliary object is usually a Killing vector
(or some generalization thereof, such as conformal Killing vectors, Killing
tensors, etcetera). The idea was explained in some detail when deriving
(\ref{divems}). As remarked there, the symmetry property of the
divergence-free tensor is crucial. Imagine, for instance, that we take a
tensor such as that of Sachs (or Collinson), which is divergence-free
\b*
\nabla_{\alpha}T'{}^{\alpha\beta\lambda\mu}=0
\e*
in the {\it first} index, but such that this first index has no symmetry
properties with the other indices (in fact, Sachs' tensor is symmetric
in the last three indices, $T'{}^{\alpha\beta\lambda\mu}=
T'{}^{\alpha(\beta\lambda\mu)}$, but it has no symmetry involving $\alpha$).
Then, when seeking for a current $\vec{j}$, and if there is at least one
Killing vector $\vec{\xi}$, one could try
\b*
j_{\alpha}\equiv T'{}_{\alpha\beta\lambda\mu}\xi^{\beta}\xi^{\lambda}\xi^{\mu}
\e*
and then a direct computation gives, using that $T'$ has no divergence and
that $\vec{\xi}$ satisfies (\ref{kil})
\b*
\nabla_{\alpha}j^{\alpha}=3T'{}^{\alpha\beta\lambda\mu}\left(\nabla_{\alpha}
\xi_{\beta}\right)\xi_{\lambda}\xi_{\mu}=
3T'{}^{[\alpha\beta]\lambda\mu}\left(\nabla_{\alpha}
\xi_{\beta}\right)\xi_{\lambda}\xi_{\mu}\neq 0
\e*
which is not zero in general, because the tensor
$T'{}_{\alpha\beta\lambda\mu}$ is {\it not} symmetric in $(\alpha\beta)$.
The use of the completely symmetric part of $T'$ (so that
$j_{\alpha}\equiv
T'{}_{(\alpha\beta\lambda\mu)}\xi^{\beta}\xi^{\lambda}\xi^{\mu}$)
is of no help here, because the divergences of $T'{}_{\alpha\beta\lambda\mu}$
in any index other than the first is not zero.
In summary, this type of tensors do not have the required mathematical
properties which have been shown to hold in general for the s-e tensors 
(\ref{set}) and, moreover, they may not be useful regarding the
existence of conserved s-e quantities, which was the reason why they were
constructed in the first place.

Another example of this type is given by a four-index tensor presented
in spinor form in \cite{PR} as the s-e tensor of the electromagentic field which,
when combined with the Bel-Robinson tensor, was claimed to provide
the appropriate s-e tensor for the case of Einstein-Maxwell spacetimes for
$n=4$, in General Relativity. The explicit expression of this tensor is
\b*
P_{\alpha\beta\lambda\mu}\equiv 8\left(\nabla_{(\beta}F^{\rho}{}_{\lambda}
\nabla_{\mu)}F_{\rho\alpha}+\nabla_{\rho}F_{\alpha(\beta}
\nabla_{\lambda}F_{\mu)}{}^{\rho}-\frac{}{}\right.\hspace{7mm}\\
\left.-\frac{1}{4}g_{\alpha(\beta}\nabla_{\lambda}F^{\rho\sigma}
\nabla_{\mu)}F_{\rho\sigma}+\frac{1}{2}\nabla_{\sigma}F_{\rho(\beta}
\nabla^{\rho}F^{\sigma}{}_{\lambda}g_{\mu)\alpha}-\frac{1}{2}
\nabla_{\rho}F_{\sigma\alpha}\nabla^{\sigma}F^{\rho}{}_{(\beta}g_{\lambda\mu)}
\right)
\e*
which has the properties
\b*
P_{\alpha\beta\lambda\mu}=P_{\alpha(\beta\lambda\mu)}, \hspace{3mm}
P_{\alpha\beta\rho}{}^{\rho}=0 , \hspace{5mm}
\nabla_{\alpha}\left({\cal T}^{\alpha\beta\lambda\mu}+
P^{\alpha\beta\lambda\mu}\right)=0\hspace{3mm} \mbox{in $n=4$}
\e*
where in the last one the Einstein-Maxwell equations must be used and ${\cal T}$
is the Bel-Robinson tensor (\ref{BR*}). Nevertheless,
the tensor ${\cal T}+P$ has non-vanishing divergences in the last three indices,
and there are no index symmetries involving the first index, so that it does not
give divergence-free currents for {\it general} Killing vectors. Thus,
for instance, $P$ does not provide conserved currents for the rotational
Killings in {\it flat} spacetime, in contrast to what happens with (\ref{seF}).
The tensor $P_{\alpha\beta\lambda\mu}$ is quadratic in the first derivatives
$\nabla_{[1]}F_{[2]}$, but nonetheless it is not included in the
six-parameter family $\bbb{E}_{\alpha\beta\lambda\mu}$ given in (\ref{c16F}),
not even allowing for negative constants $c_1,\dots,c_6$ in (\ref{c16F}),
as can be checked. Actually, the tensor $P_{\alpha\beta\lambda\mu}$ does not
satisfy the DSEP nor the vanishing of its timelike component (which is
non-negative) implies that $\nabla_{[1]}F_{[2]}$ is zero. All in all, it is
doubtful that $P$ be a good s-e tensor for the electromagnetic field and,
in principle, the tensor (\ref{seF}) (or its generalization $\bbb{E}$) is
preferable.

Now that we know the reasons why the general divergence-free tensors may not
be good enough for the goal of s-e interchange, let us come back to the
question of how to achieve this interchange and the conserved s-e quantities
involving two fields.

\subsection{Exchange of super-energy between scalar and
gravitational fields: general divergence-free currents}
In this subsection an important theorem will be proved: if the
Einstein-Klein-Gordon equations hold, then the sum of the s-e tensors
(\ref{Bel}) and (\ref{sesm}) always provide conserved currents if there
is a Killing vector \cite{S2}. To that end, assume that there exists a
(generally massive) scalar field $\p$ in the spacetieme $(V_n,g)$ and that the
Einstein field equations
\be
R_{\mu\nu}=\nabla_{\mu}\phi\nabla_{\nu}\phi +
\frac{1}{n-2}m^2\phi^2g_{\mu\nu}
\label{ric}
\ee
hold, from where one can deduce the Klein-Gordon equation (\ref{KGm}).
Then, the matter current appearing in (\ref{divbel}) is
\b*
J_{\lambda\mu\beta}=\nabla_{\beta}\nabla_{\lambda}\p\,
\nabla_{\mu}\p-\nabla_{\beta}\nabla_{\mu}\p\, \nabla_{\lambda}\p+
\frac{2}{n-2} m^2\p
\left(g_{\beta\mu}\nabla_{\lambda}\p -g_{\beta\lambda}\nabla_{\mu}\p\right)
\e*
so that equation (\ref{divbel}) becomes in this case
\bea
\nabla_{\alpha}B^{\alpha}{}_{\beta\lambda\mu}=2\nabla_{\sigma}\p
\nabla^{\rho}\nabla_{(\lambda}\p R^{\sigma}{}_{\mu)\rho\beta}-
g_{\lambda\mu}\nabla_{\sigma}\p\nabla^{\rho}\nabla^{\tau}\p
R^{\sigma}{}_{\tau\rho\beta}+2\nabla^{\sigma}\nabla^{\rho}\p
R_{\beta\rho\sigma(\lambda}\nabla_{\mu)}\p -\nonumber\\
-\frac{2}{n-2} m^2\p\left[\frac{}{}
2\nabla_{\sigma}\p R^{\sigma}{}_{(\lambda\mu)\beta}-
2\nabla_{\beta}\p\nabla_{\lambda}\p\nabla_{\mu}\p-\nonumber\right.\\
\left. -\frac{2}{n-2} m^2\p^2 g_{\beta(\lambda}
\nabla_{\mu)}\p +g_{\lambda\mu}\nabla_{\beta}\p\left(\!
\nabla_{\rho}\p \nabla^{\rho}\p +\frac{1}{n-2}m^2\p^2\right)\right].
\label{divEKG1}
\eea
Similarly, expression (\ref{divsesm}) for the divergence of the basic
super-energy tensor of the scalar field becomes now
\bea
\nabla_{\alpha}{\cal S}^{\alpha}{}_{\beta\lambda\mu}=-2\nabla_{\sigma}\p
\nabla^{\rho}\nabla_{(\lambda}\p R^{\sigma}{}_{\mu)\rho\beta}+
g_{\lambda\mu}\nabla_{\sigma}\p\nabla^{\rho}\nabla^{\tau}\p
R^{\sigma}{}_{\tau\rho\beta}+\nonumber \\
+\left(\!\nabla_{\rho}\p \nabla^{\rho}\p +\frac{1}{n-2}m^2\p^2\right)
\left(2\nabla_{\beta}\nabla_{(\lambda}\p\,\nabla_{\mu)}\p -g_{\lambda\mu}
\nabla_{\beta}\nabla_{\rho}\p\,\nabla^{\rho}\p\right) .\label{divEKG2}
\eea
Therefore, the direct sum of tensors (\ref{Bel}) and (\ref{sesm}) is {\it not}
divergence-free in general. But this is of no importance for our purposes,
as will be proved in what follows.

As remarked in the previous subsection, in order to find a conserved
current one always needs an auxiliary object, usually given by a Killing
vector, {\it independently} of whether the tensor providing the current is
divergence-free or not. Thus, let us assume that ${\vec \xi}$ is a Killing
vector in our spacetime. A classical result \cite{Pa,Sh} ensures then that
the massive scalar field is invariant under $\vec{\xi}$, that is
\be
\xi^{\mu}\nabla_{\mu}\phi =0, \hspace{1cm} \mbox{(if $m\neq 0$)}.\label{xp}
\ee
If the scalar field is massless, then in fact one has
$\xi^{\mu}\nabla_{\mu}\phi =$const.\, see \cite{Pa,Sh}. Whether or not the
mass is zero, it always follows that
\be
\xi^{\beta} \nabla^{\rho}\p\, \nabla_{\beta}\nabla_{\rho}\p  = 0 \label{xpp}
\ee
where (\ref{kil}) and the derivative of (\ref{xp}) have been used. Contracting
(\ref{divEKG1}) and (\ref{divEKG2}) with any three Killing vectors $\vec{\xi}_1,
\vec{\xi}_2,\vec{\xi}_3$ (or with three copies of the same if only one Killing
vector exists) and using (\ref{xp}-\ref{xpp}) one finds
\b*
\xi^{\beta}_1\xi^{\lambda}_2\xi^{\mu}_3\,
\nabla_{\alpha}B^{\alpha}{}_{(\beta\lambda\mu)} 
=\nabla_{\sigma}\phi 
\left(2\nabla_{\rho}\nabla_{(\lambda}\phi
R^{\sigma}{}_{\mu}{}^{\rho}{}_{\beta)}
+g_{(\lambda\mu}R^{\sigma\rho}{}_{\beta)}{}^{\tau}
\nabla_{\rho}\nabla_{\tau}\phi\right)\xi^{\beta}_1\xi^{\lambda}_2\xi^{\mu}_3,\\
\xi^{\beta}_1\xi^{\lambda}_2\xi^{\mu}_3\,
\nabla_{\alpha}{\cal S}^{\alpha}{}_{(\beta\lambda\mu)} 
=-\nabla_{\sigma}\phi 
\left(2\nabla_{\rho}\nabla_{(\lambda}\phi
R^{\sigma}{}_{\mu}{}^{\rho}{}_{\beta)}
+g_{(\lambda\mu}R^{\sigma\rho}{}_{\beta)}{}^{\tau}
\nabla_{\rho}\nabla_{\tau}\phi\right)\xi^{\beta}_1\xi^{\lambda}_2\xi^{\mu}_3
\e*
and, in general, neither of these quantities is zero. However, as is obvious
\b*
\xi^{\beta}_1\xi^{\lambda}_2\xi^{\mu}_3\nabla_{\alpha}\left(
B^{\alpha}{}_{(\beta\lambda\mu)} 
+{\cal S}^{\alpha}{}_{(\beta\lambda\mu)}\right)=0.
\e*
This still does not give a divergence-free vector, but now using again the
symmetry properties of (\ref{Bel}) and (\ref{sesm}) (which reveal themselves
as {\it essential} in this calculation), one can rewrite the last formula as
follows
\b*
0=\xi^{\beta}_1\xi^{\lambda}_2\xi^{\mu}_3\nabla^{\alpha}\left(
B_{(\alpha\beta\lambda\mu)} +{\cal S}_{(\alpha\beta\lambda\mu)}\right)
=\nabla^{\alpha}\left[\left(B_{(\alpha\beta\lambda\mu)} 
+{\cal S}_{(\alpha\beta\lambda\mu)}\right)
\xi^{\beta}_1\xi^{\lambda}_2\xi^{\mu}_3\right]=0
\e*
where (\ref{kil}) has been used. Therefore, the super-energy currents
\be
j_{\alpha}\left(R_{[2],[2]}\leftrightarrow \nabla_{[1]}\nabla_{[1]}\p ;
\vec{\xi}_1,\vec{\xi}_2,\vec{\xi}_3\right)\equiv
\left(B_{(\alpha\beta\lambda\mu)}+{\cal S}_{(\alpha\beta\lambda\mu)}\right)
\xi^{\beta}_1\xi^{\lambda}_2\xi^{\mu}_3 \label{gsc}
\ee
are divergence-free. Observe that only the completely symmetric part of the
tensor is relevant here and thus, coming back to the discussion of section
\ref{sec:general}, perhaps only the symmetric part of the general tensor
(\ref{gset}) is meaningful for physical applications. Notice that the vector
(\ref{gsc}) reduces to the corresponding
conserved current of the scalar field in flat spacetimes and to the
vector (\ref{gc}) if there is no scalar field. Thus, one can consider
the vector field (\ref{gsc}) as the appropriate generalization of those
currents to the case involving both fields at the same time. 
\begin{theo}
\label{gordo}
In any $(V_n,g)$ satisfying the Einstein-Klein-Gordon equations (\ref{ric})
and containing at least a Killing vector, the current (\ref{gsc}) is identically
divergence-free.\N
\end{theo}
As always, this produces conserved quantities by means of the Gauss theorem
and the adequate integrals. The theorem also holds when $m=0$ for the case
of the Killing vectors being either parallel or orthogonal to $\nabla_{[1]}\phi$.
The relevance of Theorem \ref{gordo} is that shows, explicitly, the interchange
of super-energy properties between the gravitational and scalar fields, because
{\it neither}
$B_{(\alpha\beta\lambda\mu)}\xi^{\beta}_1\xi^{\lambda}_2\xi^{\mu}_3$ nor
${\cal S}_{(\alpha\beta\lambda\mu)}\xi^{\beta}_1\xi^{\lambda}_2\xi^{\mu}_3$ are
divergence-free separately in the general mixed case, but each of them reduces
to the corresponding conserved current in the absence of one of the two
fields. 

\subsection{Exchange of super-energy between electromagnetic and
gravitational fields: propagation of discontinuities}
Special cases concerning the s-e interchange for the electromagnetic and
gravitational fields have been considered in \cite{Le,L,WZ}.
Due to the difficulty to handle the general case of an electromagnetic
field with a Killing vector, mainly because the different possibilities arising 
(null or non-null electromagnetic field, with the Killing vector being also a
symmetry of the electromagnetic field or not, etcetera \cite{Coll,Fa,Mc,MW,RT})
must be treated separately, in this paper only the illustrative and important
case of the propagation of discontinuities of the fields will be presented.
This will be enough to prove the interchange of s-e quantities between the
electromagnetic and the gravitational field. A thorough treatment of
the propagation of discontinuities in general (including also other fields)
will be given elsewhere \cite{S3} (see also \cite{S,S2}), with the result that
(super)$^k$-energy quantities and conservation laws arise naturally in this
type of situations where the field has a `wave-front'. 

For our present purposes, let us consider the case when there is an
electromagnetic field $F_{[2]}$ propagating in a background spacetime
so that there is a {\it necessarily null} hypersurface of discontinuity $\sigma$
\cite{Fri,L,S3} (a `characteristic'). Let us denote by $[V]_{\sigma}$ the
discontinuity of any object $V$ across $\sigma$. Using the classical Hadamard
theory \cite{Had,Fri,L} or equivalent procedures \cite{MS,S3}, one proves the
existence of a 1-form $\bm{c}$ defined only on $\sigma$ and such that
\b*
\left[F_{\mu\nu}\right]_{\sigma}=n_{\mu}c_{\nu}-n_{\nu}c_{\mu}, \hspace{5mm}
n_{\rho}c^{\rho}=0
\e*
where $\bm{n}$ is a 1-form normal to $\sigma$. Obviously, given that
$\sigma$ is null, it follows that $\bm{n}$ is null
\b*
n_{\mu}n^{\mu}=0
\e*
so that ${\vec n}$ is in fact a vector tangent to $\sigma$ \cite{MS}. Besides,
$\bm{n}$ cannot be normalized, and thus it is defined up to a transformation
of the form
\be
\bm{n} \hspace{5mm} \longrightarrow \hspace{5mm} \g \bm{n}\, ,
\hspace{15mm} \g >0 \label{free}
\ee
where $\g$ is any positive function defined on $\sigma$. The null curves
tangent to $\vec{n}$ are necessarily null geodesics (`bicharacteristics')
contained in $\sigma$ \cite{Fri,L,S3}
\b*
n^{\mu}\nabla_{\mu}n^{\nu}=\Psi \,  n^{\nu}.
\e*
Obviously, an affine parameter could be chosen, using (\ref{free}), so that
$\Psi $ can be set equal to zero for some $\bm{n}$, but nevertheless the
freedom (\ref{free}) still remains partly. Observe that $\bm{c}$ is also
affected by the freedom (\ref{free}) as follows
\b*
\bm{c} \hspace{5mm} \longrightarrow \hspace{5mm} \frac{1}{\g} \, \bm{c}.
\e*

Let $\bar{g}$ denote the first fundamental form of $\sigma$, which is a 
degenerate metric in the {\it null} $\sigma$ because $\vec{n}$ is an eigenvector
of $\bar{g}$ with zero eigenvalue, \cite{HE,MS,Sc,S0}. Then, 
the second fundamental form of $\sigma$ relative to $\bm{n}$ can be defined
by means of
\b*
K_{ij}\equiv \frac{1}{2} \pounds_{\vec{n}} \, \bar{g}_{ij}
\e*
where $\pounds_{\vec{n}}$ denotes the Lie derivative \cite{Sc} with
respect to $\vec{n}$ within $\sigma$. $K_{ij}$ is a 2-covariant symmetric
tensor intrinsic to the null $\sigma$ and it shares the same degeneracy than
$\bar{g}$, that is, $K_{ij}n^i=0$ \cite{Sc,MS}. Due to this property, even
though there is no inverse metric for $\bar{g}$, one can get the trace
of the second fundamental form by contracting with the inverse of the
metric defined on the quotient spaces defined by the
tangent spaces to $\sigma$ modulus the degeneration vector $\vec{n}$.
This trace will be denoted by $\vartheta$ and has the following interpretation:
let $S\subset \sigma$ be any spacelike cut of $\sigma$, that is,
a spacelike $(n-2)$-surface living in $\sigma$ and orthogonal to $\bm{n}$.
Then, $\vartheta$ is a measure of the expansion of these cuts along the
null generators of $\sigma$ as it can be easily related to the derivative
along $\vec{n}$ of the $(n-2)$-volume element of $S$ \cite{S0}.

Making use of the Maxwell equations (\ref{Max}), the following propagation
law (also called a `transport equation' \cite{Fri}) is found \cite{L,L2,S2,S3}
\b*
n^{\mu}\partial_{\mu}c^2+c^2(\vartheta +2\Psi)=0,\hspace{5mm}
c^2\equiv c_{\mu}c^{\mu}\geq 0.
\e*
This propagation equation ensures that if $c_{\mu}=0$ at any point $x\in \sigma$,
then $c_{\mu}=0$ along the whole null geodesic originated at $x$ and tangent
to $\vec{n}$. In fact, for arbitrary conformal Killing vectors
$\vec{\z}_1,\vec{\z}_2$ (or for two copies of the same if there is only one),
the above propagation law allows to prove that \cite{Fri,S2,S3}
\be
\int_{S} c^2 n_{\mu}n_{\nu}\z^{\mu}_1\z^{\nu}_2\,
\bm{\omega}|_{S} \label{cq}
\ee
are conserved quantities along $\sigma$, where $\bm{\omega}|_S$ is the canonical
volume element $(n-2)$-form of $S$. What is meant by this is that
\b*
\int_{S} c^2 n_{\mu}n_{\nu}\z^{\mu}_1\z^{\nu}_2\,
\bm{\omega}|_{S}=
\int_{\hat{S}} c^2 n_{\mu}n_{\nu}\z^{\mu}_1\z^{\nu}_2\,
 \bm{\omega}|_{\hat{S}}
\e*
for any two spacelike cuts $S$ and $\hat{S}$ of $\sigma$. Of course,
these conserved quantities can be easily related to the energy-momentum
properties of the electromagnetic field because
$T_{\mu\nu}\left\{[F_{[2]}]_{\sigma}\right\}=
c^2 n_{\mu}n_{\nu}$,
so that the energy-momentum tensor of the discontinuity $[F_{[2]}]_{\sigma}$
contracted with the conformal Killing vectors $\vec{\z}_1,\vec{\z}_2$ gives the
function integrated in (\ref{cq}).
Further, this can be even related to the discontinuity of the electromagnetic
energy-momentum tensor, because $\left[T_{\mu\nu}\{F_{[2]}\}\right]_{\sigma}=
T_{\mu\nu}\left\{[F_{[2]}]_{\sigma}\right\}$
if the background is empty. Notice also that the conserved quantities (\ref{cq})
are invariant under the transformation (\ref{free}).

However, the integral (\ref{cq}) is trivial when
$\left[F_{\mu\nu}\right]_{\sigma}=0$, in which case $c_{\mu}=0$.
When this happens, there arises a situation such that
the electromagnetic field vanishes at $\sigma$ but
its derivatives are non-zero there, so that the (super)$^k$-energy tensors
must be relevant according to Proposition \ref{prop:ceroF}. 
Using again Hadamard's classical results, now there exist a 2-covariant
symmetric tensor $B_{\mu\nu}$ and a 1-form $f_{\mu}$ defined only on $\sigma$
such that \cite{Beltesis,B2,G,Ge,Ge2,L,L2,MS,S3,Z}
\b*
\left[R_{\alpha\beta\lambda\mu}\right]_{\sigma}=
4n_{[\alpha}B_{\beta][\mu}n_{\lambda]}, \hspace{5mm} B_{\mu\nu}=B_{\nu\mu}
\hspace{5mm} 2n^{\mu}B_{\mu\beta}+g^{\mu\nu}B_{\mu\nu}n_{\beta}=0, \\
\left[\nabla_{\alpha}F_{\mu\nu}\right]_{\sigma}=2 n_{\alpha} n_{[\mu}f_{\nu]},
\hspace{5mm} n_{\rho}f^{\rho}=0 .\hspace{25mm}
\e*
Again, these objects are affected by the freedom (\ref{free}) according to
\b*
\left(f_{\mu},B_{\mu\nu}\right) \hspace{5mm} \longrightarrow \hspace{5mm}
\frac{1}{\g^2} \, \left(f_{\mu},B_{\mu\nu}\right).
\e*
Assuming that the Einstein-Maxwell equations hold, that is, (\ref{Max})
are valid and the Einstein field equations $R_{\mu\nu}-(R/2)g_{\mu\nu}=
T_{\mu\nu}\{F_{[2]}\}$ hold for the energy-momentum tensor (\ref{emF})
of the electromagnetic field, Lichnerowicz was able to prove the
following propagation laws back in 1960 \cite{L} (see also \cite{G})
\bea
n^{\mu}\partial_{\mu}B^2+B^2\left(\vartheta +4\Psi\right)-
2 \, n^{\sigma}F_{\sigma\rho}B^{\rho\tau}f_{\tau}=0, \hspace{5mm} 
B^2\equiv B_{\mu\nu}B^{\mu\nu}\geq 0, \label{prop1}\\
n^{\mu}\partial_{\mu}f^2+f^2\left(\vartheta +4\Psi\right)+
2n^{\sigma}F_{\sigma\rho}B^{\rho\tau}f_{\tau}=0, \hspace{5mm}
f^2\equiv f_{\mu}f^{\mu}\geq 0 \, .\label{prop2}
\eea
Therefore, the discontinuities of the Riemann tensor and of
$\nabla_{[1]}F_{[2]}$ are mutual sources. This is yet another fact in support
of our definition of (super)$^k$-energies and of the choice of levels:
$R_{[2],[2]}$ and $\nabla_{[1]}F_{[2]}$ are at the same super-energy level,
in this case the first $k=1$. Actually, from (\ref{prop1}) and (\ref{prop2})
one clearly sees that \cite{G,L,L2}
\b*
n^{\mu}\partial_{\mu}\left(B^2+ f^2\right)+\left(B^2+
f^2\right)\left(\vartheta +4\Psi\right)=0
\e*
so that, with the help of any conformal Killing vectors
$\vec{\z}_1,\dots,\vec{\z}_4$ (or four copies of the same one),
the following conserved s-e quantities along $\sigma$ can be built
\be
\int_{S} \left(B^2+ f^2\right)n_{\alpha}n_{\beta}n_{\mu}n_{\nu}
\z^{\alpha}_1\z^{\beta}_2\z^{\mu}_3\z^{\nu}_4\,
 \bm{\omega}|_S \label{em-g}
\ee
in the sense that this integral is independent of the spacelike cut $S$.
The fundamental lesson derived from this important formula is that
{\it both} the electromagnetic and gravitational contributions are necessary,
so that neither the integrals of
$B^2n_{\alpha}n_{\beta}n_{\mu}n_{\nu}
\z^{\alpha}_1\z^{\beta}_2\z^{\mu}_3\z^{\nu}_4$ nor
of $f^2 n_{\alpha}n_{\beta}n_{\mu}n_{\nu}
\z^{\alpha}_1\z^{\beta}_2\z^{\mu}_3\z^{\nu}_4$ are equal for
different cuts $S$ in general.
Another interesting point is that (\ref{em-g}) can be related to super-energy
tensors \cite{G,Ge,Ge2,L,S3}. The generalized Bel tensor (\ref{Bel}) of the
discontinuity $[R_{[2],[2]}]_{\sigma}$ becomes
\b*
T_{\alpha\beta\mu\nu}\left\{[R_{[2],[2]}]_{\sigma}\right\}=
2B^2 n_{\alpha}n_{\beta}n_{\mu}n_{\nu}
\e*
while the s-e tensor (\ref{seF}) of the discontinuity
$[\nabla_{[1]}F_{[2]}]_{\sigma}$ reads
\b*
T_{\alpha\beta\mu\nu}\left\{[\nabla_{[1]}F_{[2]}]_{\sigma}\right\}=
2f^2 n_{\alpha}n_{\beta}n_{\mu}n_{\nu}
\e*
Therefore, the function integrated in (\ref{em-g}) is simply
\b*
\frac{1}{2}\left(T_{\alpha\beta\mu\nu}\left\{[R_{[2],[2]}]_{\sigma}\right\}+
T_{\alpha\beta\mu\nu}\left\{[\nabla_{[1]}F_{[2]}]_{\sigma}\right\}\right)
\z^{\alpha}_1\z^{\beta}_2\z^{\mu}_3\z^{\nu}_4
\e*
which once again proves the interplay between super-energy quantities
of different fields, in this case the electromagentic and gravitational
ones. Observe that the above tensors are completely symmetric in this
case, and that they together with the conserved quantity
(\ref{em-g}) are invariant under the transformation (\ref{free}).

Actually, the exchange of higher-order (super)$^k$-energy quantities can also
be demonstrated using this type of calculations. Again, the integral
(\ref{em-g}) vanishes when $f_{\mu}=0$, $B_{\mu\nu}=0$, or equivalently when
$\left[\nabla_{\alpha}F_{\mu\nu}\right]_{\sigma}=
\left[R_{\alpha\beta\lambda\mu}\right]_{\sigma}=0$.
Thus, the (super)$^2$-energy tensors
are now relevant according to Proposition \ref{prop:ceroF}. 
Under these hypotheses, one can prove that there exist a 2-covariant
symmetric tensor $A_{\mu\nu}$ and a 1-form $Y_{\mu}$ defined only on $\sigma$
such that \cite{S3}
\bea
\left[\nabla_{\rho} R_{\alpha\beta\lambda\mu}\right]_{\sigma}=
4n_{\rho} n_{[\alpha}A_{\beta][\mu}n_{\lambda]},
\hspace{5mm} A_{\mu\nu}=A_{\nu\mu}
\hspace{5mm} 2n^{\mu}A_{\mu\beta}+g^{\mu\nu}A_{\mu\nu}n_{\beta}=0, \label{dd1}\\
\left[\nabla_{\alpha}\nabla_{\beta} F_{\mu\nu}\right]_{\sigma}=
2 n_{\alpha} n_{\beta} n_{[\mu}Y_{\nu]},
\hspace{5mm} n_{\rho}Y^{\rho}=0 .\hspace{25mm}\label{dd2}
\eea
These tensors are affected by (\ref{free}) according to
\b*
\left(Y_{\mu},A_{\mu\nu}\right) \hspace{5mm} \longrightarrow \hspace{5mm}
\frac{1}{\g^3} \, \left(Y_{\mu},A_{\mu\nu}\right).
\e*
Using again the Einstein-Maxwell equations one can prove transport equations
analogous to (\ref{prop1}-\ref{prop2}) \cite{S3}
\bea
n^{\mu}\partial_{\mu}A^2+A^2\left(\vartheta +6\Psi\right)-
2 \, n^{\sigma}F_{\sigma\rho} A^{\rho\tau}Y_{\tau}=0, \hspace{5mm} 
A^2\equiv A_{\mu\nu}A^{\mu\nu}\geq 0, \label{prop1+}\\
n^{\mu}\partial_{\mu}Y^2+Y^2\left(\vartheta +6\Psi\right)+
2n^{\sigma}F_{\sigma\rho} A^{\rho\tau}Y_{\tau}=0, \hspace{5mm}
Y^2\equiv Y_{\mu}Y^{\mu}\geq 0 \, .\label{prop2+}
\eea
Hence, the discontinuities of $\nabla_{[1]}R_{[2],[2]}$ and of
$\nabla_{[1]}\nabla_{[1]}F_{[2]}$ are sources of each other, which is again
in favour of the definition of (super)$^k$-energies and of the choice of levels,
in this case for $k=2$. From (\ref{prop1+}) and (\ref{prop2+})
\b*
n^{\mu}\partial_{\mu}\left(A^2+ Y^2\right)+\left(A^2+
Y^2\right)\left(\vartheta +6\Psi\right)=0
\e*
so that, with the help of any conformal Killing vectors
$\vec{\z}_1,\dots,\vec{\z}_6$ (which can be repeated), one finds {\it mixed}
conserved (super)$^2$-energy quantities along $\sigma$ similar to (\ref{em-g})
and which need {\it both} the electromagnetic and gravitational contributions
again. These new currents can be related to the (super)$^2$-energy tensors of
the discontinuities $[\nabla_{[1]}R_{[2],[2]}]_{\sigma}$ and
$[\nabla_{[1]}\nabla_{[1]}F_{[2]}]_{\sigma}$ because, using
(\ref{sseg}) and (\ref{sseF}) together with (\ref{dd1}-\ref{dd2}), one has
\b*
T_{\alpha\beta\lambda\mu\tau\nu}
\left\{[\nabla_{[1]}R_{[2],[2]}]_{\sigma}\right\}=
4A^2 n_{\alpha}n_{\beta}n_{\lambda}n_{\mu}n_{\tau}n_{\nu},\\
T_{\alpha\beta\lambda\mu\tau\nu}
\left\{[\nabla_{[1]}\nabla_{[1]}F_{[2]}]_{\sigma}\right\}=
4Y^2 n_{\alpha}n_{\beta}n_{\lambda}n_{\mu}n_{\tau}n_{\nu}.
\e*
In consequence, the integration over $\sigma$ of the functions 
\b*
\frac{1}{4}\left(T_{\alpha\beta\lambda\mu\tau\nu}
\left\{[\nabla_{[1]}R_{[2],[2]}]_{\sigma}\right\}+
T_{\alpha\beta\lambda\mu\tau\nu}
\left\{[\nabla_{[1]}\nabla_{[1]}F_{[2]}]_{\sigma}\right\}\right)
\z^{\alpha}_1\z^{\beta}_2\z^{\lambda}_3\z^{\mu}_4\z^{\tau}_5\z^{\nu}_6
\e*
provides the conserved (super)$^2$-energy currents. Notice that this last
expression, as well as the corresponding conserved quantity,
is invariant under the transformation (\ref{free}). Similar results can be
derived for the general (super)$^k$-energy tensors of the discontinuities.

\section{Concluding remarks}
Very briefly, let us simply remark that the Definition \ref{def:set} of basic
s-e tensors has many good properties. From the mathematical point
of view, they provide an appropriate way of defining scalars (the super-energy
densities) and vectors (the s-e flux vectors) with the required
mathematical properties, and especially they always have a 
definiteness property (the dominant property), which is strict for timelike
vectors. Furthermore, the construction (\ref{set}) is, on the one hand,
{\it universal} (valid for any given tensor, in arbitrary dimensions,
independent of any field equations), and thus all these good properties can be
used with full generality; and on the other hand, it is fairly {\it unique}
(apart from the index permutations studied in Section \ref{sec:general}), as
proved by the Lemma \ref{r=1} and the Corollary \ref{uniqueness}. The full
uniqueness can be achieved by simply choosing the completely symmetric part,
which in fact is the only one of relevance in the physical applications
of Section \ref{sec:physics}.

Many mathematical applications are possible, such as i) the study
of the causal propagation of fields and the uniqueness of the solutions, as
noticed in Section \ref{sec:dsep}, see \cite{ACY,BerS,Bo,BS};
ii) the use of s-e estimates to control the behaviour of the
components of any given field and its derivatives \cite{ACY,BerS};
iii) the analysis of the
non-linear stability of given spacetimes $(V_n,g)$, or of any possible
field therein contained, or both of them together. This may serve to improve
and generalize the results of \cite{CK}; iv) the construction of
adimensional positive scalars (such as the $q_1,q_2,q_3$ of subsection
\ref{subsec:grav}) measuring the quality of approximate solutions
\cite{calidad,BS2}, or some physical feature as time-arrows or entropies
\cite{Bon,P6,W3}, or the departure from some well-defined property of a given
field; and many others.

Of greater importance is that Definition \ref{def:set} not {\it only} has
mathematical relevance, but it is also physically very promising. The s-e
tensors can be organized in a hierarchy of (super)$^k$-energy levels, but such
that the tensors belonging to some level $k$ for a physical field can be put in
physical correspondance with the tensors of the same level for another
field. Hence, the traditional super-energy Bel-Robinson tensor can be
compared with the corresponding super-energy tensors of the electromagnetic
or scalar fields. Thus, the correct physical dimensions for any given
(super)$^k$-energy density seem to be those of energy density times $L^{-2k}$.
Furthermore, the analysis of the `strength' of a field at points where its
energy density vanishes but such that every neighbourhood of it 
contains the field needs of the (super)$^k$-energy concepts, and this is
why the s-e Bel-Robinson and Bel tensors arise {\it naturally} in
General Relativity, where the energy density of the gravitational field can be
always made to vanish at {\it any point} by appropriate choice of the reference
system due to the equivalence principle. Analogously, the `wave-fronts', shock
waves and similar discontinuities must be analyzed from the
(super)$^k$-energy viewpoint, specially those propagating causally along
null hypersurfaces \cite{G,Ge,Ge2,L,L2,S3,Z}.

Finally, the most important point is that all the (super)$^k$-energy tensors
give rise to divergence-free currents if the field generating the s-e tensors
is isolated (that is, no other fields interact with it), and these currents can
be {\it combined} to produce {\it divergence-free currents mixing} the s-e
tensors of different fields, as explicitly shown in Section \ref{sec:cons}
in several outstanding cases. Therefore, the interchange of super-energy
properties (in a wide sense: they can be super-momentum, or super-stresses
etcetera) does happen, and the super-energy characteristics can be
transferred from one field to another, such as energy properties do. 
A systematic study of the s-e exchanges between several fields in 
many different theories must be performed, and the explicit expressions
of some of the conserved mixed s-e quantities should
be analyzed in order to gain further insight into the possibilities opened
by this new type of conservation physical laws.

\section*{Acknowledgements}
I am grateful to Llu\'{\i}s Bel, G\"oran Bergqvist, Ruth Lazkoz,
Roy Maartens, Marc Mars, Pierre Teyssandier and
Ra\"ul Vera for their comments and suggestions.

\section*{Appendix}
\setcounter{equation}{0}
\renewcommand{\theequation}{A.\arabic{equation}}
Let $(V_n,g)$ be any orientable $n$-dimensional differentiable manifold with
a metric $g$ of Lorentzian signature (--,+,\dots ,+).
The alternating Levi-Civita symbols are denoted and defined by
\b*
\delta_{\mu_1\dots\mu_n}=\delta_{[\mu_1\dots\mu_n]}, \hspace{1cm}
\delta^{\mu_1\dots\mu_n}=\delta^{[\mu_1\dots\mu_n]}, \hspace{1cm}
\delta_{01\dots n-1}=\delta^{01\dots n-1}=+1,
\e*
and then the canonical volume element $n$-form $\bm{\eta}$ is given by
\b*
\eta_{\mu_1\dots\mu_n}=\eta_{[\mu_1\dots\mu_n]}\equiv
\epsilon \sqrt{\left|\mbox{det}\, g\right|}\, \delta_{\mu_1\dots\mu_n}
\e*
where $\epsilon$ is a sign to be chosen according to the orientation.
From here one can easily deduce
\b*
\eta^{\mu_1\dots\mu_n}=\eta^{[\mu_1\dots\mu_n]}= \epsilon
\frac{\sqrt{\left|\mbox{det}\, g\right|}}{\mbox{det}\, g}\, 
\delta^{\mu_1\dots\mu_n}.
\e*
Therefore, one can write in general
\be
\eta_{\mu_1\dots\mu_n}\eta^{\nu_1\dots\nu_n}=
\frac{\left|\mbox{det}\, g\right|}{\mbox{det}\, g}\delta_{\mu_1\dots\mu_n}
\delta^{\nu_1\dots\nu_n}=-\delta_{\mu_1\dots\mu_n}^{\nu_1\dots\nu_n}
\label{ap:vv}
\ee
where in the last expression the hyperbolic signature has been used and the
so-called Kronecker tensor of order $n$ has been defined
\b*
\delta_{\mu_1\dots\mu_n}^{\nu_1\dots\nu_n}\equiv
\delta_{\mu_1\dots\mu_n}\delta^{\nu_1\dots\nu_n}.
\e*
In fact, one can define the Kronecker tensor of any order $p\leq n$ by means
of the following symbolic definition
\b*
\delta_{\mu_1\dots\mu_p}^{\nu_1\dots\nu_p}\equiv \left|
\begin{array}{ccc}
\delta_{\mu_1}^{\nu_1} & \dots & \delta_{\mu_1}^{\nu_p}\\
\delta_{\mu_2}^{\nu_1} & \dots & \delta_{\mu_2}^{\nu_p}\\
\dots & \dots & \dots \\
\delta_{\mu_p}^{\nu_1} & \dots & \delta_{\mu_p}^{\nu_p}
\end{array}\right|
\e*
so that
\b*
\delta_{\mu_1\dots\mu_p}^{\nu_1\dots\nu_p}=
\delta_{[\mu_1\dots\mu_p]}^{[\nu_1\dots\nu_p]}
\e*
and one immediately gets
\bea
\delta_{\mu_1\dots\mu_p\mu_{p+1}\dots\mu_{p+q}}^{\nu_1\dots
\nu_p\nu_{p+1}\dots\nu_{p+q}}=\frac{(p+q)!}{p!q!}
\delta_{[\mu_1\dots\mu_p}^{\nu_1\dots\nu_p}
\delta_{\mu_{p+1}\dots\mu_{p+q}]}^{\nu_{p+1}\dots\nu_{p+q}}=
\frac{(p+q)!}{p!q!}
\delta_{\mu_1\dots\mu_p}^{[\nu_1\dots\nu_p}
\delta_{\mu_{p+1}\dots\mu_{p+q}}^{\nu_{p+1}\dots\nu_{p+q}]}, \label{ap:dd}\\
\delta_{\mu_1\dots\mu_p\mu_{p+1}\dots\mu_{p+q}}^{\nu_1\dots
\nu_p\mu_{p+1}\dots\mu_{p+q}}=q!\, \delta_{\mu_1\dots\mu_p}^{\nu_1\dots\nu_p},
\hspace{2cm} \label{ap:cK}\\
A^{\{\Omega\}}{}_{\mu_1\dots\mu_p}=
A^{\{\Omega\}}{}_{[\mu_1\dots\mu_p]} \hspace{5mm} \Longrightarrow \hspace{5mm}
\frac{1}{p!}\delta_{\mu_1\dots\mu_p}^{\nu_1\dots\nu_p}
A^{\{\Omega\}}{}_{\nu_1\dots\nu_p}=A^{\{\Omega\}}{}_{\mu_1\dots\mu_p}
\label{ap:dA}
\eea
where $A^{\{\Omega\}}{}_{\mu_1\dots\mu_p}$ denotes any tensor containing
a set of $p$ completely antisymmetric indices $\mu_1\dots\mu_p$.

With these ingredients the fundamental formulas (\ref{**1}-\ref{**2}) can be
proved. As in the beginning of section \ref{sec:EH}, let
$A^{\{\Omega\}}{}_{\mu_1\dots\mu_p}{}_{\nu_1\dots\nu_q}=
A^{\{\Omega\}}{}_{[\mu_1\dots\mu_p]}{}_{[\nu_1\dots\nu_q]}$
denote any tensor with an arbitrary number of indices schematically denoted by
$\{\Omega\}$, plus two sets of $p$ and $q$ {\it completely antisymmetric}
indices $\mu_1\dots\mu_p$ and $\nu_1\dots\nu_q$. Then, forming the duals
with respect to these two sets of indices according to definition (\ref{*})
\b*
A^{\{\Omega\}}{}_{\stackrel{*}{\mu_{p+1}\dots
\mu_n}}{}^{\stackrel{*}{\nu_{q+1}\dots\nu_n}}\equiv
\frac{1}{p!q!}\eta_{\mu_1\dots\mu_n}\eta^{\nu_1\dots\nu_n}
A^{\{\Omega\}}{}^{\mu_1\dots\mu_p}{}_{\nu_1\dots\nu_q}
\e*
and using here (\ref{ap:vv}) and (\ref{ap:dd})
\b*
A^{\{\Omega\}}{}_{\stackrel{*}{\mu_{p+1}\dots
\mu_n}}{}^{\stackrel{*}{\nu_{q+1}\dots\nu_n}}=-\frac{1}{p!q!}
\delta_{\mu_1\dots\mu_n}^{\nu_1\dots\nu_n}
A^{\{\Omega\}}{}^{\mu_1\dots\mu_p}{}_{\nu_1\dots\nu_q}=\\
=-\frac{n!}{p!(n-q)!(q!)^2}
\delta_{[\mu_1\dots\mu_q}^{\nu_1\dots\nu_q}
\delta_{\mu_{q+1}\dots\mu_n]}^{\nu_{q+1}\dots\nu_n}
A^{\{\Omega\}}{}^{\mu_1\dots\mu_p}{}_{\nu_1\dots\nu_q}
\e*
from where, with the help of (\ref{ap:dA}), one arrives at
\b*
A^{\{\Omega\}}{}_{\stackrel{*}{\mu_{p+1}\dots
\mu_n}}{}^{\stackrel{*}{\nu_{q+1}\dots\nu_n}}=-\frac{n!}{p!(n-q)!q!}
A^{\{\Omega\}}{}^{\mu_1\dots\mu_p}{}_{[\mu_1\dots\mu_q}
\delta_{\mu_{q+1}\dots\mu_n]}^{\nu_{q+1}\dots\nu_n}
\e*
which is (\ref{**2}). A similar calculation leads to (\ref{**1}).

Some particular cases of this identity are used in the main text. Let us
start by considering the case in which $p=q$, and let us contract
$\mu_{p+2}\dots\mu_{n}$ with $\nu_{p+2}\dots\nu_{n}$, that is
\b*
A^{\{\Omega\}}{}_{\stackrel{*}{\lambda\mu_{p+2}\dots
\mu_n}}{}^{\stackrel{*}{\mu\mu_{p+2}\dots\mu_n}}=-\frac{n!}{(n-p)!(p!)^2}
A^{\{\Omega\}}{}^{\mu_1\dots\mu_p}{}_{[\mu_1\dots\mu_p}
\delta_{\lambda\mu_{p+2}\dots\mu_n]}^{\mu\mu_{p+2}\dots\mu_n}
\e*
which after use of (\ref{ap:dd}) becomes
\b*
A^{\{\Omega\}}{}_{\stackrel{*}{\lambda\mu_{p+2}\dots
\mu_n}}{}^{\stackrel{*}{\mu\mu_{p+2}\dots\mu_n}}=-\frac{n!}{(n-p-1)!(p!)^2}
A^{\{\Omega\}}{}^{\mu_1\dots\mu_p}{}_{[\mu_1\dots\mu_p}
\delta_{\lambda}^{[\mu}\delta_{\mu_{p+2}\dots\mu_n]}^{\mu_{p+2}\dots\mu_n]}
\e*
and now, with the help of (\ref{ap:cK}), this can be finally written as
\be
\frac{1}{(p-1)!}A^{\{\Omega\}}{}^{\mu\sigma_2\dots\sigma_p}
{}_{\lambda\sigma_2\dots\sigma_p}-
\frac{1}{(n-p-1)!}A^{\{\Omega\}}{}_{\stackrel{*}{\lambda\mu_{p+2}\dots
\mu_n}}{}^{\stackrel{*}{\mu\mu_{p+2}\dots\mu_n}}=\frac{1}{p!}
\delta_{\lambda}^{\mu}A^{\{\Omega\}}{}^{\sigma_1\sigma_2\dots\sigma_p}
{}_{\sigma_1\sigma_2\dots\sigma_p}. \label{ap:form}
\ee
As a particular application of this general formula, we derive
\bea
\frac{1}{(p-1)!}A^{\{\Omega\}}{}_{\lambda\sigma_2\dots\sigma_p,}
{}_{\mu}{}^{\sigma_2\dots\sigma_p}+
\frac{1}{(n-p-1)!}A^{\{\Omega\}}{}_{\stackrel{*}{\lambda\mu_{p+2}\dots
\mu_n},\,\mu}{}^{\stackrel{*}{\mu_{p+2}\dots\mu_n}}=\nonumber\\
=\frac{1}{(p-1)!}\left(A^{\{\Omega\}}{}_{\lambda\sigma_2\dots\sigma_p,}
{}_{\mu}{}^{\sigma_2\dots\sigma_p}+
A^{\{\Omega\}}{}_{\mu}{}^{\sigma_2\dots\sigma_p}
{}_{\lambda\sigma_2\dots\sigma_p}-\frac{1}{p}g_{\lambda\mu}
A^{\{\Omega\}}{}^{\sigma_1\sigma_2\dots\sigma_p}
{}_{\sigma_1\sigma_2\dots\sigma_p}\right)\label{ap:sym}
\eea
which has a righthand side symmetric in $(\lambda\mu)$. Another application
of (\ref{ap:form}) arises contracting $\lambda$ with $\mu$, which leads to
\be
\frac{1}{p!}
A^{\{\Omega\}}{}^{\sigma_1\sigma_2\dots\sigma_p}
{}_{\sigma_1\sigma_2\dots\sigma_p} +
\frac{1}{(n-p)!}A^{\{\Omega\}}{}_{\stackrel{*}{\mu_{p+1}\mu_{p+2}\dots
\mu_n}}{}^{\stackrel{*}{\mu_{p+1}\mu_{p+2}\dots\mu_n}}=0, \label{ap:form2}
\ee
while contraction of $\lambda$ with $\mu$ in (\ref{ap:sym}) gives
\bea
\frac{1}{(p-1)!}
A^{\{\Omega\}}{}^{\sigma_1\sigma_2\dots\sigma_p}
{}_{\sigma_1\sigma_2\dots\sigma_p} +
\frac{1}{(n-p-1)!}A^{\{\Omega\}}{}_{\stackrel{*}{\mu_{p+1}\mu_{p+2}\dots
\mu_n}}{}^{\stackrel{*}{\mu_{p+1}\mu_{p+2}\dots\mu_n}}=\nonumber\\
=\frac{(2p-n)}{p!}
A^{\{\Omega\}}{}^{\sigma_1\sigma_2\dots\sigma_p}
{}_{\sigma_1\sigma_2\dots\sigma_p} \hspace{2cm} \label{ap:tr}
\eea
so that the lefthand side of (\ref{ap:sym}) is traceless in $(\lambda\mu)$ if
$n=2p$.

Consider now any $p$-form $\bm{\S}$ and let us apply (\ref{ap:form}) to the
double $(p,p)$-form given by $\S_{\mu_1\dots\mu_p}\S_{\nu_1\dots\nu_p}$.
One readily obtains
\be
\frac{1}{(p-1)!}\S_{\lambda\sigma_2\dots\sigma_p}\S^{\mu\sigma_2\dots\sigma_p}-
\frac{1}{(n-p-1)!}\S_{\stackrel{*}{\lambda\mu_{p+2}\dots
\mu_n}}\S^{\stackrel{*}{\mu\mu_{p+2}\dots\mu_n}}=\frac{1}{p!}
\delta_{\lambda}^{\mu}\S_{\sigma_1\sigma_2\dots\sigma_p}
\S^{\sigma_1\sigma_2\dots\sigma_p}. \label{ap:pp}
\ee

Let us finally derive some formulas concerning any double $(p,q)$-form
$K_{\mu_1\dots\mu_p,\nu_1\dots\nu_q}=K_{[\mu_1\dots\mu_p],[\nu_1\dots\nu_q]}$.
From (\ref{ap:form}) it follows
\bea
\frac{1}{(n-p-1)!}K_{\stackrel{*}{\alpha\rho_{p+2}\dots\rho_n},\lambda
\sigma_2\dots\sigma_q}K^{\stackrel{*}{\beta\rho_{p+2}\dots\rho_n},\mu
\sigma_2\dots\sigma_q}=\nonumber \hspace{2cm}\\
=\frac{1}{(p-1)!}K^{\beta\rho_2\dots\rho_p,}{}_{\lambda\sigma_2\dots\sigma_q}
K_{\alpha\rho_2\dots\rho_p,}{}^{\mu\sigma_2\dots\sigma_q}-
\frac{1}{p!}\delta_{\alpha}^{\beta}
K_{\rho_1\rho_2\dots\rho_p,\lambda\sigma_2\dots\sigma_q}
K^{\rho_1\rho_2\dots\rho_p,\mu\sigma_2\dots\sigma_q},\label{ap:1}\\
\frac{1}{(n-q-1)!}K_{\alpha\rho_{2}\dots\rho_p,\stackrel{*}{\lambda
\sigma_{q+2}\dots\sigma_n}}K^{\beta\rho_{2}\dots\rho_p,\stackrel{*}{\mu
\sigma_{q+2}\dots\sigma_n}}=\nonumber \hspace{2cm}\\
=\frac{1}{(q-1)!}K_{\alpha\rho_2\dots\rho_p,}{}^{\mu\sigma_2\dots\sigma_q}
K^{\beta\rho_2\dots\rho_p,}{}_{\lambda\sigma_2\dots\sigma_q}-
\frac{1}{q!}\delta_{\lambda}^{\mu}
K_{\alpha\rho_2\dots\rho_p,\sigma_1\dots\sigma_q}
K^{\beta\rho_2\dots\rho_p,\sigma_1\dots\sigma_q},\label{ap:2}
\eea
and double application of (\ref{ap:form}) gives also
\bea
\frac{1}{(n-q-1)!(n-p-1)!}K_{\stackrel{*}{\alpha\rho_{p+2}\dots\rho_n},
\stackrel{*}{\lambda\sigma_{q+2}\dots\sigma_n}}
K^{\stackrel{*}{\beta\rho_{p+2}\dots\rho_n},
\stackrel{*}{\mu\sigma_{q+2}\dots\sigma_n}}=\nonumber\hspace{1cm}\\
=\frac{1}{(p-1)!(q-1)!}\left(K^{\beta\rho_2\dots\rho_p,\mu\sigma_2\dots\sigma_q}
K_{\alpha\rho_2\dots\rho_p,\lambda\sigma_2\dots\sigma_q}
-\frac{1}{q}\delta_{\lambda}^{\mu}
K^{\beta\rho_2\dots\rho_p,\sigma_1\dots\sigma_q}
K_{\alpha\rho_2\dots\rho_p,\sigma_1\dots\sigma_q}\right)-\nonumber\\
-\frac{1}{p!(q-1)!}\delta_{\alpha}^{\beta}\left(
K^{\rho_1\dots\rho_p,\mu\sigma_2\dots\sigma_q}
K_{\rho_1\dots\rho_p,\lambda\sigma_2\dots\sigma_q}
-\frac{1}{q}\delta_{\lambda}^{\mu}
K^{\rho_1\dots\rho_p,\sigma_1\dots\sigma_q}
K_{\rho_1\dots\rho_p,\sigma_1\dots\sigma_q}\right).\hspace{7mm}\label{ap:3}
\eea
These three formulas (\ref{ap:1}-\ref{ap:3}) are useful in finding the
explicit expression of the super-energy tensor of $K_{[p],[q]}$.

Let $L_{\alpha\beta,\lambda\mu}=L_{[\alpha\beta],[\lambda\mu]}$ be any double
$(2,2)$-form (such as the Riemann or Weyl tensors). From (\ref{**2}) one has
\be
L_{\stackrel{*}{\mu_{3}\dots\mu_n}}{}^{\stackrel{*}{\nu_{3}\dots\nu_n}}=
-\frac{n!}{2!(n-2)!2!}
L^{\mu_1\mu_2}{}_{[\mu_1\mu_2}
\delta_{\mu_{3}\dots\mu_n]}^{\nu_{3}\dots\nu_n}\label{ap:slanczos}
\ee
and contracting here $\mu_5\dots\mu_n$ with $\nu_5\dots\nu_n$ (if $n\geq 5$,
of course), the following identity is found
\be
\frac{1}{(n-4)!}
L_{\stackrel{*}{\alpha\beta\rho_{5}\dots\rho_n}}
{}^{\stackrel{*}{\lambda\mu\rho_{5}\dots\rho_n}}+L^{\lambda\mu}{}_{\alpha\beta}
=\hat{L}^{\lambda}{}_{\alpha}\delta^{\mu}_{\beta}-
\hat{L}^{\mu}{}_{\alpha}\delta^{\lambda}_{\beta}-
\hat{L}^{\lambda}{}_{\beta}\delta^{\mu}_{\alpha}+
\hat{L}^{\mu}{}_{\beta}\delta^{\lambda}_{\alpha}\label{ap:glanczos}
\ee
where 
\b*
\hat{L}_{\alpha\lambda}\equiv L_{\alpha\lambda}-\frac{1}{4}\,g_{\alpha\lambda}\,
g^{\rho\sigma}L_{\rho\sigma}, \hspace{2cm} L_{\alpha\lambda}\equiv
L_{\alpha}{}^{\rho}{}_{\lambda\rho}.
\e*
Eq.(\ref{ap:glanczos}) is a generalization of the classical Lanczos identity
\cite{Lan}, and provides a direct relation between the double dual of
$L_{[2],[2]}$, $L_{[2],[2]}$ itself and its traces. Notice that, if the
double $(2,2)$-form $L_{[2],[2]}$ is traceless, then the righthand side
of (\ref{ap:glanczos}) vanishes.
The classical identity of Lanczos' \cite{Lan} is for $n=4$ and simply reads
\be
L_{\stackrel{*}{\alpha\beta}}
{}^{\stackrel{*}{\lambda\mu}}+L^{\lambda\mu}{}_{\alpha\beta}
=\hat{L}^{\lambda}{}_{\alpha}\delta^{\mu}_{\beta}-
\hat{L}^{\mu}{}_{\alpha}\delta^{\lambda}_{\beta}-
\hat{L}^{\lambda}{}_{\beta}\delta^{\mu}_{\alpha}+
\hat{L}^{\mu}{}_{\beta}\delta^{\lambda}_{\alpha},
\hspace{1cm} \mbox{if $n=4$}.\label{ap:lanczos}
\ee

\end{document}